\documentclass[journal]{IEEEtran}
\usepackage{graphicx}
\usepackage{amssymb,amsmath}
\usepackage{enumerate}
\usepackage{multirow}
\usepackage{subfig}
\usepackage{url}
\usepackage{hyperref}
\usepackage{color}

\let\oldnl\nl
\newcommand{\nonl}{\renewcommand{\nl}{\let\nl\oldnl}}

\usepackage{cite}

\newcommand{\bA}{{\boldsymbol A}}

\newcommand{\bP}{{\boldsymbol P}}

\newcommand{\bW}{{\boldsymbol W}}

\newcommand{\bb}{{\boldsymbol b}}

\newcommand{\be}{{\boldsymbol e}}

\newcommand{\bu}{{\boldsymbol u}}
\newcommand{\bv}{{\boldsymbol v}}

\newcommand{\bx}{{\boldsymbol x}}

\newcommand{\by}{{\boldsymbol y}}
\newcommand{\bz}{{\boldsymbol z}}

\newcommand{\E}{{\mathbb E}}
\newcommand{\R}{{\mathbb R}}

\newcommand{\Pb}{{\mathbb P}}

\newcommand{\cN}{{\mathcal{N}}}

\begin{document}

\title{Error Correction Codes for COVID-19 Virus and Antibody Testing: Using Pooled Testing to Increase Test Reliability}





\author{Jirong Yi,
        Myung Cho, Xiaodong Wu,
        Weiyu~Xu,
        and~Raghu~Mudumbai
\thanks{Jirong Yi, Raghu Mudumbai, Xiaodong Wu and Weiyu Xu are from Department of Electrical and Computer Engineering, University of Iowa, Iowa City, IA, 52242.}
\thanks{Myung Cho is from Department of Electrical and Computer Engineering, Penn State Behrend, Erie, PA, 16563.}
\thanks{Email: rmudumbai@engineering.uiowa.edu, weiyu-xu@uiowa.edu}
}

\maketitle
\begin{abstract}

We consider a novel method to increase the reliability of COVID-19 virus or antibody tests by using specially designed pooled testings. Specifically, to increase test reliability,  instead of testing nasal swab or blood samples from individual persons, we propose to test mixtures of samples from many individuals. Group testing has traditionally been used for the purpose of reducing the number of tests required to diagnose a large number of individuals, but, in contrast, the pooled sample testing method proposed in this paper also serves a different purpose: for increasing test reliability and providing accurate diagnoses even if the tests themselves are not very accurate. Our method uses ideas from compressed sensing and error-correction coding to correct for a certain number of errors in the test results. The intuition is that when each individual's sample is part of many pooled sample mixtures, the test results from all of the sample mixtures contain redundant information about each individual's diagnosis, which can be exploited to automatically correct for wrong test results in exactly the same way that error correction codes correct errors introduced in noisy communication channels. While such redundancy can also be achieved by simply testing each individual's sample multiple times, we present simulations and theoretical arguments that show that our method is significantly more efficient in increasing diagnostic accuracy. In contrast to group testing and compressed sensing which aim to reduce the number of required tests, this proposed error correction code idea purposefully uses pooled testing to increase test accuracy, and works not only in the ``undersampling'' regime, but also in the ``oversampling'' regime, where the number of tests is bigger than the number of subjects. The results in this paper run against traditional beliefs that, ``even though pooled testing increased test capacity, pooled testings were less reliable than testing individuals separately.''

\end{abstract}

\section{Introduction}\label{Sec:Introduction}

In the absence of a vaccine to the SARS-CoV-2 coronavirus, the experience of public health authorities in several countries has shown that large-scale shutdowns can (only) be safely ended if a systematic ``test and trace" program \cite{lee_interrupting_2020,salathe_covid-19_2020} is put in place to control the spread of the virus. This, in turn, is predicated on the widespread availability of mass diagnostic testing. However, most countries including the US are currently experiencing a scarcity \cite{ranney_critical_2020} of various medical resources including tests \cite{emanuel_fair_2020}. Pooled sample testing has been proposed as a method for increasing the effective capacity of existing testing infrastructure using the classical method of group testing. However, group testing requires highly accurate test results to be effective; a single false negative test result can potentially cause many infected individuals to be incorrectly diagnosed which could lead to further propogation of the virus. Of couirse, test accuracy can be increased by testing each sample multiple times, but this defeats the main purpose of group testing which is to reduce the number of tests.

In this paper, we propose a more sophisticated version of pooled sample testing that also has the ability to increase the diagnostic accuracy of existing tests even if the individual tests are not highly accurate without requiring an increase in the number of tests. In other words, our proposed method of pooled sample testing can deliver highly accurate diagnostic results for individuals with very low rates of false positives and false negatives, even if the tests themselves are highly error-prone. Our method achieves this using mathematical ideas from the theories of compressed sensing and error-correction coding.

\subsection{Background: COVID-19 virus and antibody tests}

The most common tests for the COVID-19 virus currently used in the US and recommended by the CDC are swab tests. These tests use the Reverse Transcription Polymerase Chain Reaction (RT-PCR) process to selectively amplify DNA strands produced by viral RNA specific to the Covid-19 virus. The RT-PCR process which is considered the gold standard for the detection of mRNA consists of three distinct steps: (1) reverse transcription of RNA into cDNA, (2) selective amplification of a target DNA fragment using the Polymerase Chain Reaction (PCR), and (3) detection of the amplification product. While the simple ``end-point" version of PCR only allows binary detection (presence or absence) of a target RNA sequence, the real-time or quantitative version of the PCR process (qPCR) \cite{gibson_novel_1996} or recent innovation digital PCR (dPCR) also allows the quantification of the RNA i.e. it produces an estimate of the quantity of the RNA material present in the sample \cite{nolan_quantification_2006}.

Some researchers \cite{lamb_rapid_2020} have proposed the Reverse Transcription Loop-Mediated Isothermal Amplification (RT-LAMP) as a potentially cheaper and faster alternative to RT-PCR for swab tests. While we focus on tests based on the RT-qPCR process, the methods proposed in this paper are also compatible with RT-LAMP \cite{schmid-burgk_lamp-seq_2020} and other DNA amplification methods.

The PCR-based virus tests are highly {\it sensitive} (i.e. have low rates of false negatives) as well as {\it specific} (i.e. successfully differentiates between the Covid-19 virus and other pathogens and therefore shows low false positive rates). However, pooled sampling methods require sample dilution and additional preparation that may potentially result in degraded sensitivity as well as specificity.

In addition to tests for an active COVID-19 viral infection, there has also been interest in testing for the presence of antibodies to the COVID-19 virus. The antibody tests might show whether a person in the past was infected with the COVID-19 virus. Virus and antibody tests complement each other.

Antibody tests typically use blood samples (unlike virus tests that use nasal swabs), and can use an enzyme immunoassay process such as ELISA (enzyme-linked immunosorbent assay) \cite{lequin_enzyme_2005}. ELISA's tests typically show high sensitivity; however, some of the early antibody tests that were commercially introduced for COVID-19 may have issues with selectivity \cite{lequin_enzyme_2005}.

\subsection{Group Testing for Increasing Testing Capacity}

One simple method to increase the effective testing capacity is by testing pooled samples of a number of test subjects collectively instead of testing samples from each person individually. In a simple version of this ``group testing" \cite{dorfman_detection_1943} idea, a single negative test result on a pooled sample immediately shows that all individuals in that pool are infection-free. Thus, individual tests only need to be performed when a specific pooled test sample yields a positive test result. When the rate of infection in the population is low, this method allows us to reduce the total number of tests {\it per subject} so the throughput of the existing testing infrastructure is increased \cite{hanel_boosting_2020}. Pooling tests have been successfully used for diagnostic testing for infectious diseases in the past \cite{taylor_high-throughput_2010,arnold_evaluation_2013}.

The current testing bottleneck in the COVID-19 crisis has led to a resurgence of interest in using group testing methods for COVID-19 diagnosis. Specifically, there have been recent studies \cite{sinnott-armstrong_evaluation_2020,shani-narkiss_efficient_2020,hogan_sample_2020} into adapting pooling methods similar to \cite{dorfman_detection_1943} for Covid-19 testing. Preliminary studies on the COVID-19 virus also show that pooling samples \cite{yelin_evaluation_2020} can be effective with existing RT-PCR tests.

In a recent work \cite{yi_low-cost_2020}, we proposed a different approach based on the compressed sensing theory \cite{candes_near_2006,candes_decoding_2005,donoho_compressed_2006} for detection of viruses and antibodies using quantitative PCR test results (for example, from qPCR)  for pooled sample testing. A compressed sensing approach for virus detection was also reported in \cite{ghosh_compressed_2020}. The compressed sensing method can potentially achieve higher efficiencies and better performance than group testing. In fact, group testing can be seen as a special case of the more general compressed sensing method, where the test result is more than just binary.

The basic idea behind the compressed sensing pooled sampling method is to prepare a set of mixtures of several individuals' swab specimens, where the mixtures are carefully chosen to be different from each other in such a way that, under the assumption that only a small fraction of the individual samples have non-zero viral RNA, each individual's diagnostic status can be determined by testing a number of mixtures much smaller than the number of individuals.

\subsection{Pooled Sample Testing for Increasing Testing Accuracy}

Our simulations in \cite{yi_low-cost_2020} show that the compressed sensing method is effective in achieving a significant increase in testing capacity. In this paper, we take this idea further and show that the compressed sensing method can also increase the accuracy of diagnostic tests by taking advantage of redundancy in the pooled sample test results to correct for some number of incorrect test results.

To motivate this idea, consider a population of $N$ individuals. Let $b_i \in \{ 0, 1\},~i=1 \dots N$ represent the infection status of the $i$'th individual in the population i.e. $b_i = 1$ indicates individual $i$ is infected with the virus. The information vector $\bb \doteq [b_1~b_2~\dots~b_N ] \in \{0,1 \}^N$ represents the infection status of the population as a whole.

Let $p$ denote the infection rate in the population: $p \doteq \E \left( \frac{1}{N}\sum_{i=1}^N b_i  \right) $. While the information vector $\bb$ can be represented by the $N$ information bits $b_i,~i=1 \dots N$, an elementary result from information theory shows that the entropy of the information vector is much smaller than $N$ bits, when the infection rate is low:
\begin{align}
h(\bb) &\equiv -Np \log_2(p) - N(1-p) \log_2(1-p) \nonumber \\
	& \ll N,~\mathrm{if}~p \ll 1 \label{eq:entropy1}
\end{align}
where we assumed that each individual in the population independently has a probability $p$ of being infected. The entropy $h(\bb)$ represents the number of bits required to losslessly represent the information in $\bb$.

Thus, (\ref{eq:entropy1}) can be interpreted as a theoretical justification for pooled sample testing: in theory, we only need tests that deliver a total of $N_t = h(\bb)$ bits of information in order to fully recover the infection status $b_i$ of every individual in the pool. If the tests are binary i.e. only indicate positive/negative infection status and are completely error-free, then in theory we can fully diagnose all $N$ individuals with as few as $h(\bb)$ such tests. If the test provide richer non-binary results (e.g. quantification of viral RNA concentration from RT-qPCR tests), in theory the number of tests needed may be much smaller than $h(\bb)$.

In this sense, pooled sample testing methods such as the compressed sensing method, can be thought of as data compression codes. However, the tools of information theory allow us to design codes that have much more powerful capabilities than just lossless data compression. In particular, we can generalize from lossless data compression to codes that can perform
error corrections.   {\it In the context of virus testing, this means a class of pooled sample testing techniques that can achieve accurate diagnostic results even with tests that are individually highly error-prone.}

We show in this paper a class of pooled sample testing methods that do exactly this: increased diagnostic accuracy (error correction) without requiring an increased number of tests. In other words, we demonstrate a method of pooled sample testing that requires no larger number of tests in aggregate, yet delivers more accurate diagnostic results than separately testing each individual.  In this paper, our focus is not on increasing test capacity by reducing the number of required tests, but is instead on \emph{purposefully} creating pooled samples to increase test reliability or to increase tests' robustness against test errors. There are several recognizable distinctions between this work on error correction and recent group testing/compressed sensing works for virus/antibody detection: 1) The main purpose of error correcting pooled testing is to increase test reliability, not to reduce required test numbers as in group testing/compressed sensing, even though this paper's proposed approach provides error correction capability also in the case of using a reduced number of tests; 2) In our proposed error correction codes using pooled testing, the testings can operate in an ``oversampling'' regime where the number of performed tests is larger than the number of subjects; 3) In addition, in error correction codes using pooled testing, we do not necessarily require the prevalence to be low or require the involved signal to be sparse: the signal considered can be a fully dense signal.

 We remark that the results in this paper run against traditional wisdoms that grouping samples together for testing would lower test accuracy or reliability compared with individual separate testings, due to factors such as sample dilutions and pipetting errors, even though pooled testings could increase test capacity. Our results show this conventional wisdom can be wrong, and, in some cases, we can \emph{purposefully} perform pooled testing to significantly increase, rather than decrease,  test accuracy or reliability.

\subsection{Related Works}
 Our work is most related to \cite{candes_decoding_2005}, one of the seminal papers in compressed sensing, which uses linear programming to perform decoding under channel errors. Compared with \cite{candes_decoding_2005}, in this paper,  we purposefully design pooled testing matrices with `0' or `1' elements to increase test reliability,  instead of being given a particular channel (linear transformation) as in \cite{candes_decoding_2005}. In addition, we are working with non-negative signals, which bring additional structures for sensing and inference \cite{khajehnejad_sparse_2011,nonnegativeunique}.  Compared with recent works which aim to boost test robustness against noisy measurements in group testing and compressed sensing \cite{zhu_noisy_2020,petersen_practical_2020}, our work considers not only  ``undersampling'' regime, but also ``oversampling'' regime; not only sparse signals, but also dense signals. We purposefully design/use pooled testing to increase test reliability, instead of trying to increase the reliability of group testing/compressed sensing used in the ``undersampling'' regime mainly for increasing test capacity.

\section{Problem Statement}\label{Sec:ProblemStatement}

In this section, we will give a mathematical formulation of performing robust virus testing through error correction code.  We will focus on describing the idea of error correction code for virus testing through quantitative pooled testing,  even though the idea of error correction code can be extended to traditional qualitative pooled testing. We also focus our description on virus testing, and the mathematical principles extend to antibody testing.


The quantitative modeling of the pooled testing problem requires the application of real-time  quantitative polymerase chain reaction (real-time qPCR) which is built on top of the PCR and conducted in a thermal cycler. The real-time qPCR can give quantitative measurements of the amplified DNA copies by using fluorescent reporters in multiple PCR cycles.  In each PCR cycle, the DNA template can be doubled, and the strength of the signal from fluorescent reporter is proportional to the number of amplified DNA molecules.  One can use the threshold cycle $C_t$, which is  defined as the number of cycles required for the fluorescent signal to cross a threshold,  to determine the quantity of DNAs in the qPCR.

Assume that we get  $n$ samples for $n$ subjects with one sample for each, and we will perform $m $ tests to determine the quantities of COVID-19 viruses in these samples. We denote by $\bx \in [0,\infty)^n$ the quantity of the DNA that can be generated from the subjects' viral RNAs.  In each of the $m$ tests, we will create a pooled sample by mixing the samples from multiple subjects. We use a matrix $\bP \in \{0,1\}^{m \times n}$ to denote the participation of $n$ samples in $m$ tests, i.e. the sample of the $j$-th ($1\leq j \leq n$) subject participates in the $i$-th ($1\leq i \leq m$) test if $P_{ij}=1$, and it will not be used in the $j$-th test if $P_{ij}=0$. This means that the number of $1$'s in the $i$-th column of $P$ is the number of tests that the sample of $i$-th subject will participate in. We will model the allocation of the subjects' samples by a matrix $\bW \in [0,1]^{m\times n}$, and each $W_{ij}$ is the fraction of the $j$-th sample used in the $i$-th test, meaning the sum of each column in $W$ is no bigger than 1. With those setup above, we get a measurement matrix as
\begin{align}\label{Defn:MeasurementMatrix}
\bA:=\bP \odot \bW,
\end{align}
where $\odot$ represents the Hadamard multiplication.

The corresponding $m$ mixed samples will go through $m$  quantitative PCR to quantify the amount of DNAs. Due to potential background noises and gross errors caused by factors such as dilutions, sample and reagent contamination, and operational mistakes, the final quantitative measurements $\by \in \R^m$ from the real-time PCR can be modeled as
\begin{align}\label{Defn:GeneralQuantitativeMeasurement}
	\by = f(\bA \bx) + \bv + \be,
\end{align}
where $f(\cdot):\R^n \rightarrow \R^m$, $\bv \in \R^m$, and $\be \in \R^m$ represent the true signal, the observation noise, and the possible gross error. For example,  in the ideal case   $f(\bA \bx) =A\bx$, we have
\begin{align}\label{Defn:QuantitativeMeasurementsWithSameAmplificationSpeed}
\by = \bA \bx +\be + \bv.
\end{align}
Our goal is to recover the sample measurements $\bx \in [0,\infty)^n$ for $n$ subjects from $m$ tests measurements $ \by \in \R^m$ under possible outliers.


\begin{figure}[!htb]
\centering
\includegraphics[width=0.45\textwidth]{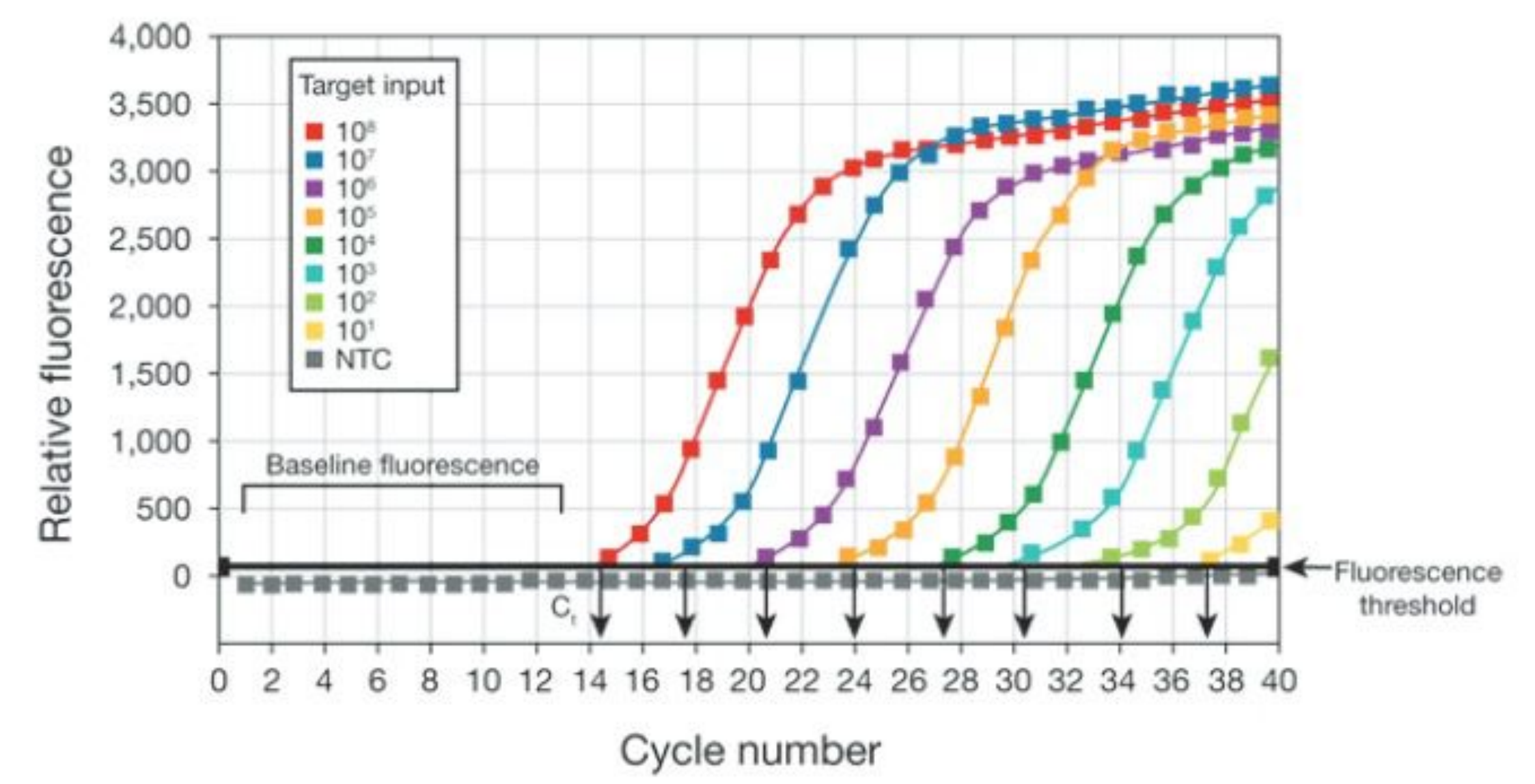}
\caption{Quantitative relation between relative fluorescence and cycle number \cite{scientific_real-time_2014}.}\label{QuatitativeMeasurementCycleNumber}
\end{figure}

In practice, according to \cite{foucart_mathematical_2013,yi_low-cost_2020}, a measurement matrix from the expander bipartite graph can achieve good practical performance with theoretical justifications, and we can specify the matrix $\bP$ as such matrices, i.e., a sparse binary matrix. The sparsity of matrix $\bP$ is determined by taking practical considerations such as reducing the operational complexity of pooling, and reducing the risk of dilution resulting from pooling. Due to the above constraints, one can design the matrix $\bP$ based on bipartite expander graphs \cite{xu_efficient_2007,jafarpour_efficient_2008}. Though one has the freedom to design the allocation matrix $\bW$, one can use an even-allocation scheme. Thus, if the $j$-th subject is involved in $c$ tests, then the $j$-th column of $\bP$ has only $c$ 1's, and the $j$-th column of $\bA$ will have nonzero values at the corresponding locations being $\frac{1}{c}$.

The low prevalence among population in practice allows us to assume that the signal $\bx \in [0,\infty)^n$ is sparse or approximately sparse, i.e., most of its entries are zero (or extremely close to zero). We also assume that that the gross error $\bv \in \R^m$ is sparse, due to relative rareness of operational mistakes,  chemical reaction failure, and sample dilution.

Under all these assumptions, we can formulate the problem of recovering $\bx \in \R^n$ from $\by \in \R^m$ (we remark here that $m$ can be bigger than $n$) as
\begin{align}\label{Defn:L0-L0Minimization}
& {\rm minimize\ } \|\bz\|_0 + \lambda \|\by - \bA\bz - \bu\|_0, \nonumber\\
&{\rm subject\ to\ }\|\bu\|_2 \leq \epsilon,\nonumber\\
&\quad\quad\quad\quad\quad \bz\geq 0,
\end{align}
where $\|\bz\|_0$ is the number of nonzero elements in $\bz$, $\lambda \in \R$ is a tuning parameter for controlling the tradeoff between $\|\bz\|_0$ and $\|\bA\bz - \by - \bu\|_0$, the $\|\bu\|_2$ is the $\ell_2$ norm of $\bu$, $\epsilon\geq 0$ is a parameter controlling the tolerance for noise, and the $\bx \geq 0$ means that every element of $\bx$ is nonnegative. In \eqref{Defn:L0-L0Minimization}, we used $\bz$ as an estimate for $\bx$ and $\bu$ as an estimate for $\bv$, and $\by - \bA\bz - \bu$ is an estimate for $\be$.

Due to combinatorial nature of $\| \cdot \|_0$, solving \eqref{Defn:L0-L0Minimization} is sometimes computationally challenging,  and the norm $\|\cdot\|_1$ can be used as a relaxation technique instead in practice to achieve good performance without much computational complexity \cite{foucart_mathematical_2013,yi_low-cost_2020}. Thus, we can reformulate \eqref{Defn:L0-L0Minimization} as
\begin{align}\label{Defn:L1-L1Minimization}
& {\rm minimize\ } \|\bz\|_1 + \lambda \|\by - \bA\bz - \bu\|_1, \nonumber\\
&{\rm subject\ to\ }\|\bu\|_2 \leq \epsilon,\nonumber\\
&\quad\quad\quad\quad\quad \bz\geq 0,
\end{align}
where $\|\bz\|_1$ is the sum of the absolute value of all the elements in $\bz$, and we will refer \eqref{Defn:L1-L1Minimization} as $\ell_1-\ell_1$ minimization. Once the estimate for $\bx$ is obtained, 
If $\bz_j \geq \tau$ where $\tau$ is the threshold value, then we claim the $j$-th subject is infected and tests ``positive''; otherwise, we declare ``negative'' test result for the subject.

There is a large volume of literature which proposed algorithms for solving \eqref{Defn:L1-L1Minimization} under certain conditions such as the restricted isometry property and the null space condition. These ideas range from using off-the-shelf softwares such as CVX \cite{grant_cvx:_2008}, to algorithms specifically designed for $\ell_1-\ell_1$ minimization such as the homotopy method and iteratively reweighted least square algorithm \cite{foucart_mathematical_2013}. In this paper, we will use the CVX \cite{grant_cvx:_2008}. 

\section{Numerical Experiments}
\label{sec:simulation}
In this section, we conduct numerical experiments in order to evaluate the performance of our proposed method, which is the error correcting pooled testing introduced in \eqref{Defn:L1-L1Minimization}. We focus on models for COVID-19 virus testing. In pooled testing, we use Bernoulli matrices each element of which is either 1 or 0. The numbers of people tested are set to 25 and 40, i.e., $n=25$ and $40$. We consider a scenario where $k$ out of $n$ people have COVID-19 virus by setting randomly chosen $k$ elements in $\bx \in \R^{n}$ to be positive and other $n-k$ elements to zero. The value of each of the non-zero elements in $\bx$ is chosen uniformly at random from  $[5,10]$. We consider the sparsity level $k$ from 1 to 6 in the simulations.

In pooled testing, the Gaussian noise vector $\bv$ in \eqref{Defn:QuantitativeMeasurementsWithSameAmplificationSpeed} is generated i.i.d. according to the Gaussian distribution $\cN(0,\sigma^2)$, where the noise level $\sigma^2$ is varied from 5e-1 to 2e0.  For the outliers, we generate the sparse outlier vector $\be$ as in \eqref{Defn:QuantitativeMeasurementsWithSameAmplificationSpeed}  by having each element of  $\be$ be a non-trivial  (non-zero) outlier  with probability $\Pb_{out}$. 
If an outlier indeed happens at the $i$-th measurement, we generate the outliers for the $i$-the measurement in the following way: 1) If the corresponding $(\bA\bx)_{i}$ is positive,  with 95\% probability,  we set the outlier $\be_{i}$ as $-(\bA\bx)_{i}$  and reset $\bv_{i}=0$ such that $\by_{i}=0$ (this is to simulate a ``false negative'' measurement); with the other 5\% probability, we set the outlier $\be_{i}$ as $5\times q+2$, where $q$ follows the standard Gaussian distribution $\cN(0,1)$,  and keep the originally generated $\bv_{i}$;  2) if the corresponding $(\bA\bx)_{i}$ is equal to 0,   we will set  $\be_{i}$ as $5\times |q|+2$, where $q$ follows distribution  $\cN(0,1)$, and keep the originally generated $\bv_{i}$.  If there is no non-trivial outlier for the $i$-the measurement,  and  $(\bA\bx)_{i}=0$, we set $\bv_{i}=0$ and $\be_{i}=0$, to simulate this test as a ``negative'' test revealing 0 virus.   Since the measurement results $\by$ must be a non-negative vector, for the $i$-th test, we let $\by_{i}=\max\{(\bA\bx)_{i}+\be_{i}+\bv_{i},0\}$ to make sure the generated  measurement $\by$ be a non-negative vector (to avoid randomly generated $\be$ and $\bv$ being too small dragging $(\bA\bx)_{i}+\be_{i}+\bv_{i}$ to negative region).  In the numerical evaluations, we respectively consider three probabilities of the outlier error, denoted by $\Pb_{out}$, to be $1\%$, $5\%$, and $15\%$.  For pooled testing, we use \eqref{Defn:L1-L1Minimization} to recover $\bx$, and use threshold $\tau=1$ to decide whether a subject tests positive or negative. We introduce this threshold to suppress the noise in $\bx$ may caused by outlier error and Gaussian noise.

We compare the pooled testing against the individual testing model, where the individual samples of subjects are tested separately (possible multiple times).  In the individual testing, the $i$-th test is modeled as
\begin{align}
\by_i = \bx_{\mod(i-1,n)+1} + \be_i +\bv_i, \;\;i=1,2,...,m
\end{align}
where $\by_i$ is $i$-th measurement result, $\mod(\cdot, \cdot)$  means module operation, $\be_i$ is  the outlier, and $\bv_i$ is Gaussian noise following distribution $\cN(0,\sigma^2)$.   We generate the noises and outliers in the same way as described for the pooled testing, except for in individual testing,  we generate outliers and noises based on $\bx_{\mod(i-1,n)+1}$ instead of $(\bA\bx)_{i}$.  For example, for $n=25$ and $i=27$, $\by_{i}$ is the measurement result for the 2nd subject (this subject has been tested once already in the 2nd test),  and the outlier and noise simulated for the $27$-th test is randomly generated based on $\bx_{2}$.   In individual testing,  if $m < n$,  there must be some subjects who do not get qPCR tested at all; and in those cases, and for our simulations, we consider these subjects as COVID-19 negative. Additionally, in individual testing, if a subject is tested multiple times, and as long as one result is identified as being COVID-19 positive, we consider the subject as COVID-19 positive. This comes from the motivation of not missing COVID-19 positive cases. The number of measurements, denoted by $m$, is varied from 10 to 50 in $n=25$ and from 10 to 80 in $n=40$.  Thus, in our individual testing scenario, the maximum number of tests for a subject is two.

For both the pooled testing and the individual testing, we run 100 random trials for each parameter setup, and record the False Negative Rate (FNR) and the False Positive Rate (FPR), which are computed on average out of 100 trials as follows:\\[-10pt]
\par\noindent\small
\begin{align*}
	& \text{FNR} = \frac{   \text{Number of negative cases in people with COVID-19 virus}  }{\text{Number of people having COVID-19 virus}},\\[5pt]
	& \text{FPR} = \frac{\text{Number of positive cases in people without COVID-19 virus}}{\text{Number of people not having COVID-19 Virus}}. \\[-12pt]
\end{align*}
\normalsize
Hence, the FNR represents the percentage of people identified as COVID-19 negative among people infected with COVID-19 viruses, which can be a critical error in COVID-19 testing.  For the FPR, it is interpreted as the percentage of people who are diagnosed as having COVID-19 virus among people not infected with COVID-19 viruses. The FPR and FNR can happen because of limited sensitivity, limited specificity, sample contamination, sample dilutions, failed chemical reactions, operational mistakes and other factors.  The FPR can be an important concern in COVID-19 antibody testing, while the FNR is an important concern in COVID-19 virus testing.  Lower both FPR and FNR represent the better testing performance in detecting virus and checking antibody. Additionally, if one method (say, Method A) achieves the same FNR and FPR but with a smaller number of tests than another method (say, Method B), then we deem Method A  better than Method B. This is because the number of tests is related to the throughput of testing, and a high-throughput testing allows us to increase the capacity of testing in a limited time. Therefore,  we will compare the FNR and the FPR of the pooled testing against those of the individual testing under various setups of parameters including the number of tests, noise levels and probability of outlier occurrences.

\subsection{Different probabilities of outlier errors}
\begin{figure*}[!htb]
    \centering
   \subfloat[FNR ($\Pb_{out} = 0.01$)]{\includegraphics[scale=0.14]{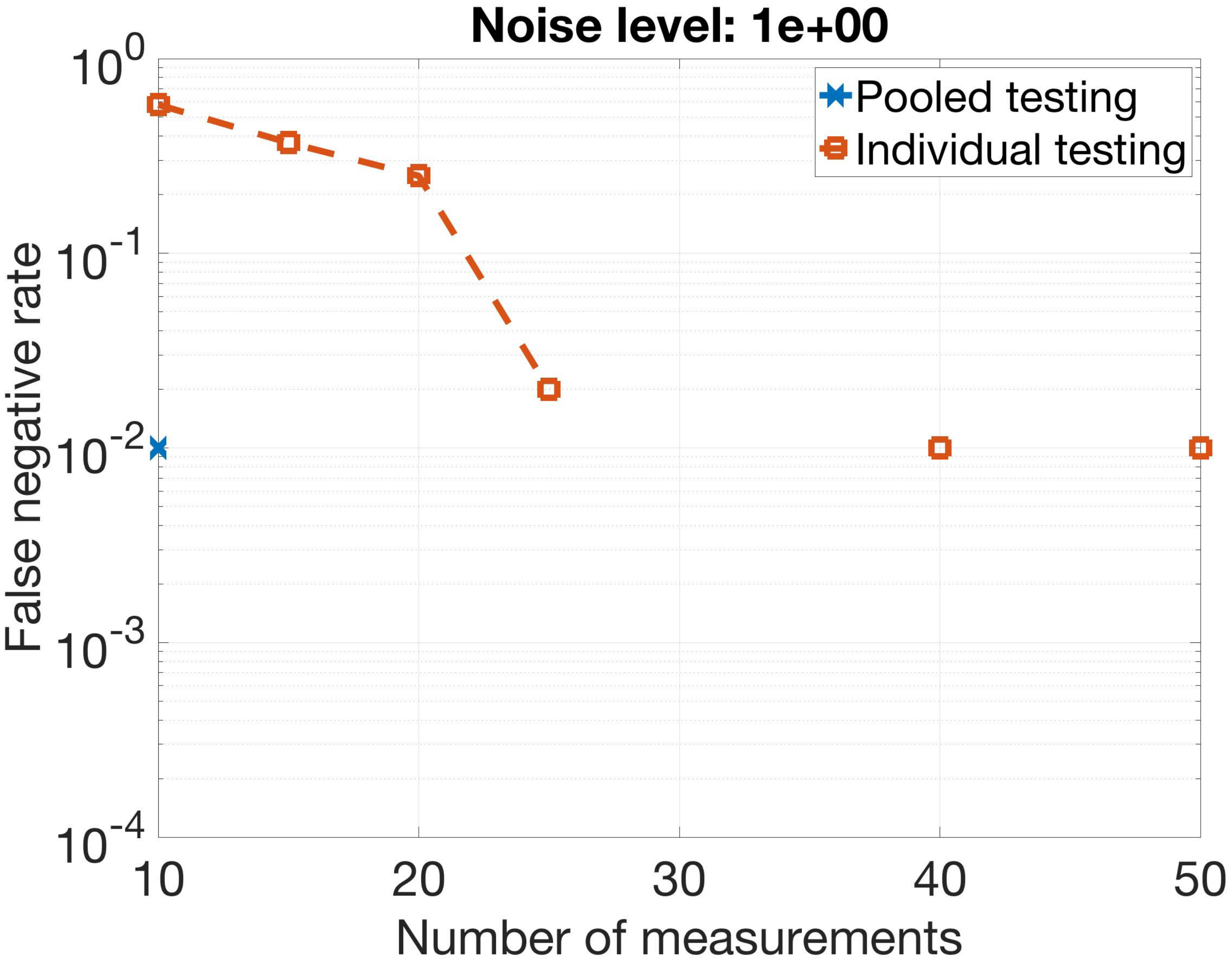}}
   \subfloat[FNR ($\Pb_{out} = 0.05$)]{\includegraphics[scale=0.14]{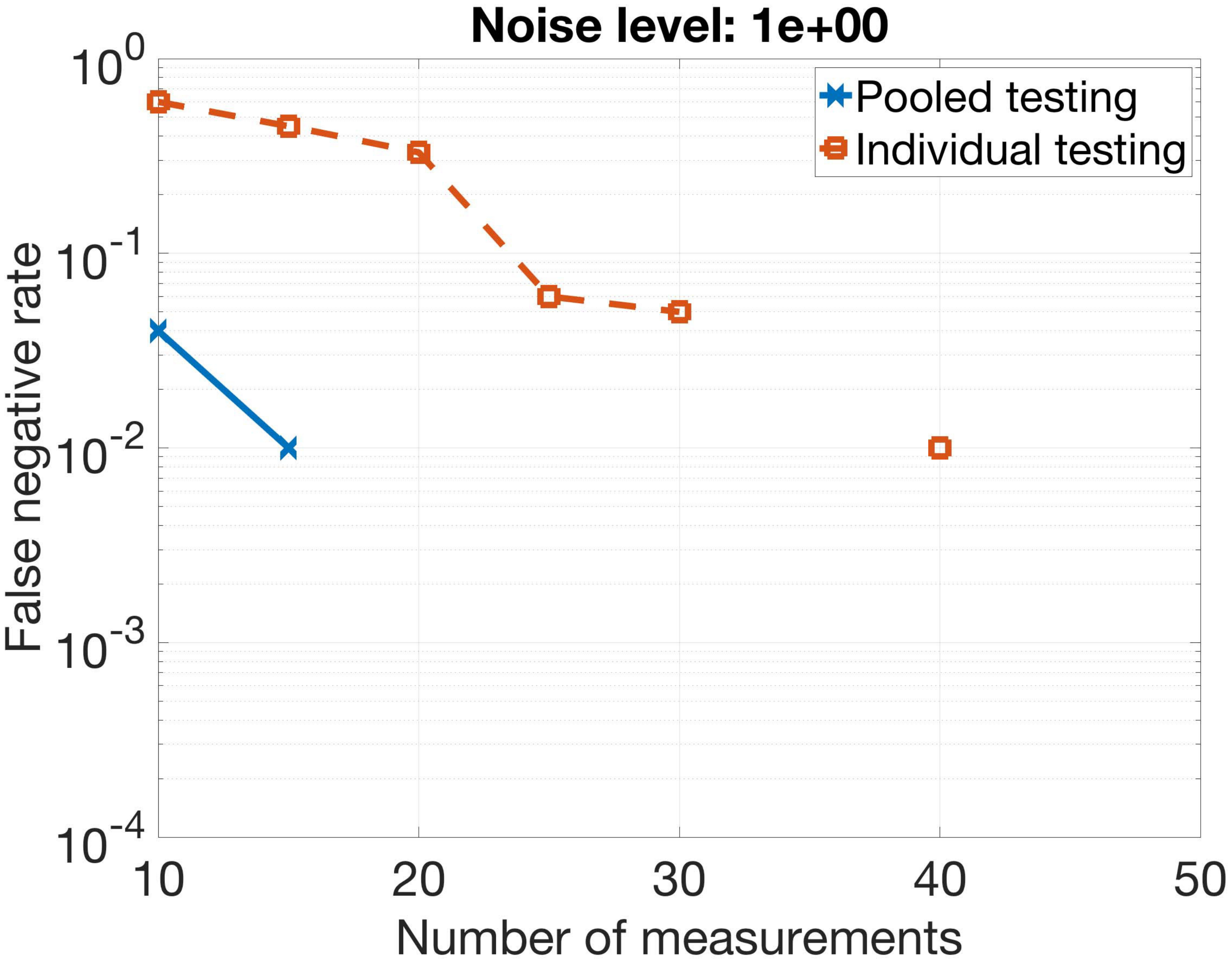}}
   \subfloat[FNR ($\Pb_{out} = 0.15$)]{\includegraphics[scale=0.14]{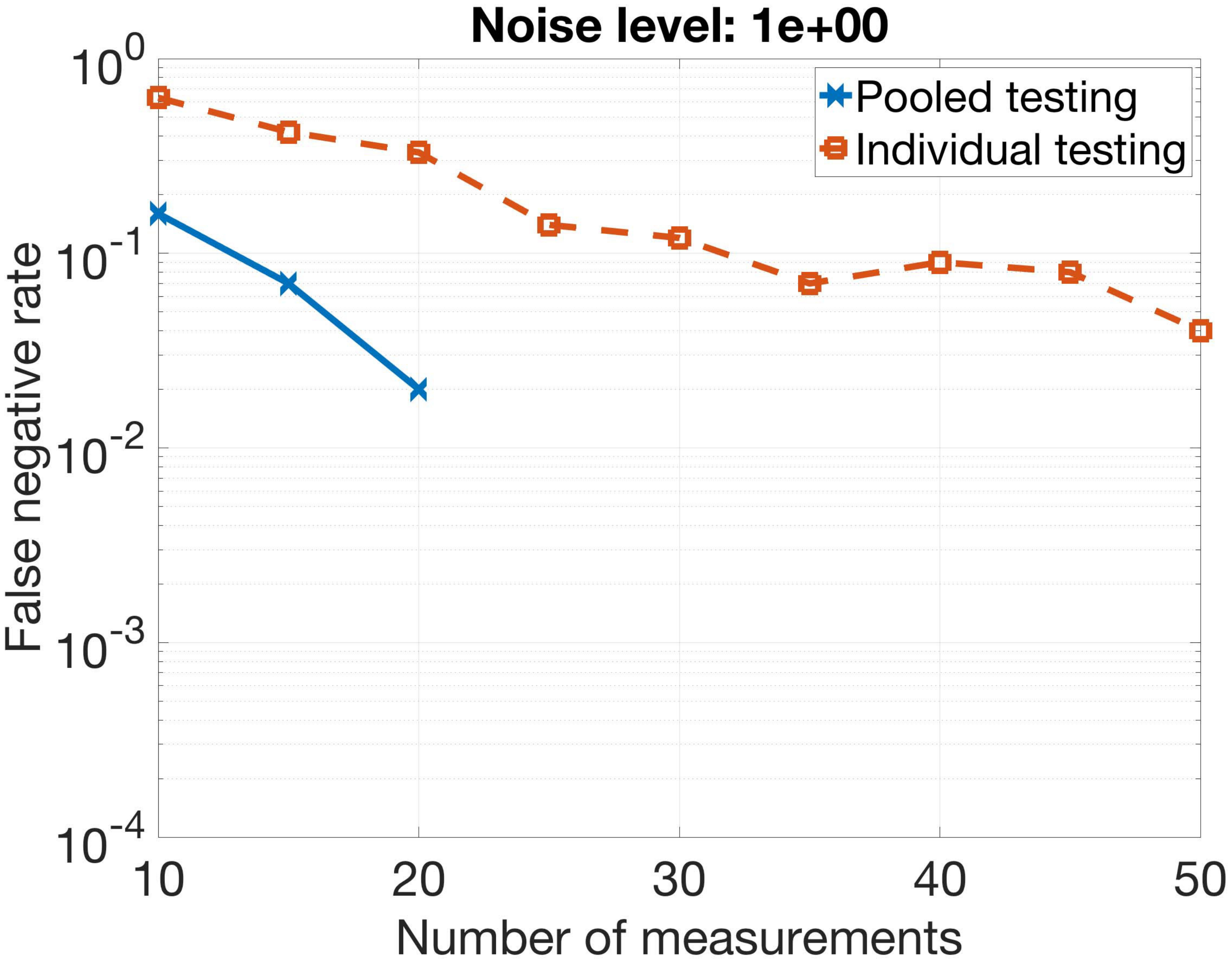}} \\
   \subfloat[FPR ($\Pb_{out} = 0.01$)]{\includegraphics[scale=0.14]{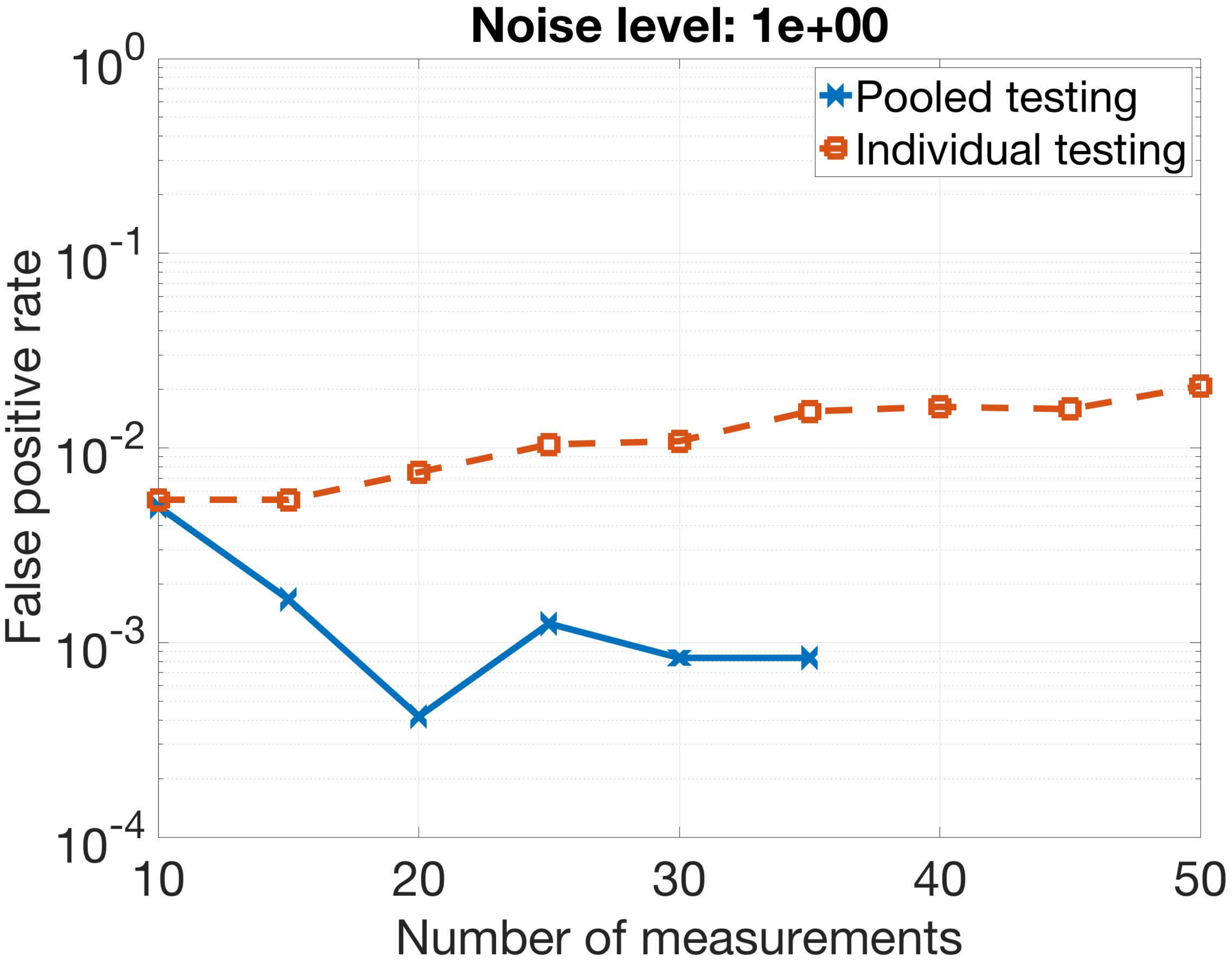}}
   \subfloat[FPR ($\Pb_{out} = 0.05$)]{\includegraphics[scale=0.14]{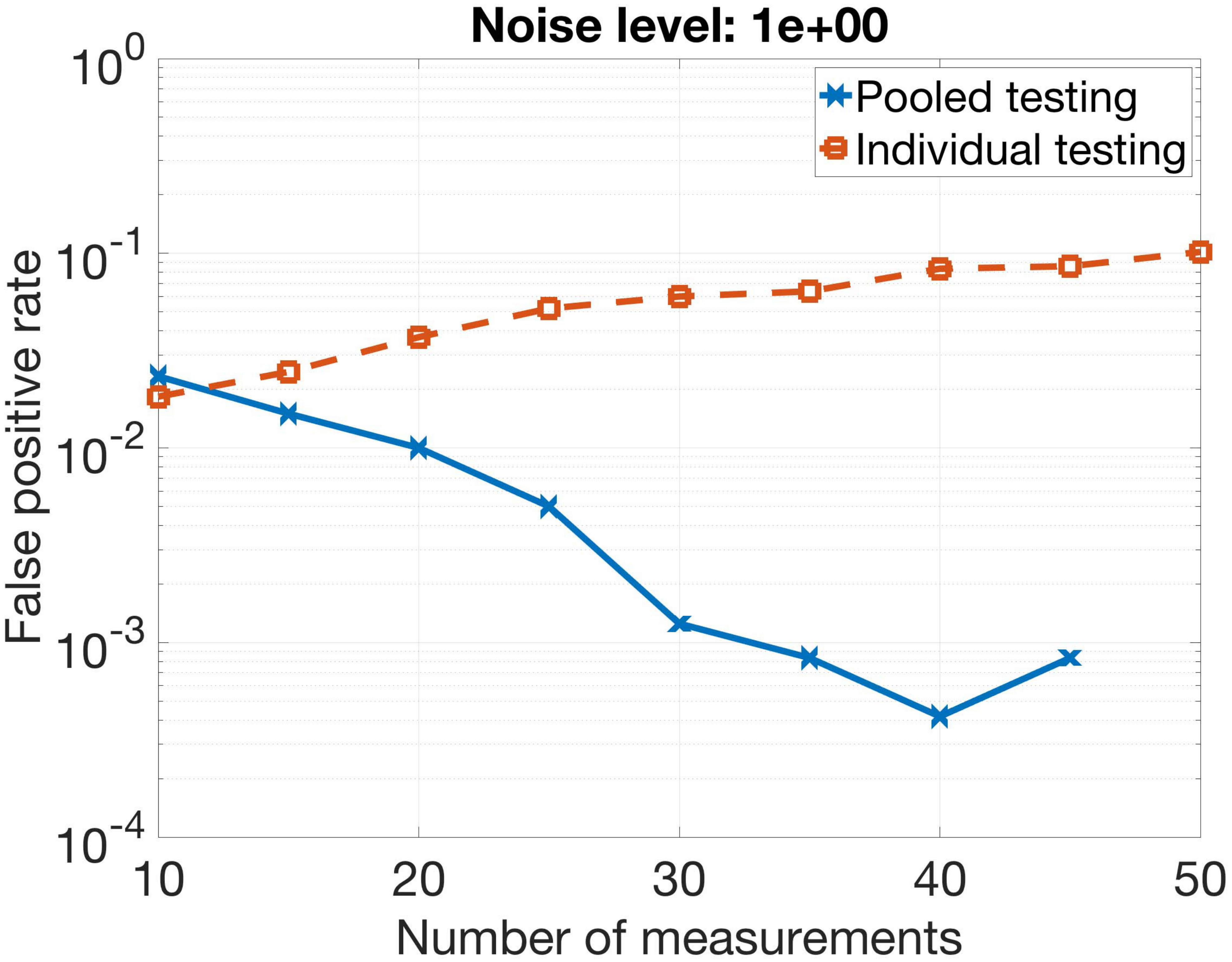}}
   \subfloat[FPR ($\Pb_{out} = 0.15$)]{\includegraphics[scale=0.14]{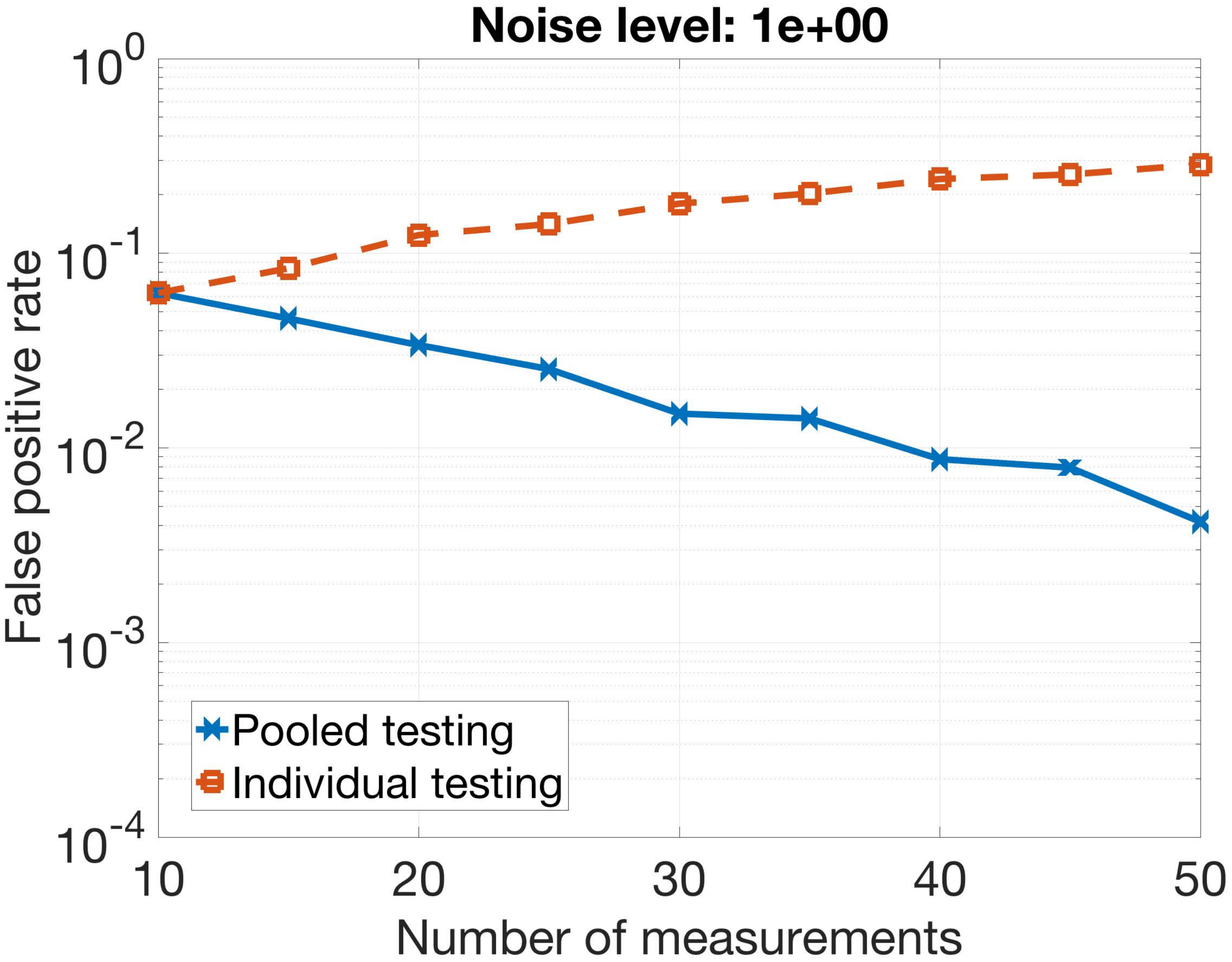}}
    \caption{\small Simulations for different probabilities of outlier errors. False Negative Rate (FNR) and the corresponding False Positive Rate (FPR) with $n=25$, $k=1$, and Gaussian noise level 1e0.}
    \label{fig:FNR_FPR_N25_K1}
\end{figure*}

\begin{figure*}[!htb]
    \centering
   \subfloat[FNR ($\Pb_{out} = 0.01$)]{\includegraphics[scale=0.14]{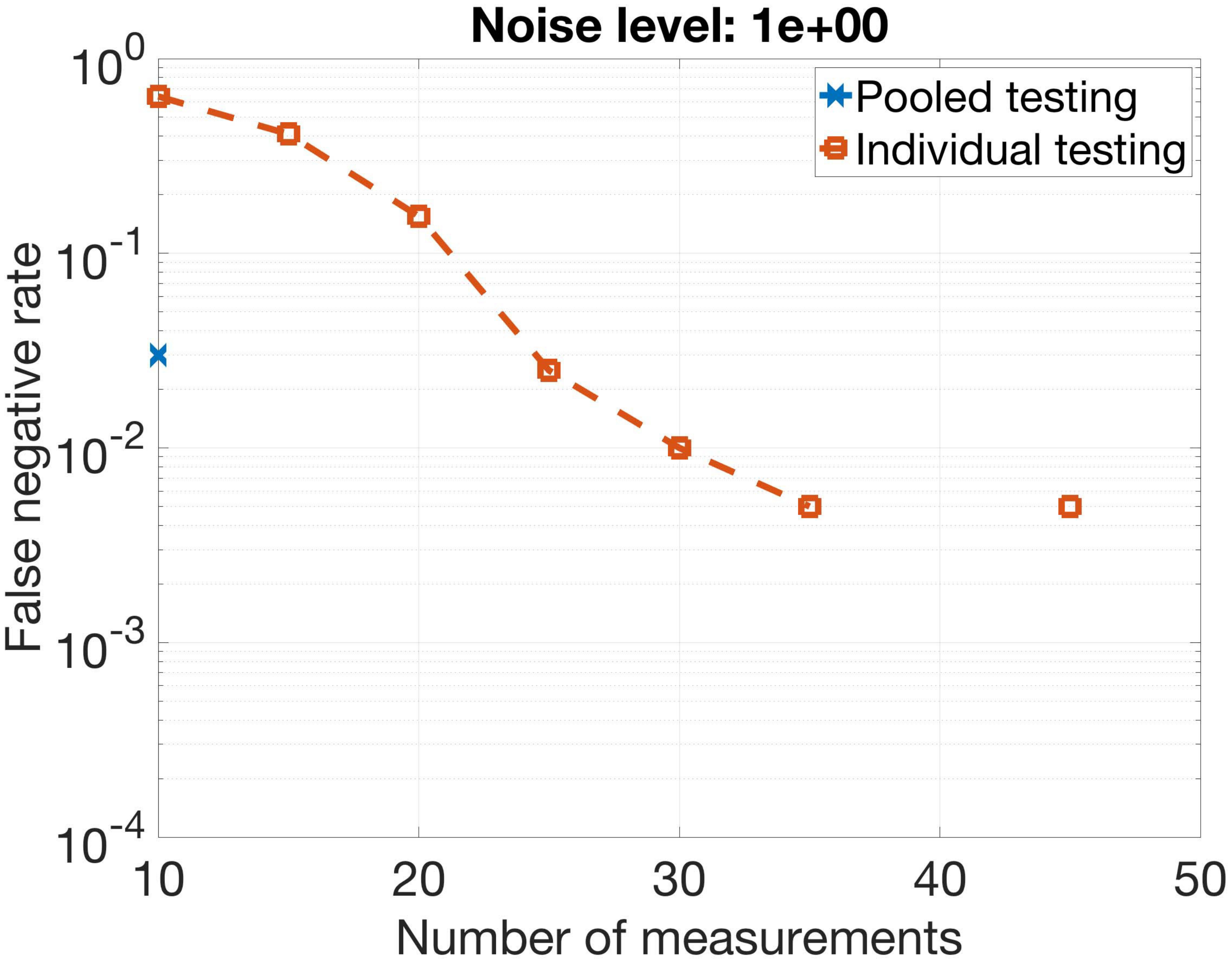}}
   \subfloat[FNR ($\Pb_{out} = 0.05$)]{\includegraphics[scale=0.14]{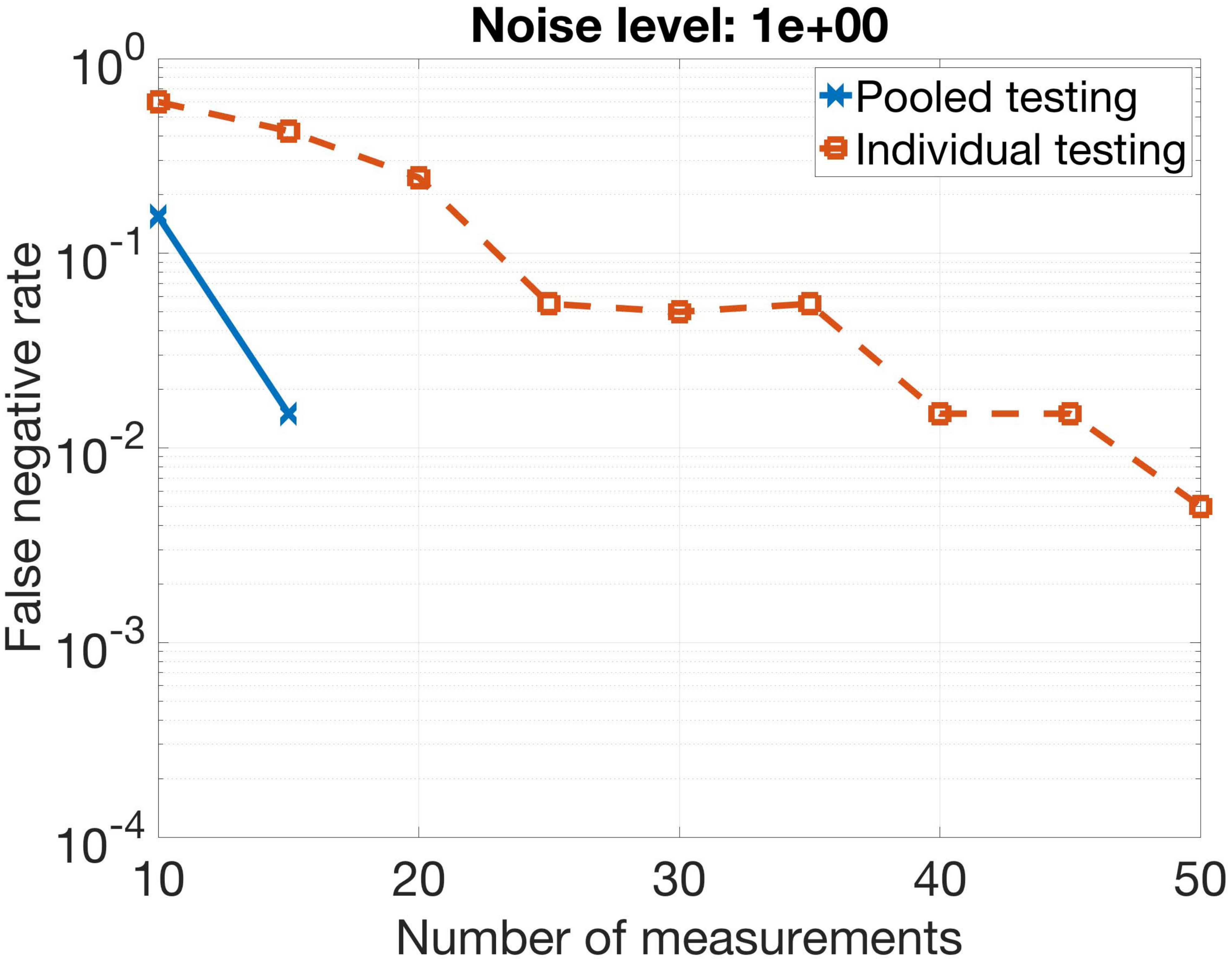}}
   \subfloat[FNR ($\Pb_{out} = 0.15$)]{\includegraphics[scale=0.14]{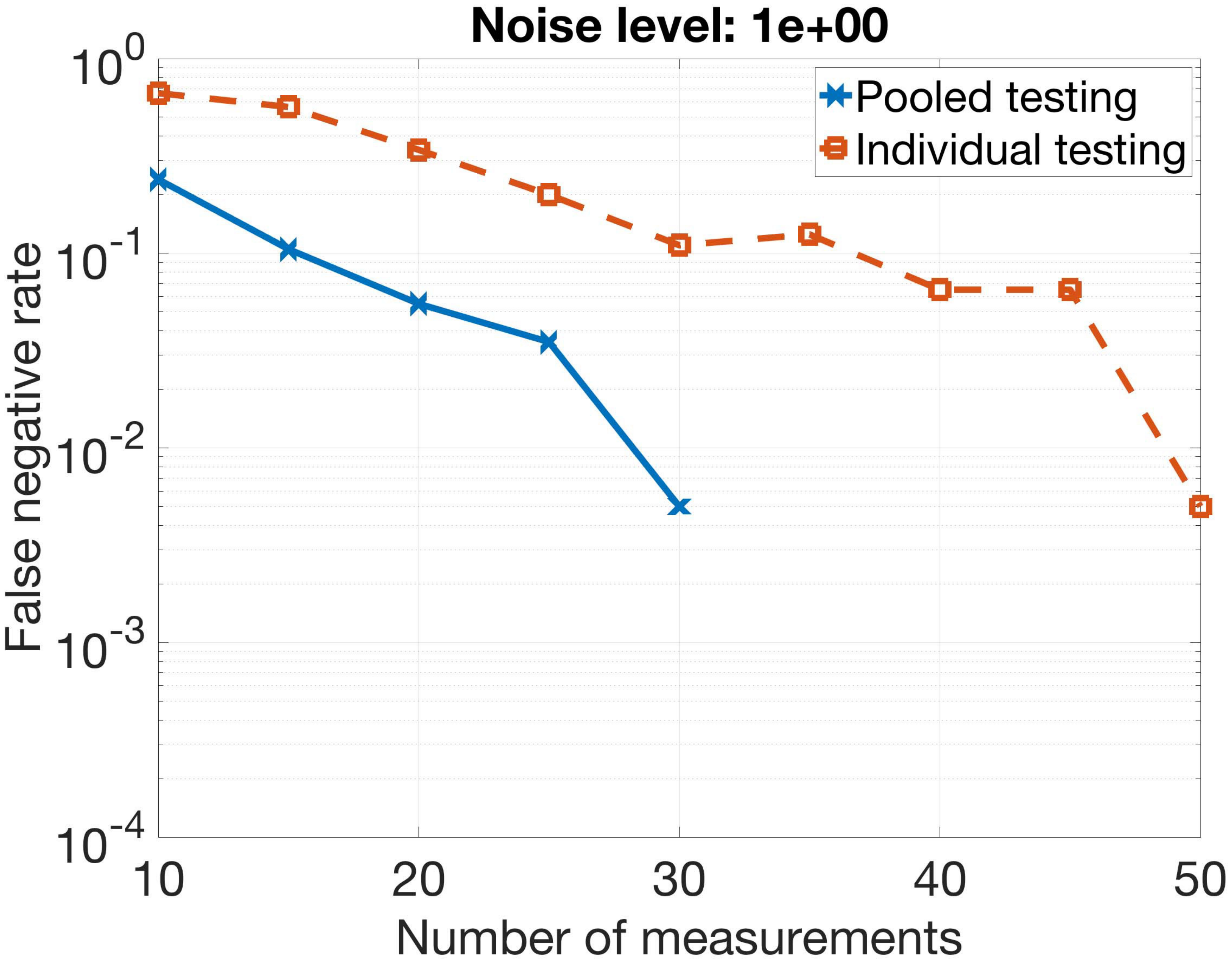}} \\
   \subfloat[FPR ($\Pb_{out} = 0.01$)]{\includegraphics[scale=0.14]{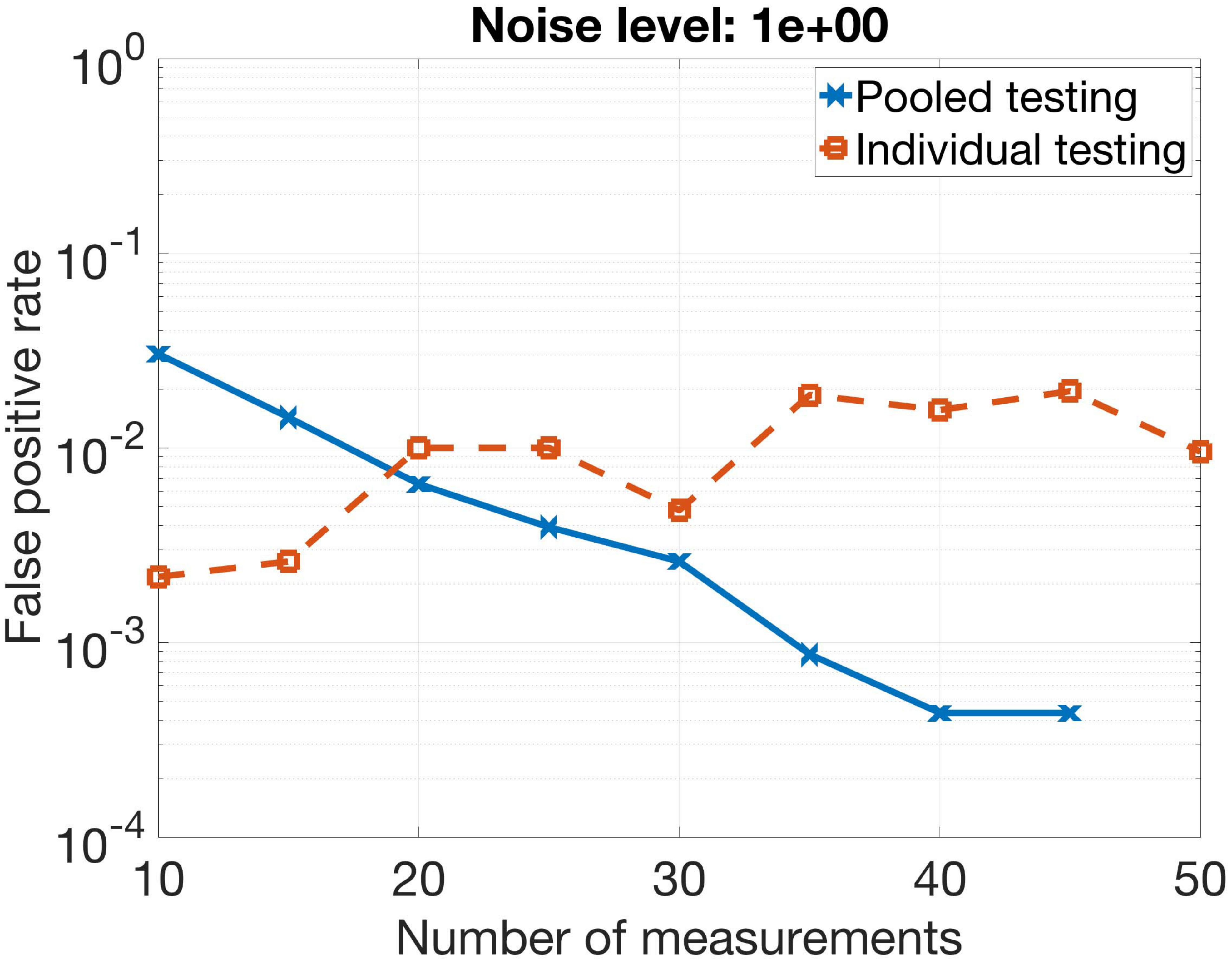}}
   \subfloat[FPR ($\Pb_{out} = 0.05$)]{\includegraphics[scale=0.14]{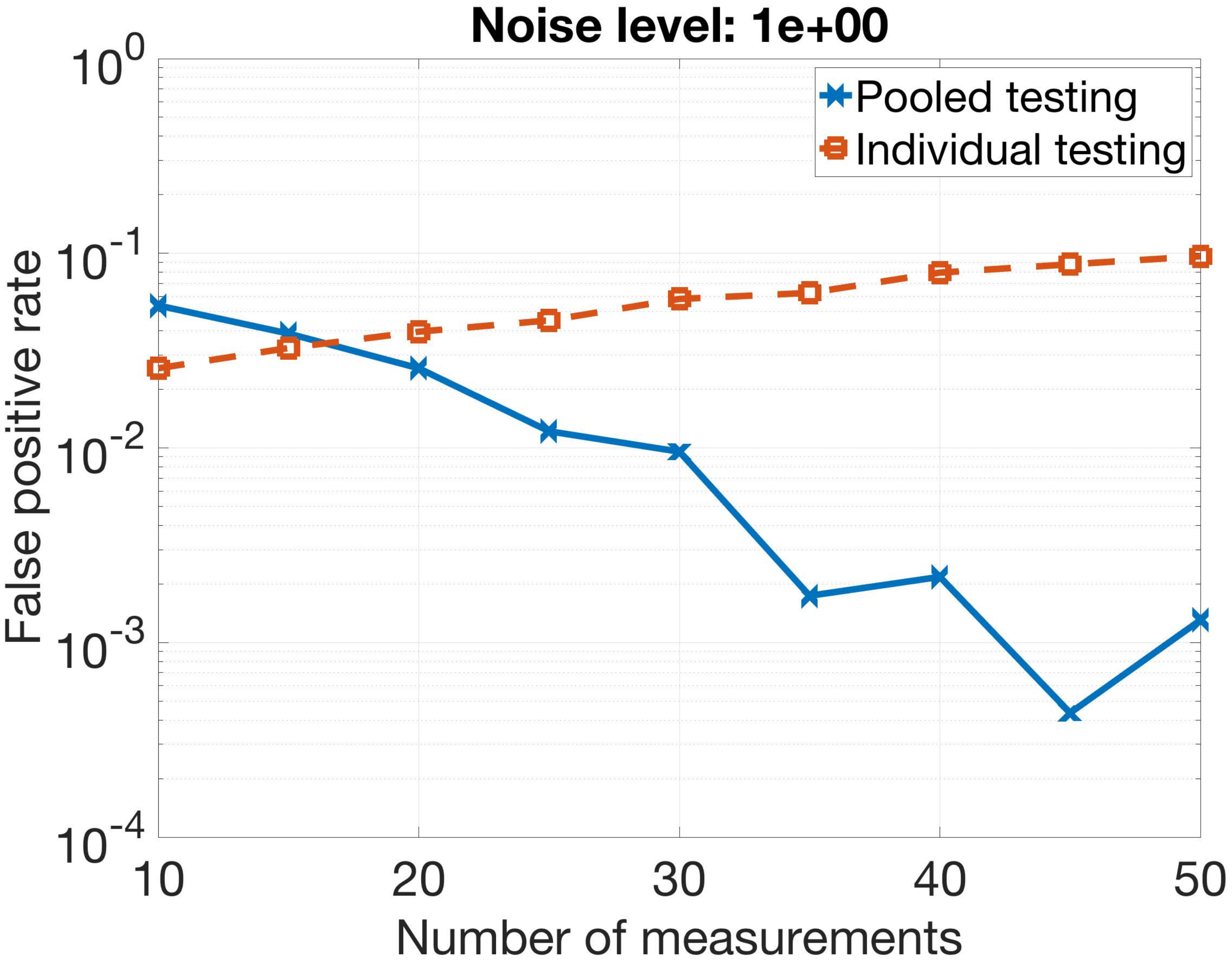}}
   \subfloat[FPR ($\Pb_{out} = 0.15$)]{\includegraphics[scale=0.14]{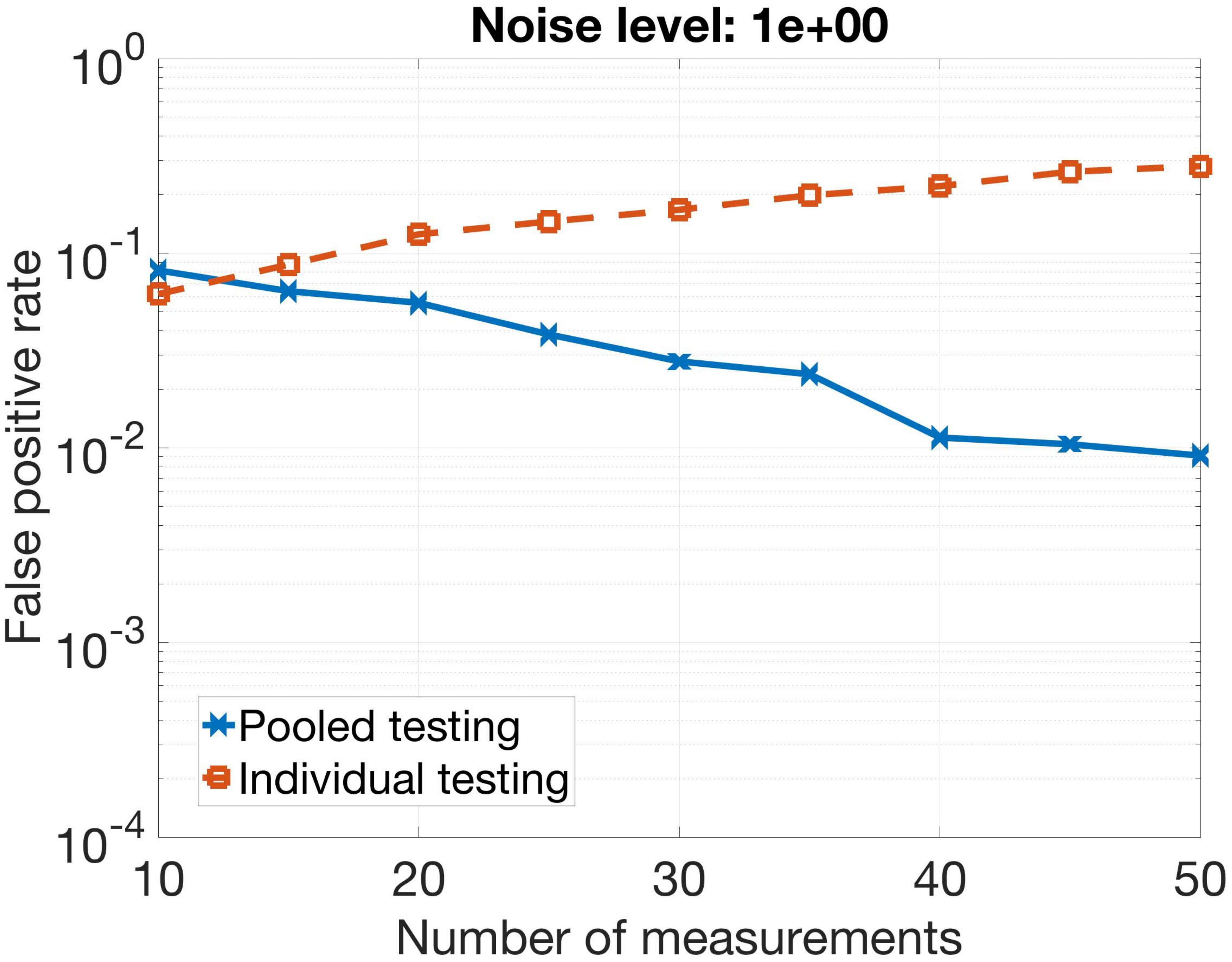}}
    \caption{\small Simulations for different probabilities of outlier errors. False Negative Rate (FNR) and the corresponding False Positive Rate (FPR) with $n=25$, $k=2$, and Gaussian noise level 1e0.}
    \label{fig:FNR_FPR_N25_K2}
\end{figure*}

\begin{figure*}[!htb]
    \centering
   \subfloat[FNR ($\Pb_{out} = 0.01$)]{\includegraphics[scale=0.14]{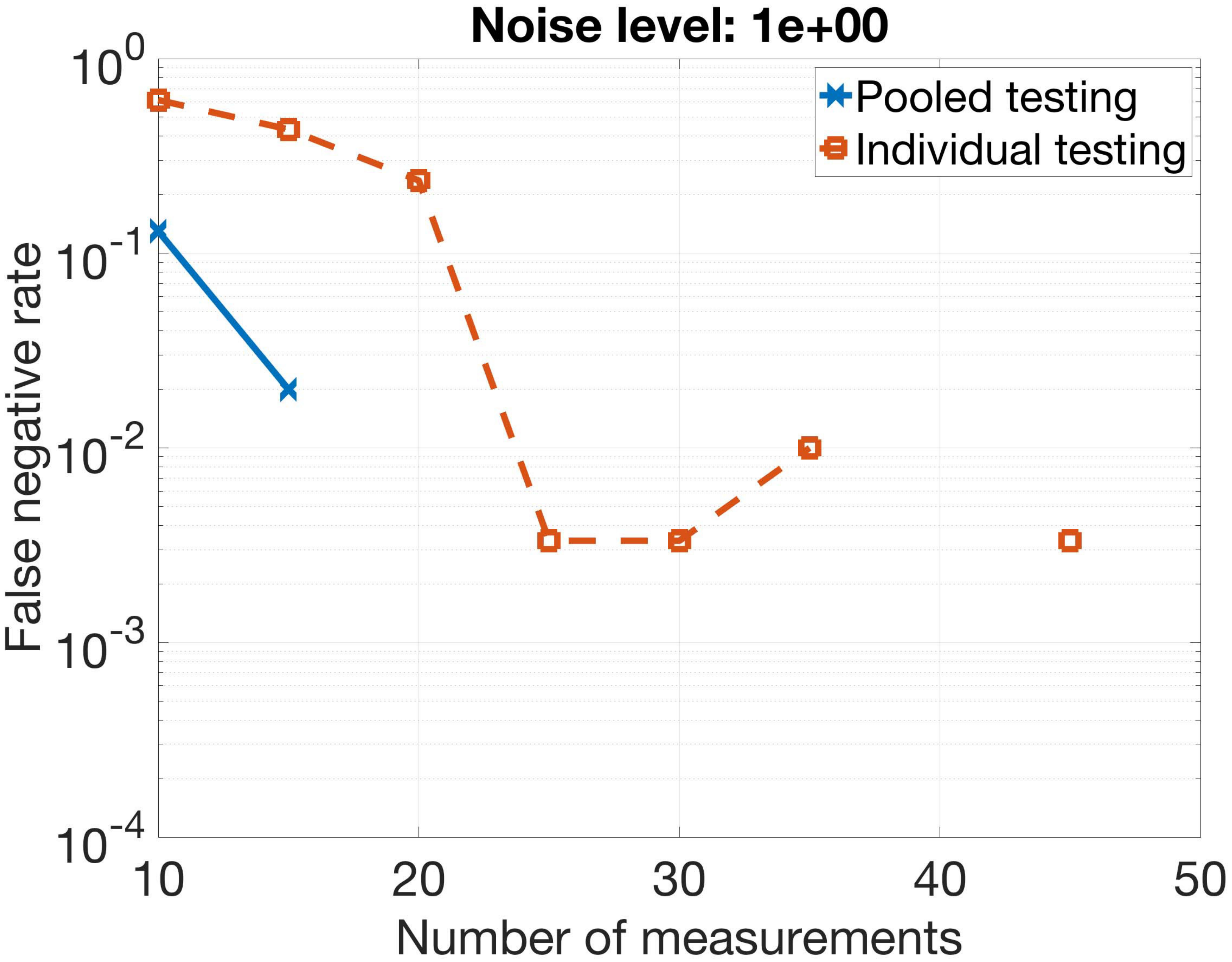}}
   \subfloat[FNR ($\Pb_{out} = 0.05$)]{\includegraphics[scale=0.14]{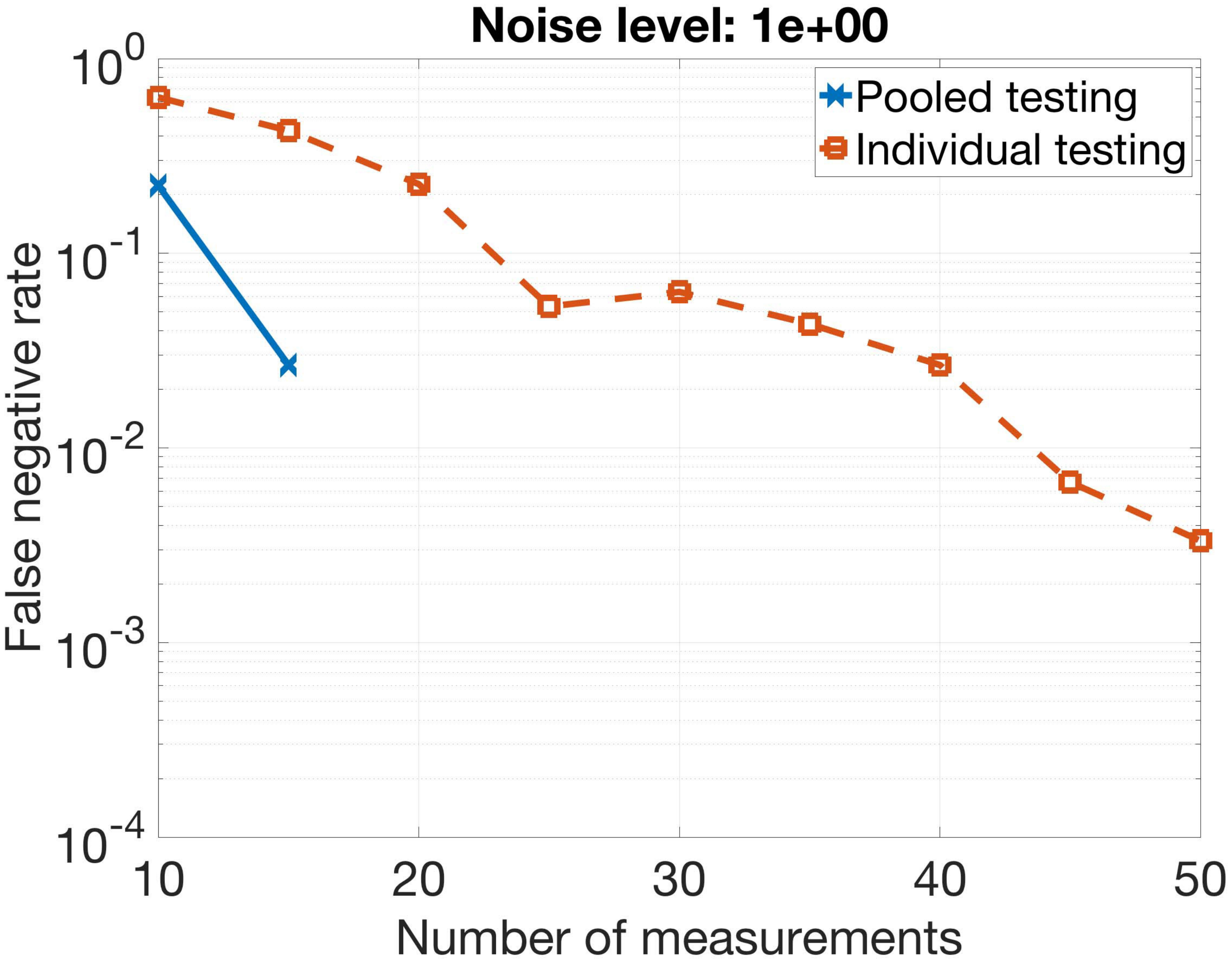}}
   \subfloat[FNR ($\Pb_{out} = 0.15$)]{\includegraphics[scale=0.14]{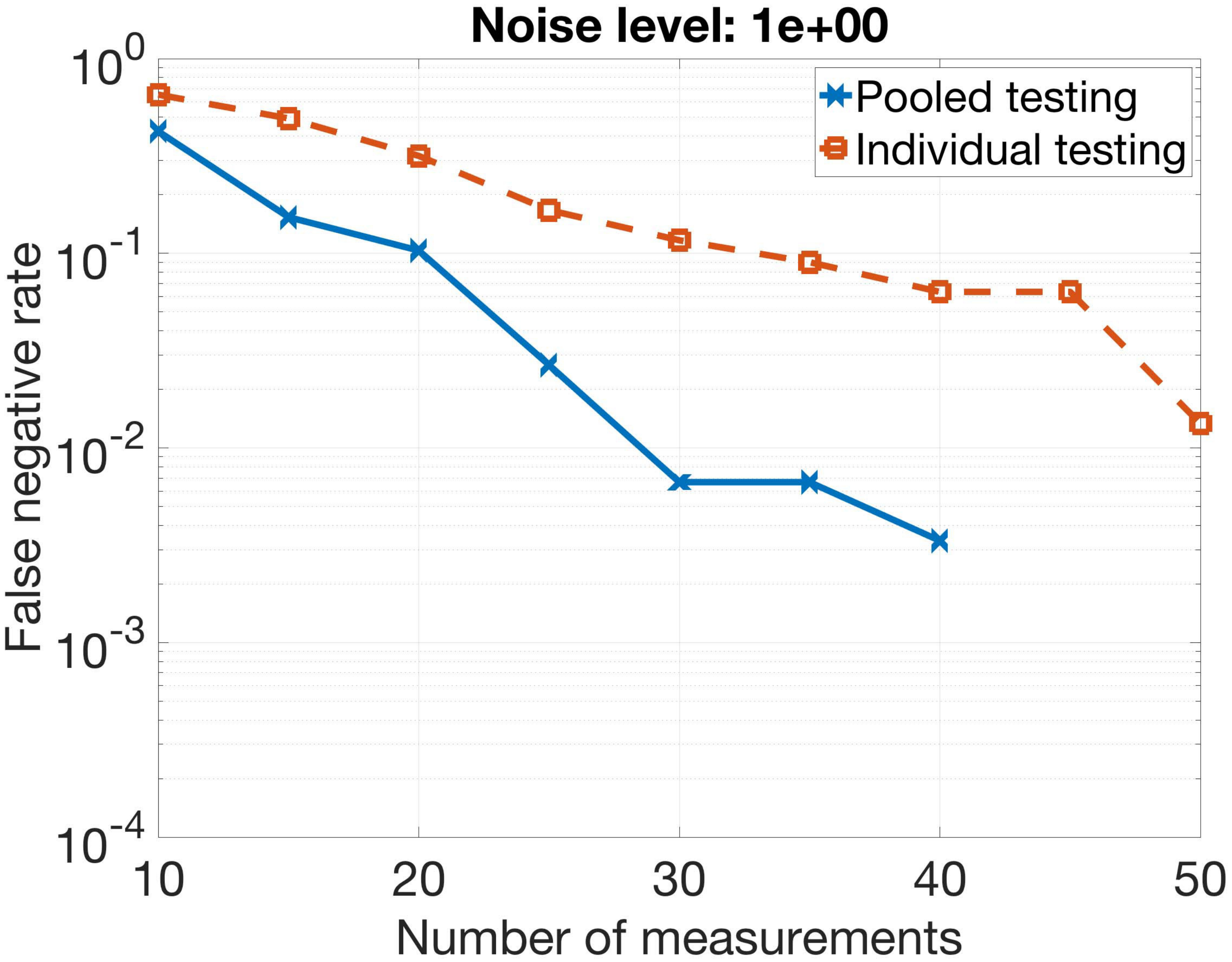}} \\
   \subfloat[FPR ($\Pb_{out} = 0.01$)]{\includegraphics[scale=0.14]{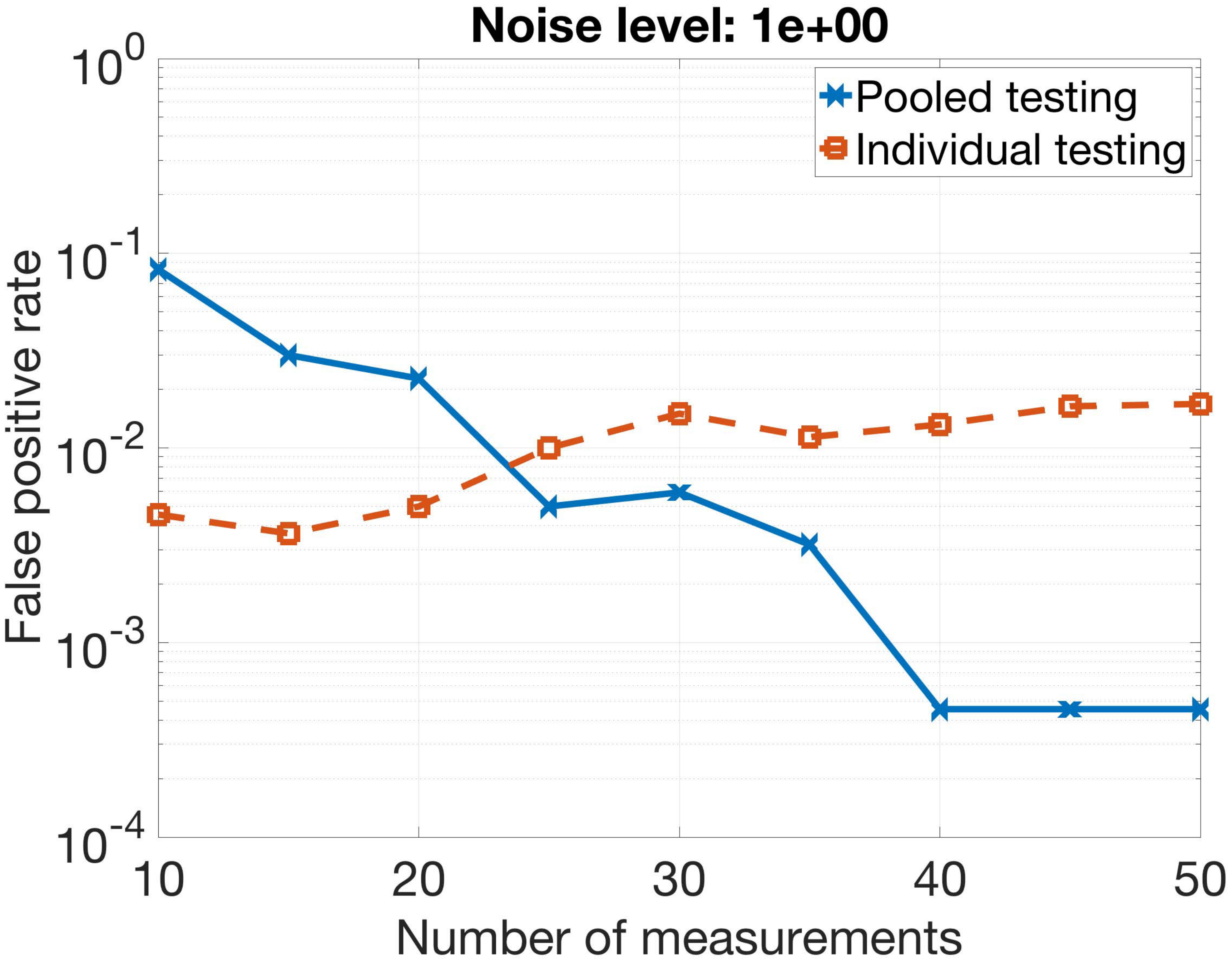}}
   \subfloat[FPR ($\Pb_{out} = 0.05$)]{\includegraphics[scale=0.14]{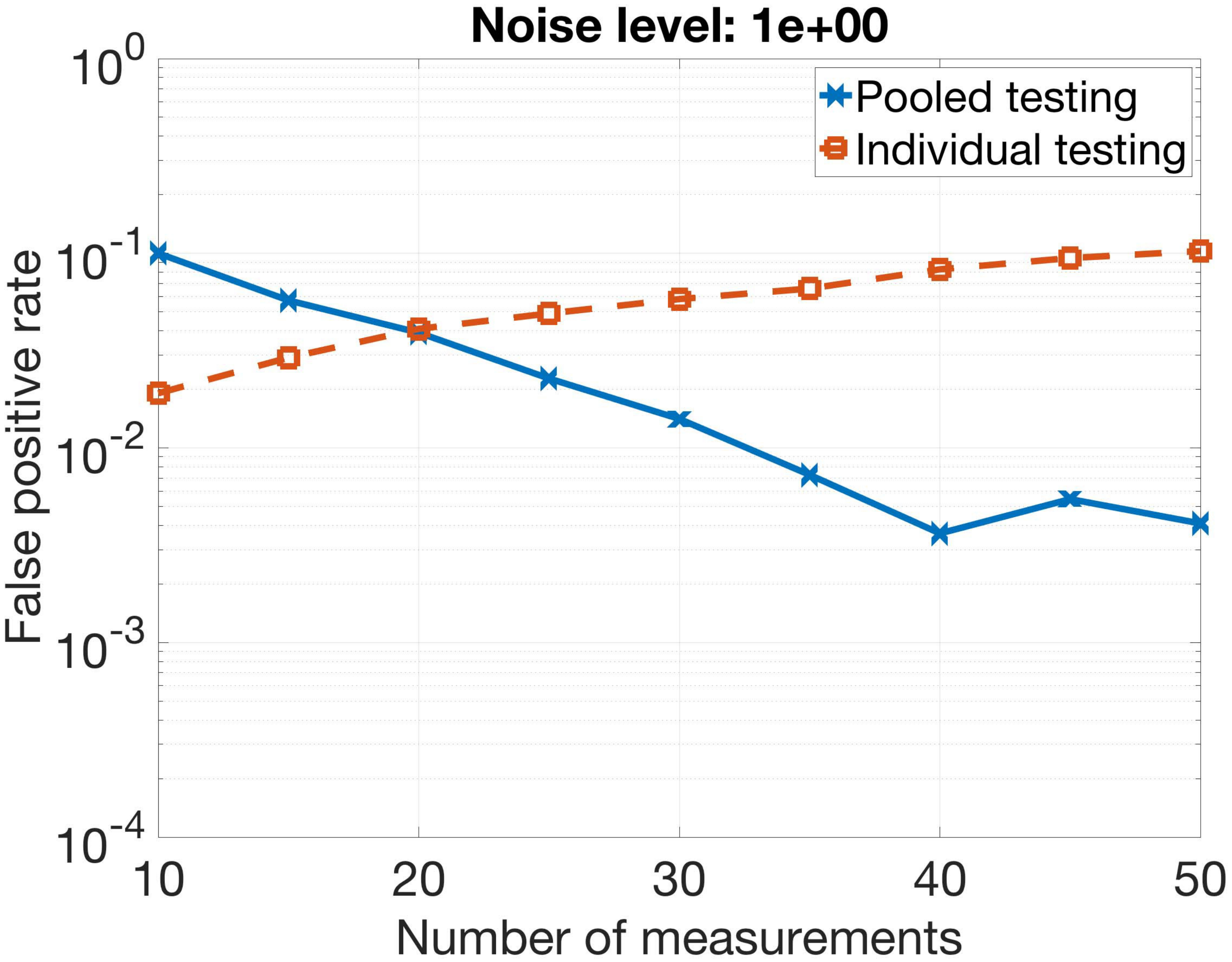}}
   \subfloat[FPR ($\Pb_{out} = 0.15$)]{\includegraphics[scale=0.14]{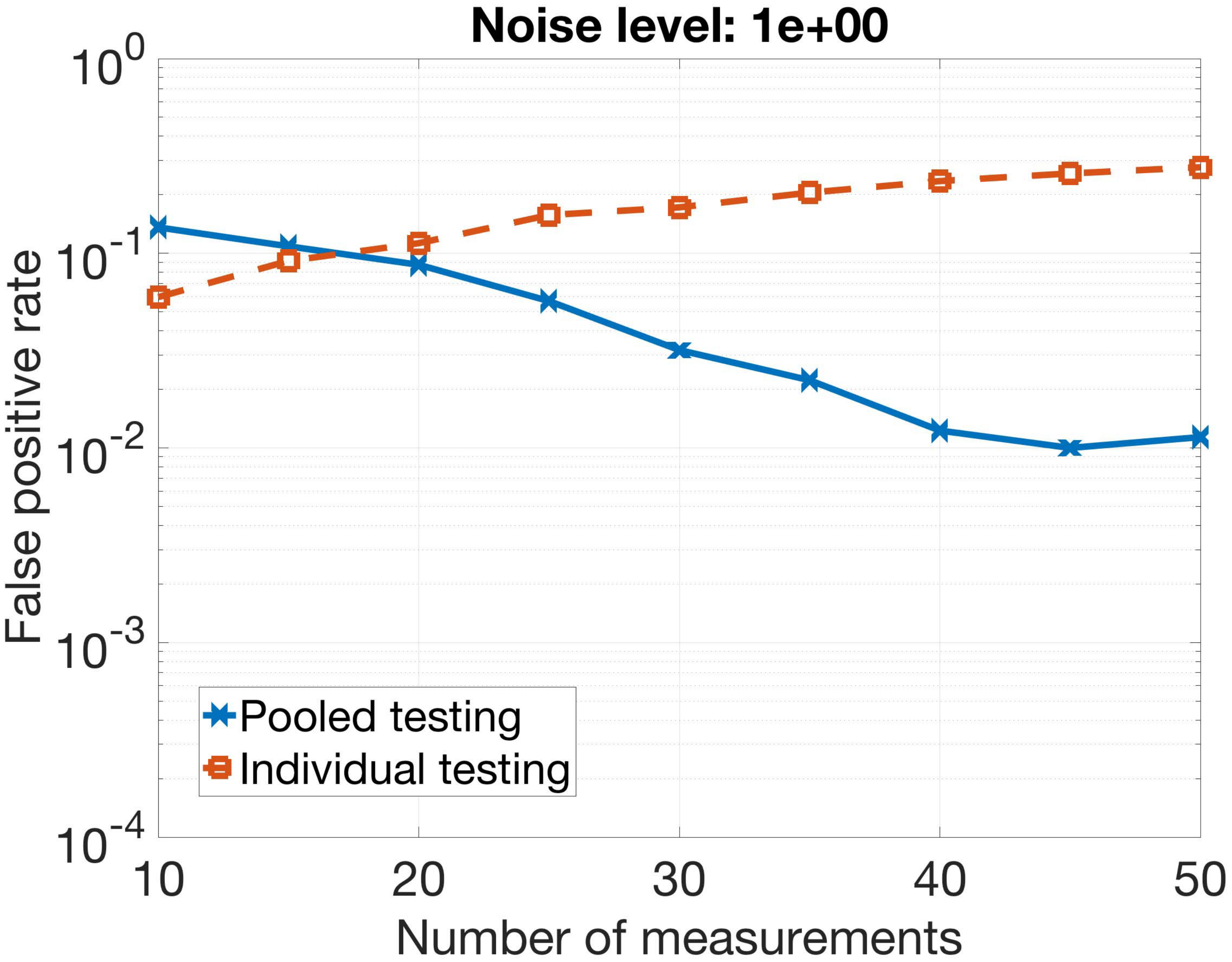}}
    \caption{\small Simulations for different probabilities of outlier errors. False Negative Rate (FNR) and False Positive Rate (FPR) with $n=25$, $k=3$, and Gaussian noise level 1e0.}
    \label{fig:FNR_FPR_N25_K3}
\end{figure*}

\begin{figure*}[!htb]
    \centering
   \subfloat[FNR ($\Pb_{out} = 0.01$)]{\includegraphics[scale=0.14]{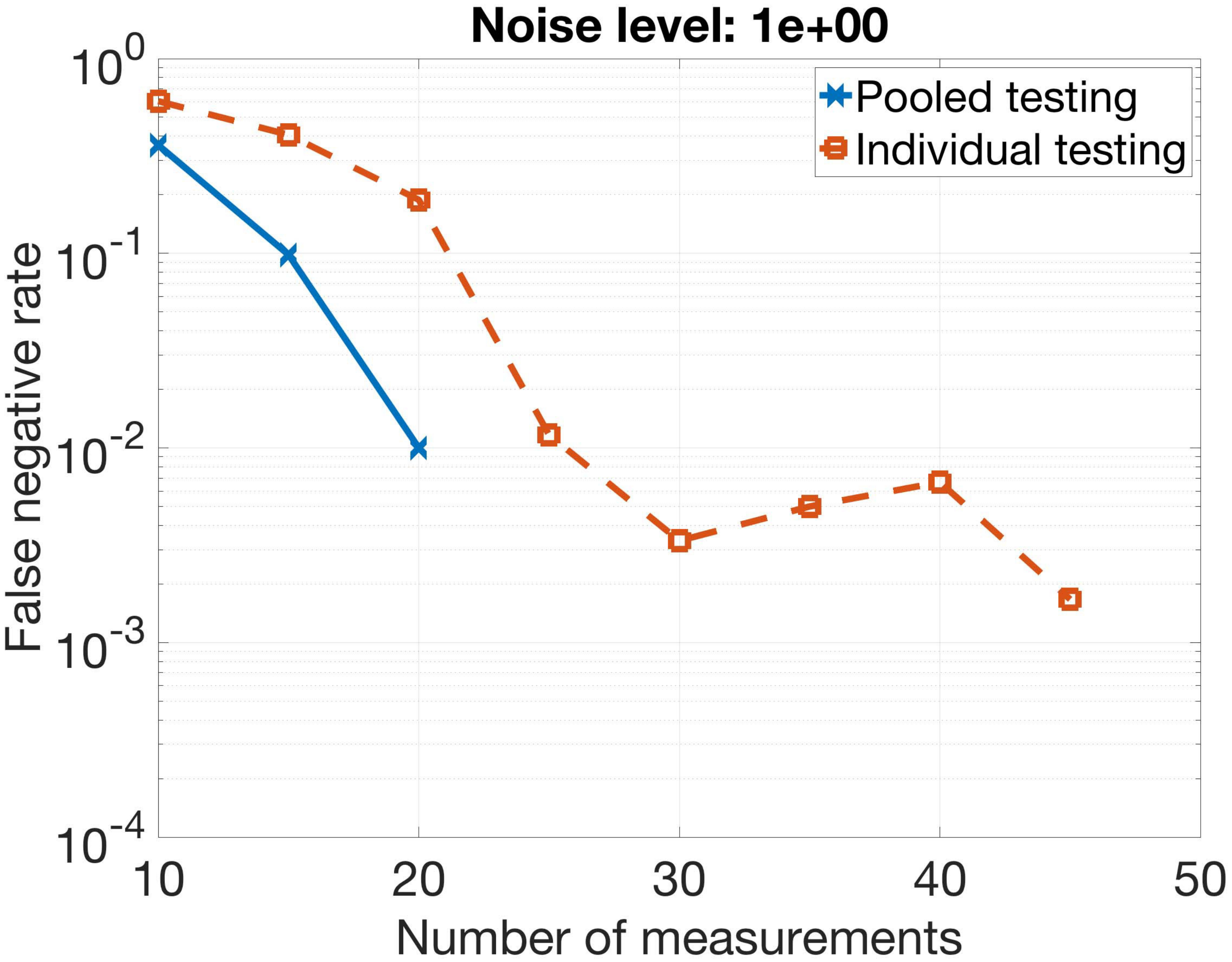}}
   \subfloat[FNR ($\Pb_{out} = 0.05$)]{\includegraphics[scale=0.14]{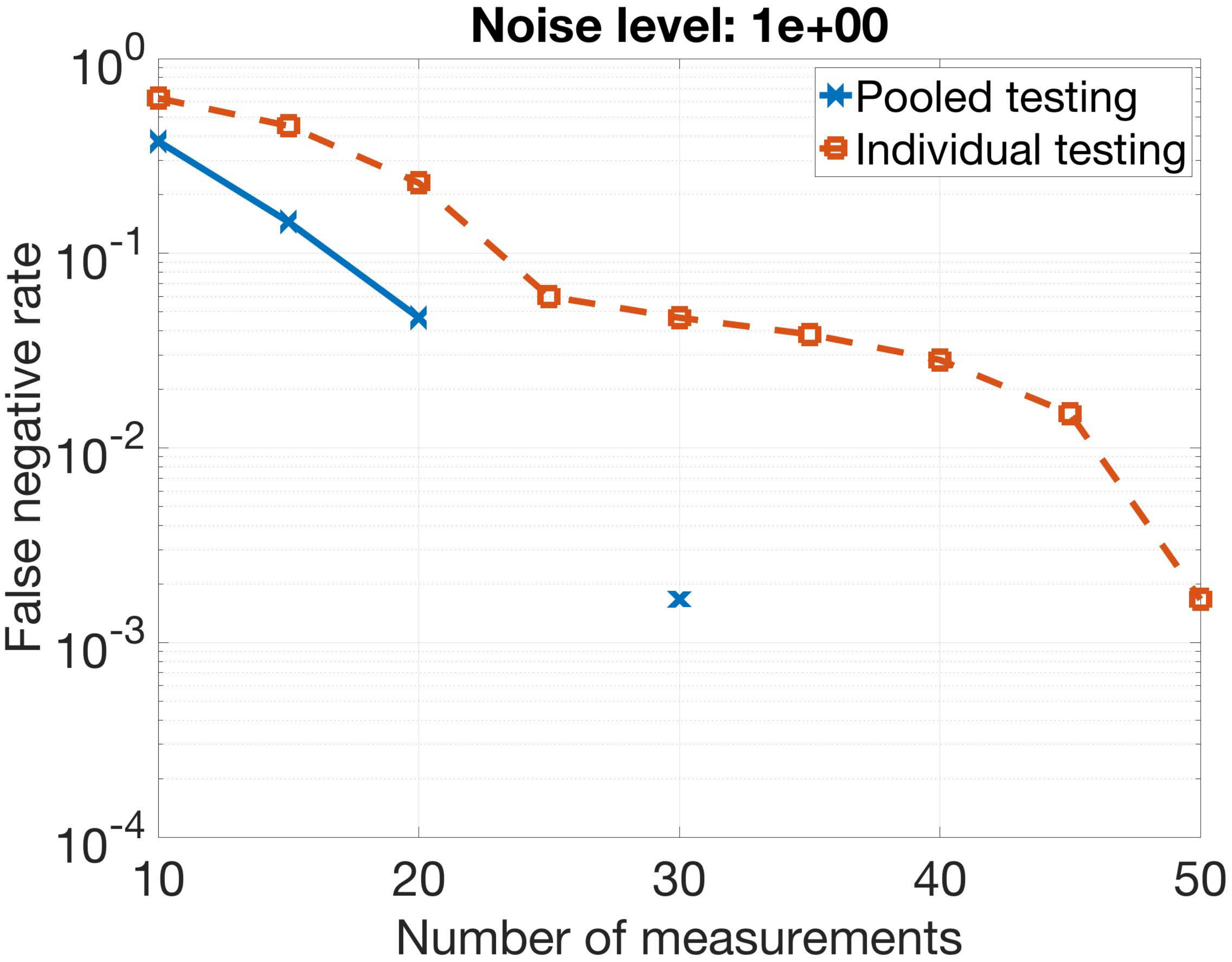}}
   \subfloat[FNR ($\Pb_{out} = 0.15$)]{\includegraphics[scale=0.14]{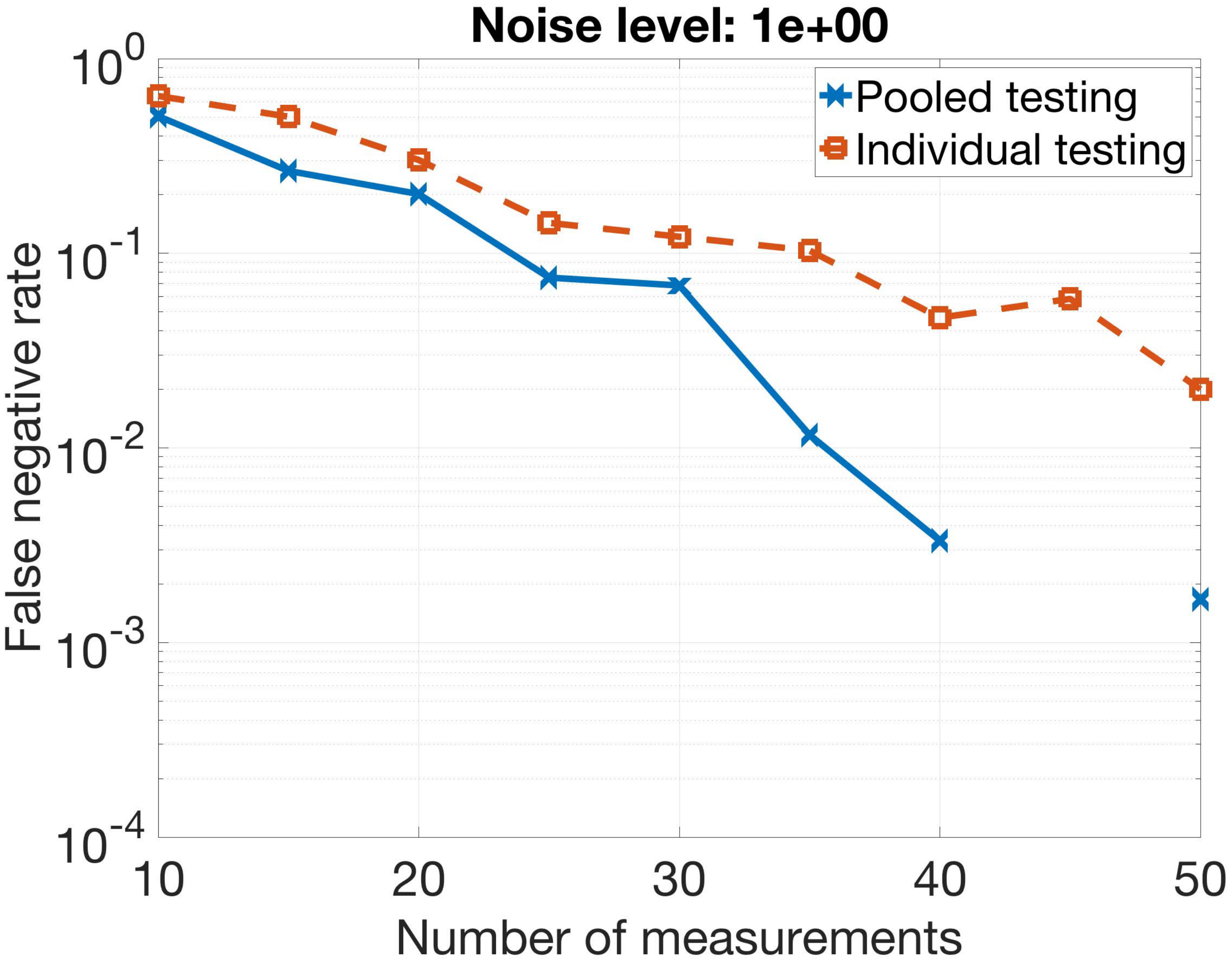}} \\
   \subfloat[FPR ($\Pb_{out} = 0.01$)]{\includegraphics[scale=0.14]{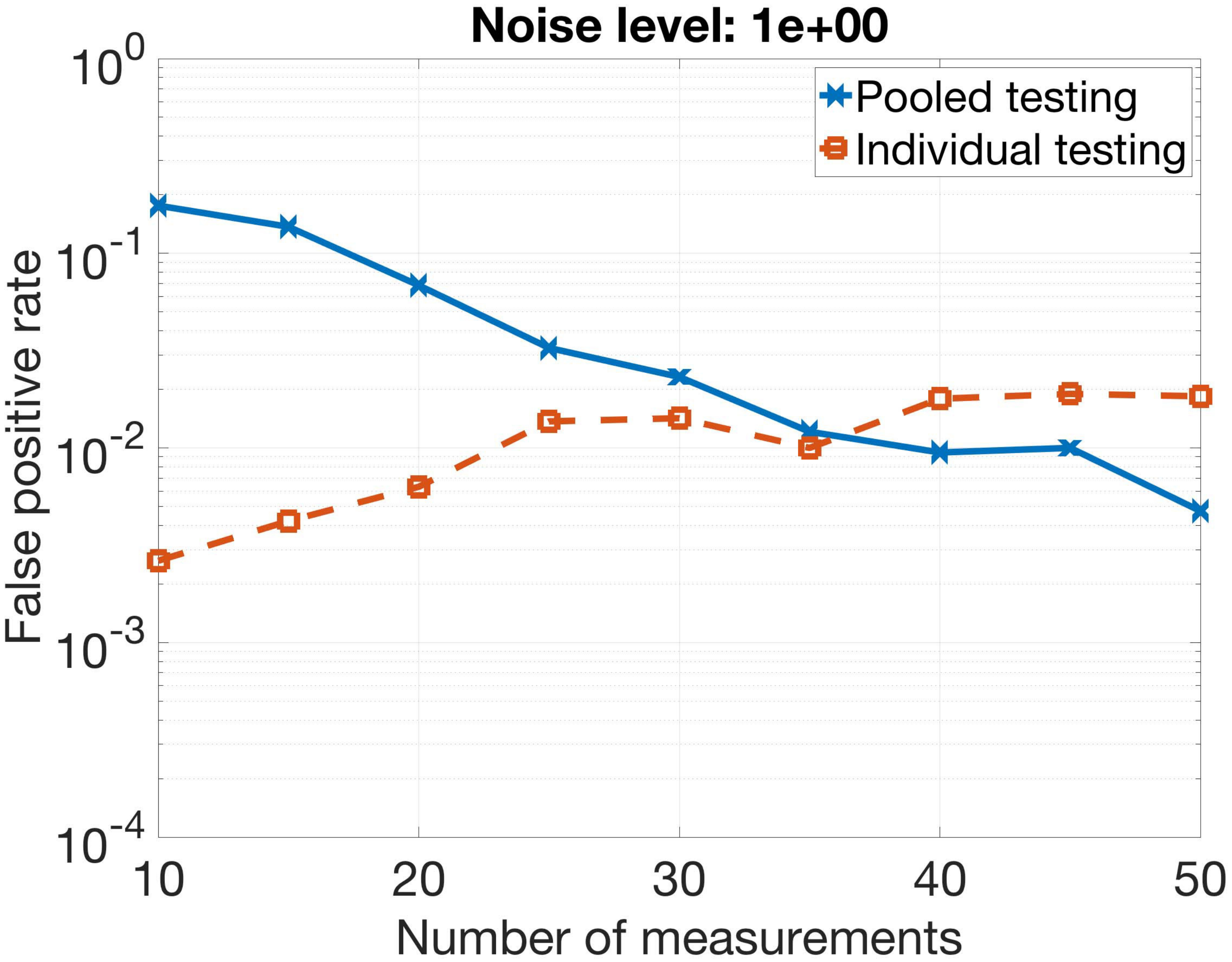}}
   \subfloat[FPR ($\Pb_{out} = 0.05$)]{\includegraphics[scale=0.14]{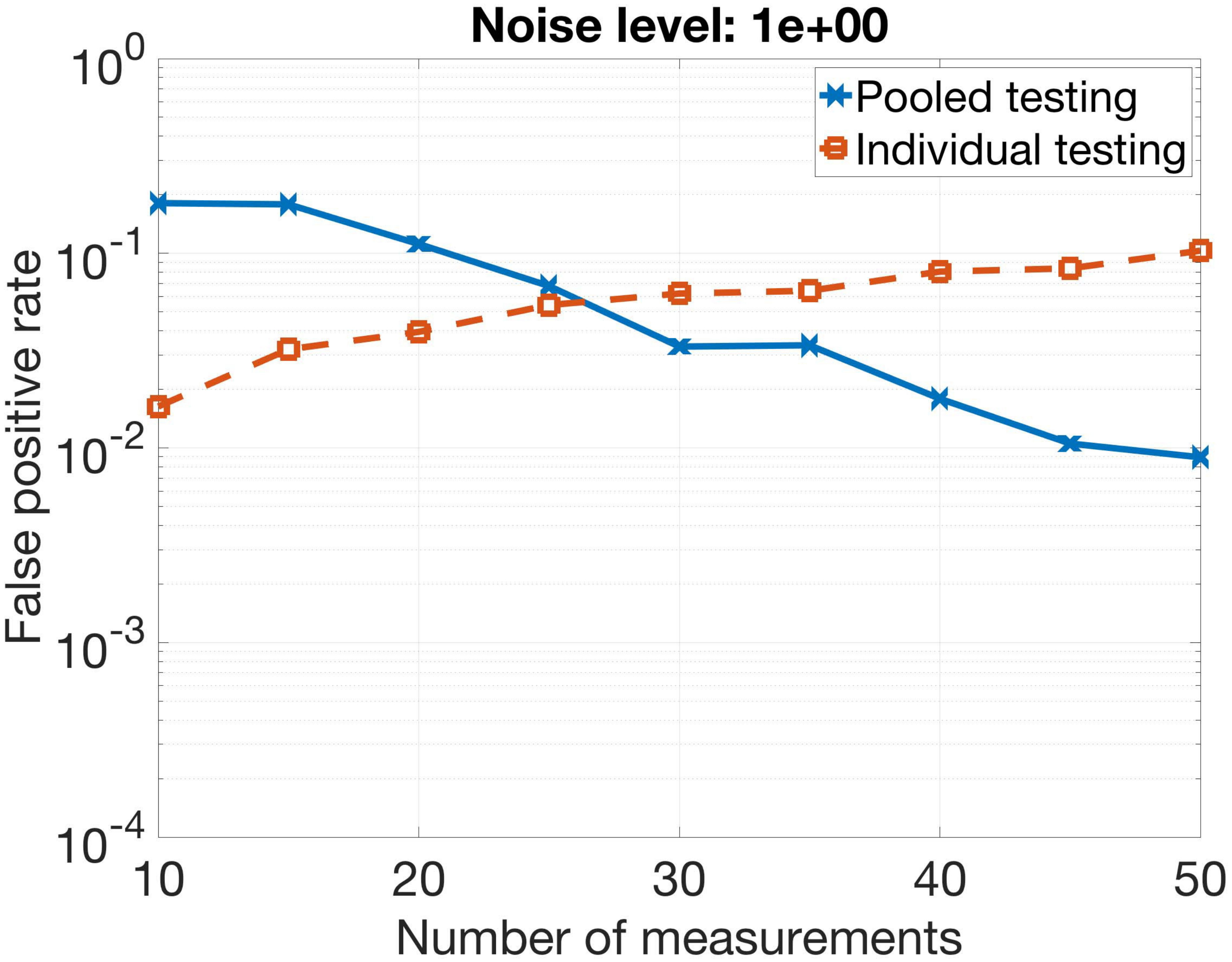}}
   \subfloat[FPR ($\Pb_{out} = 0.15$)]{\includegraphics[scale=0.14]{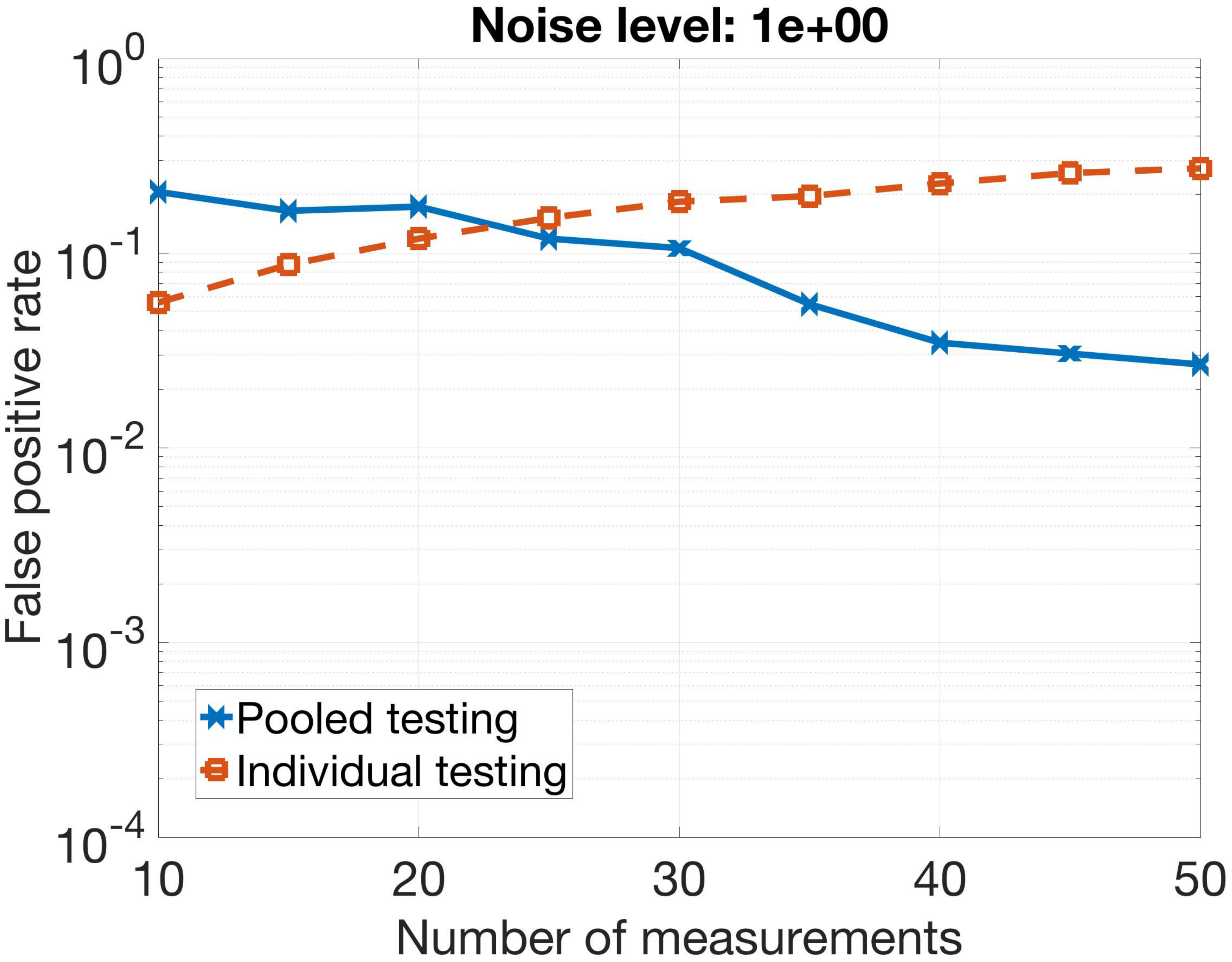}}
    \caption{\small Simulations for different probabilities of outlier errors. False Negative Rate (FNR) and the corresponding False Positive Rate (FPR) with $n=25$, $k=6$, and Gaussian noise level 1e0.}
    \label{fig:FNR_FPR_N25_K6}
\end{figure*}

In Figures from  \ref{fig:FNR_FPR_N25_K1} to \ref{fig:FNR_FPR_N25_K6}, (a), (b), and (c) show the FNR of the pooled testing and the individual testing in log-scale with different probabilities of outlier error varied from from $1\%$ to $15\%$, and (d), (e), and (f) describe the corresponding FPR. Here, the number of people tested is set to 25, i.e., $n=25$, and the number of people having COVID-19 virus is varied from 1 to 6 out of 25, i.e., $k=1, ..., 6$. The noise level is fixed to $1e0$. From various simulations as shown in Figures from \ref{fig:FNR_FPR_N25_K1} and \ref{fig:FNR_FPR_N25_K6}, the pooled testing lowers both the FNR and the FPR as the number of measurements increases. This is because as the number of measurements increases, we can recover more accurate results $\bx$ and $\be$ via $\ell_1 - \ell_1$ minimization introduced in \eqref{Defn:L1-L1Minimization}.  Unlike the pooled testing, the individual testing can reduce the FNR as the number of measurements increases,  while there is a slight increase in FPR at the same time.    This is because the individual testing diagnoses the COVID-19 virus test ``positive'' once we have one positive test result among multiple tests.

From these various simulation results with different probabilities of outlier errors, in many cases, we demonstrate that the pooled testing can simultaneously have (significant) lower FNR and FPR than those of the individual testings. In a limited number of cases under $m < n$, the individual testing provides lower (although not significantly lower) FPR than that of the pooled testing, because only a few people are tested under individual testing, and fewer false positive mistakes are made (recall that the tested subjects are assumed to be diagnosed as ``negative'' )  Additionally, since, under $m<n$, there are simply untested subjects in individual testing, and, in general,  a false negative outlier can have more impact on an individiual, the individual testing has relatively higher FNR across all the parameter setups of $m$ and $n$.


\begin{figure*}[!htb]
    \centering
   \subfloat[FNR ($\Pb_{out} = 0.01$)]{\includegraphics[scale=0.14]{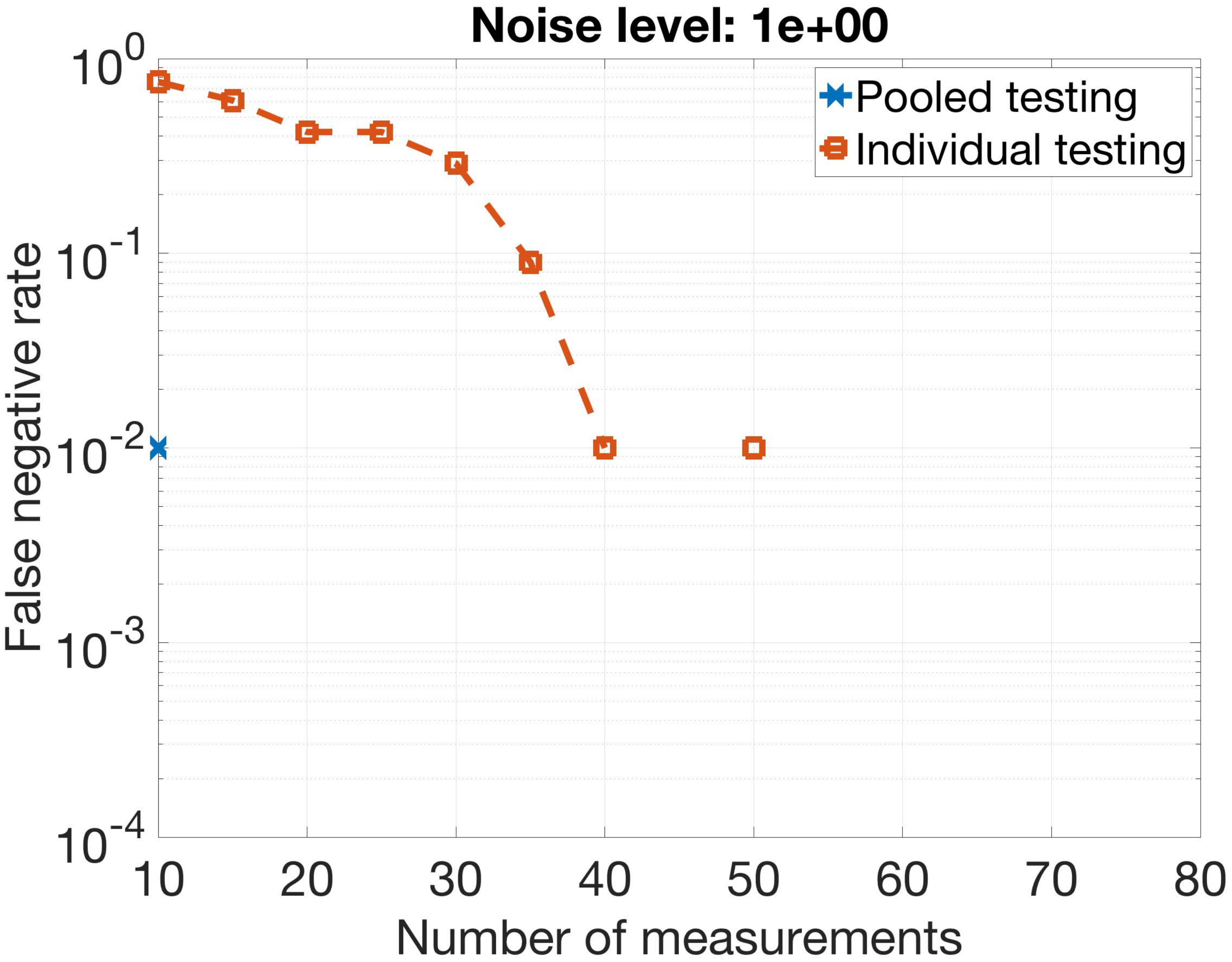}}
   \subfloat[FNR ($\Pb_{out} = 0.05$)]{\includegraphics[scale=0.14]{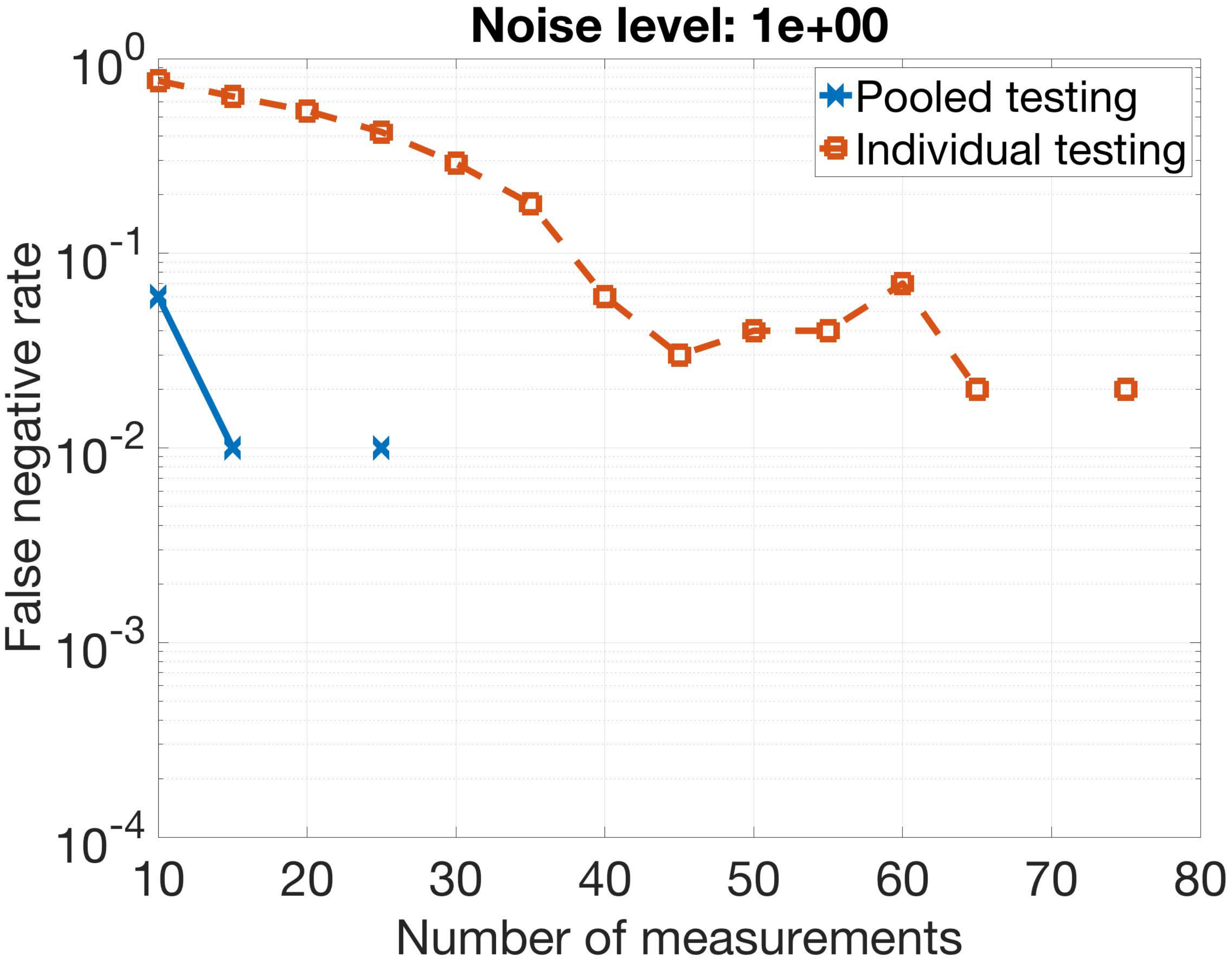}}
   \subfloat[FNR ($\Pb_{out} = 0.15$)]{\includegraphics[scale=0.14]{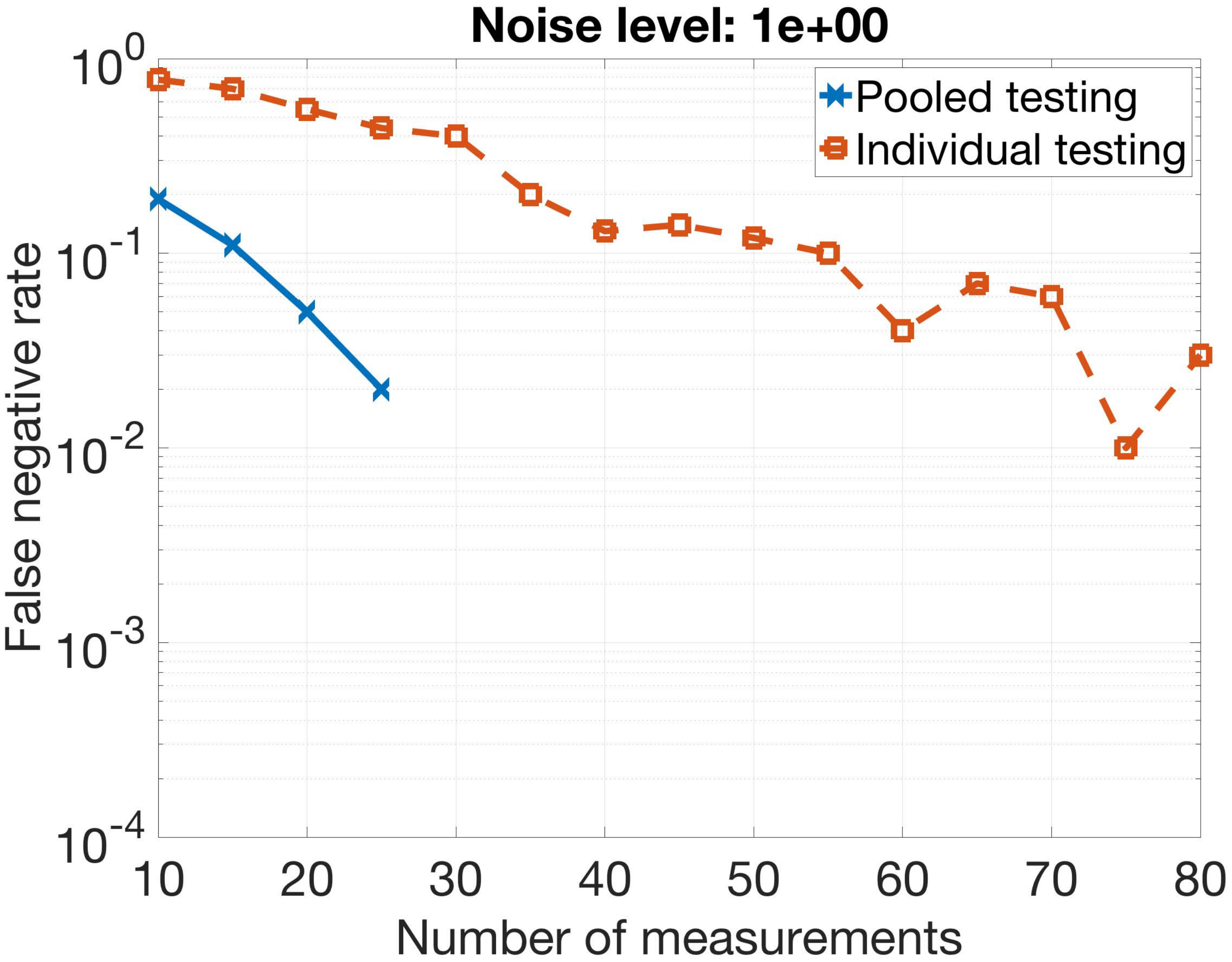}} \\
   \subfloat[FPR ($\Pb_{out} = 0.01$)]{\includegraphics[scale=0.14]{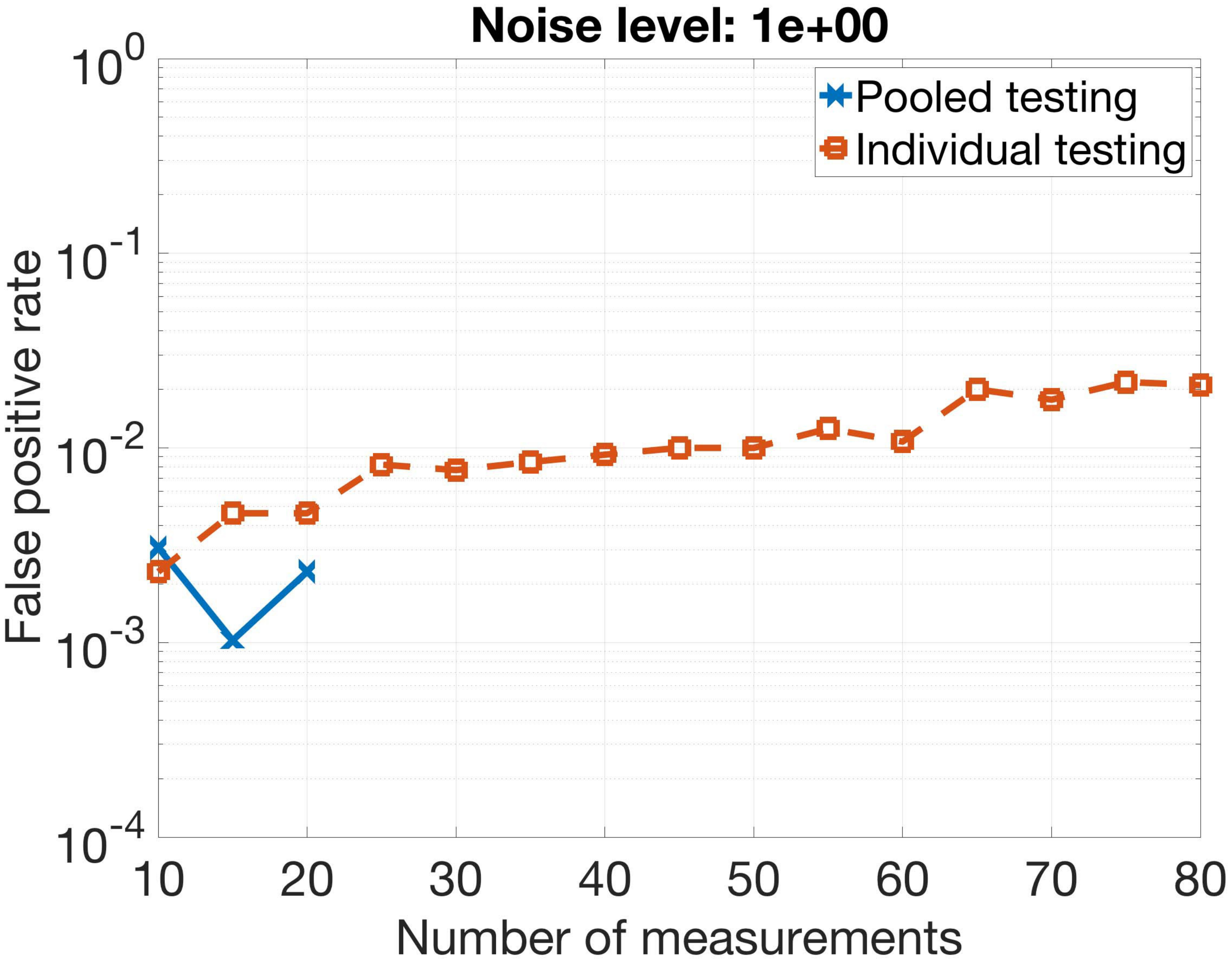}}
   \subfloat[FPR ($\Pb_{out} = 0.05$)]{\includegraphics[scale=0.14]{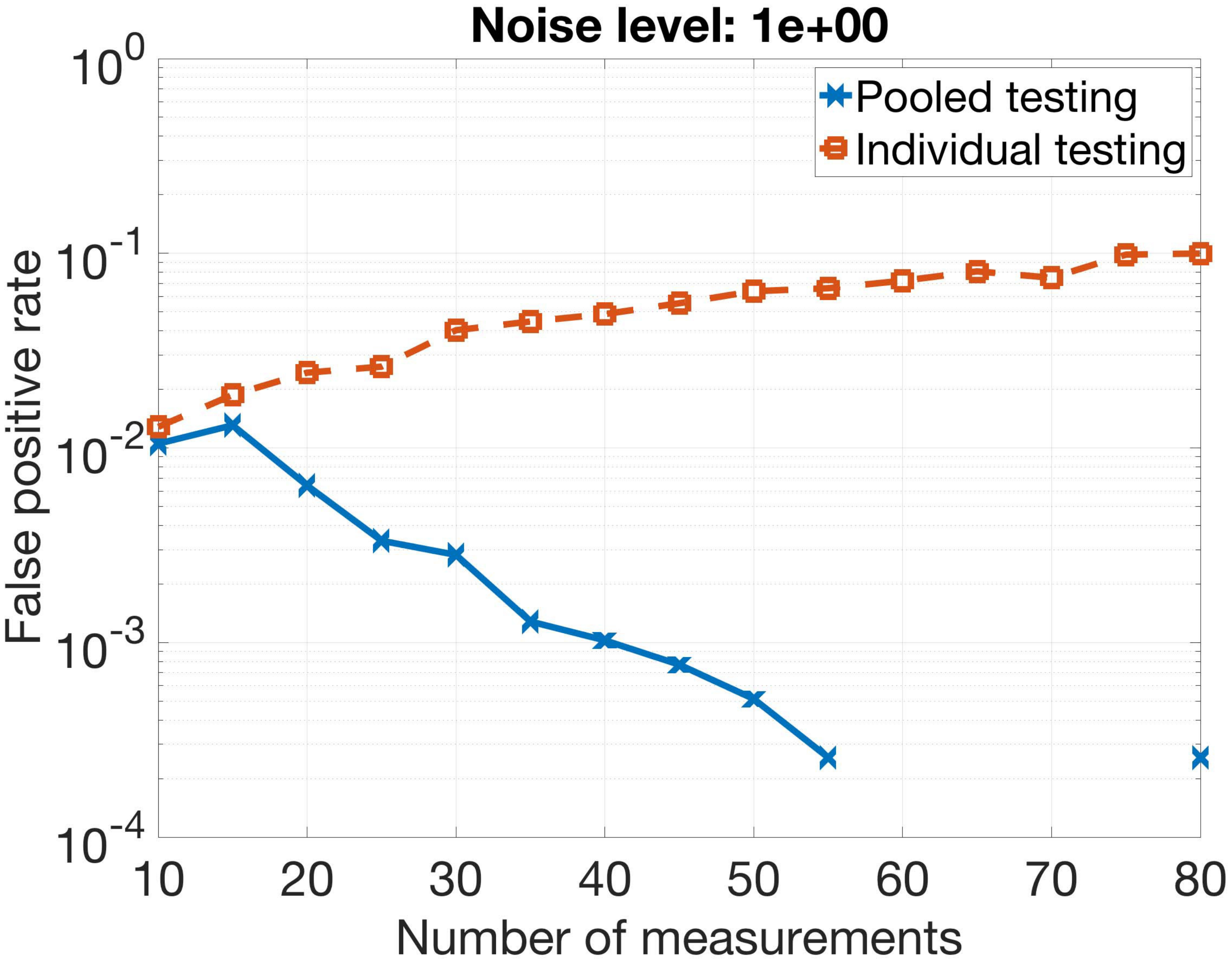}}
   \subfloat[FPR ($\Pb_{out} = 0.15$)]{\includegraphics[scale=0.14]{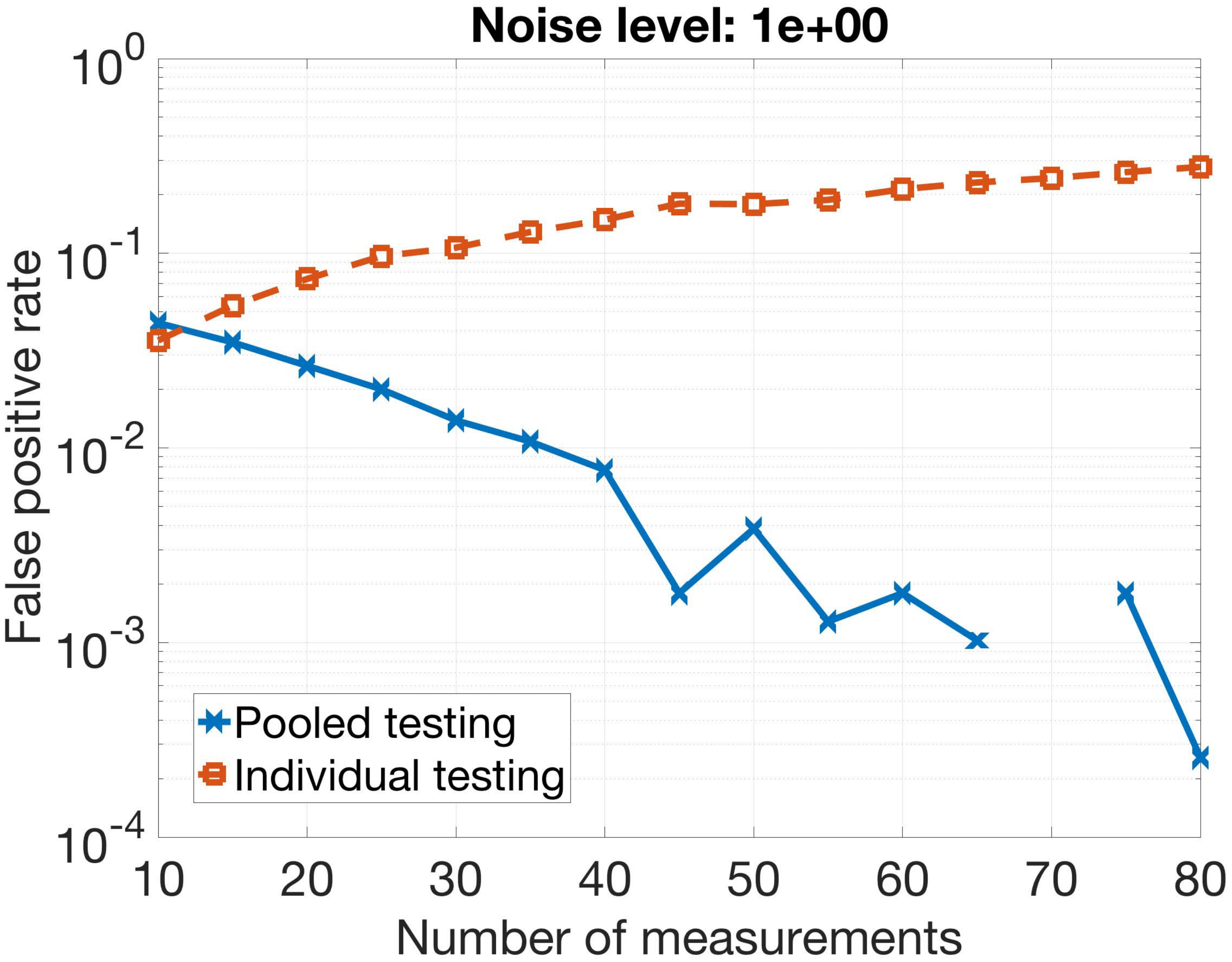}}
    \caption{\small Simulations for different probabilities of outlier errors. False Negative Rate (FNR) and the corresponding False Positive Rate (FPR) with $n=40$, $k=1$, and Gaussian noise level 1e0.}
    \label{fig:FNR_FPR_N40_K1}
\end{figure*}

\begin{figure*}[!htb]
    \centering
   \subfloat[FNR ($\Pb_{out} = 0.01$)]{\includegraphics[scale=0.14]{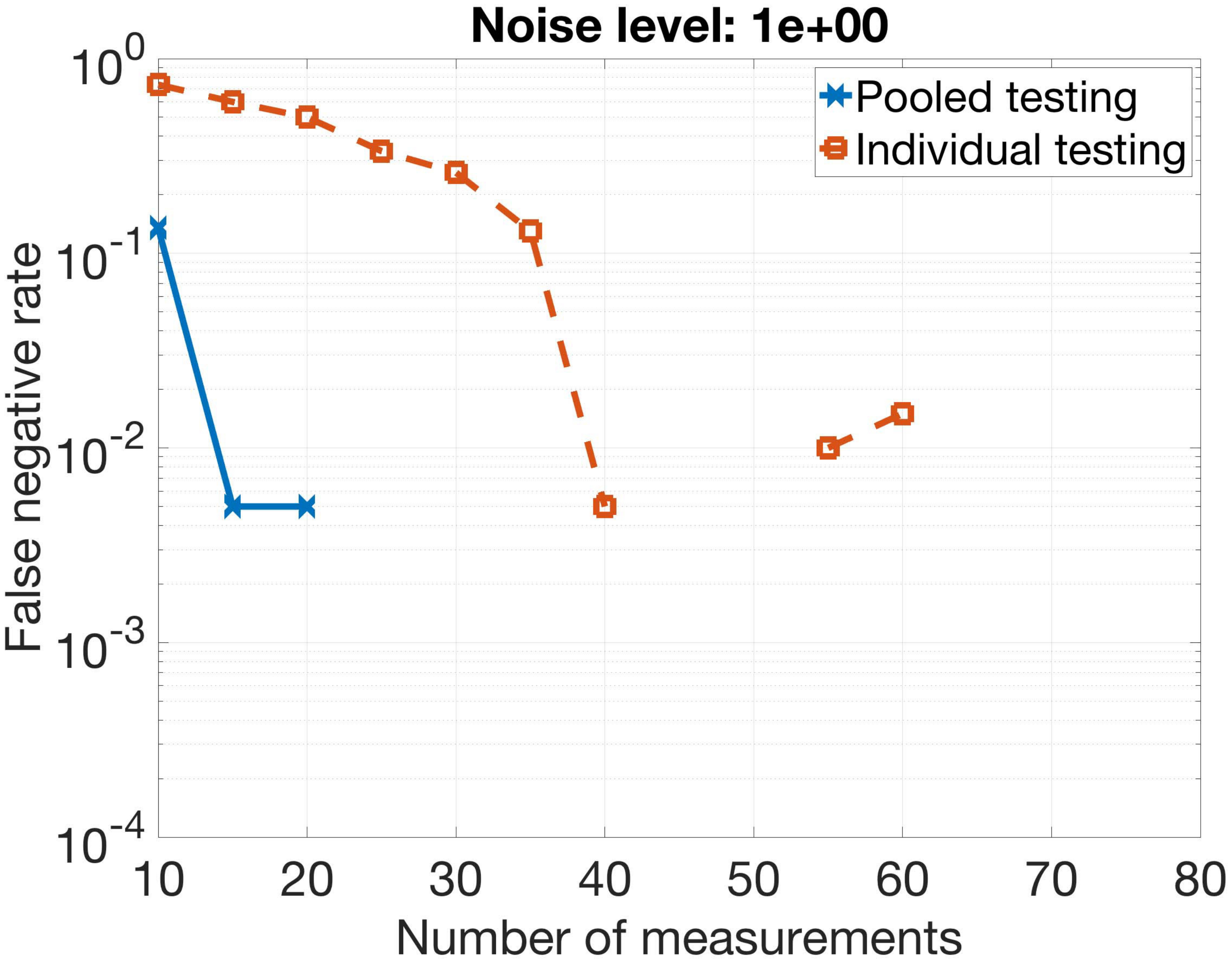}}
   \subfloat[FNR ($\Pb_{out} = 0.05$)]{\includegraphics[scale=0.14]{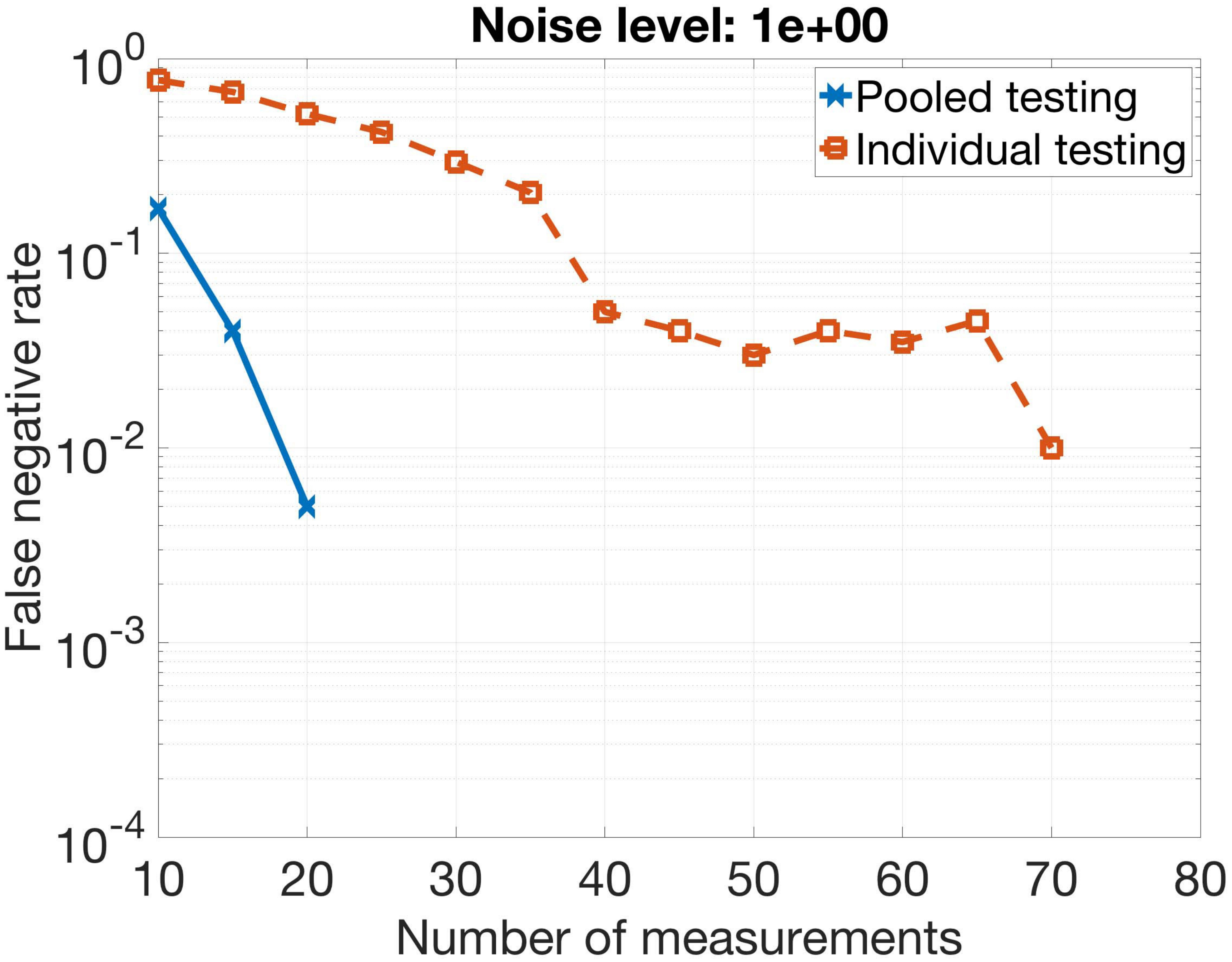}}
   \subfloat[FNR ($\Pb_{out} = 0.15$)]{\includegraphics[scale=0.14]{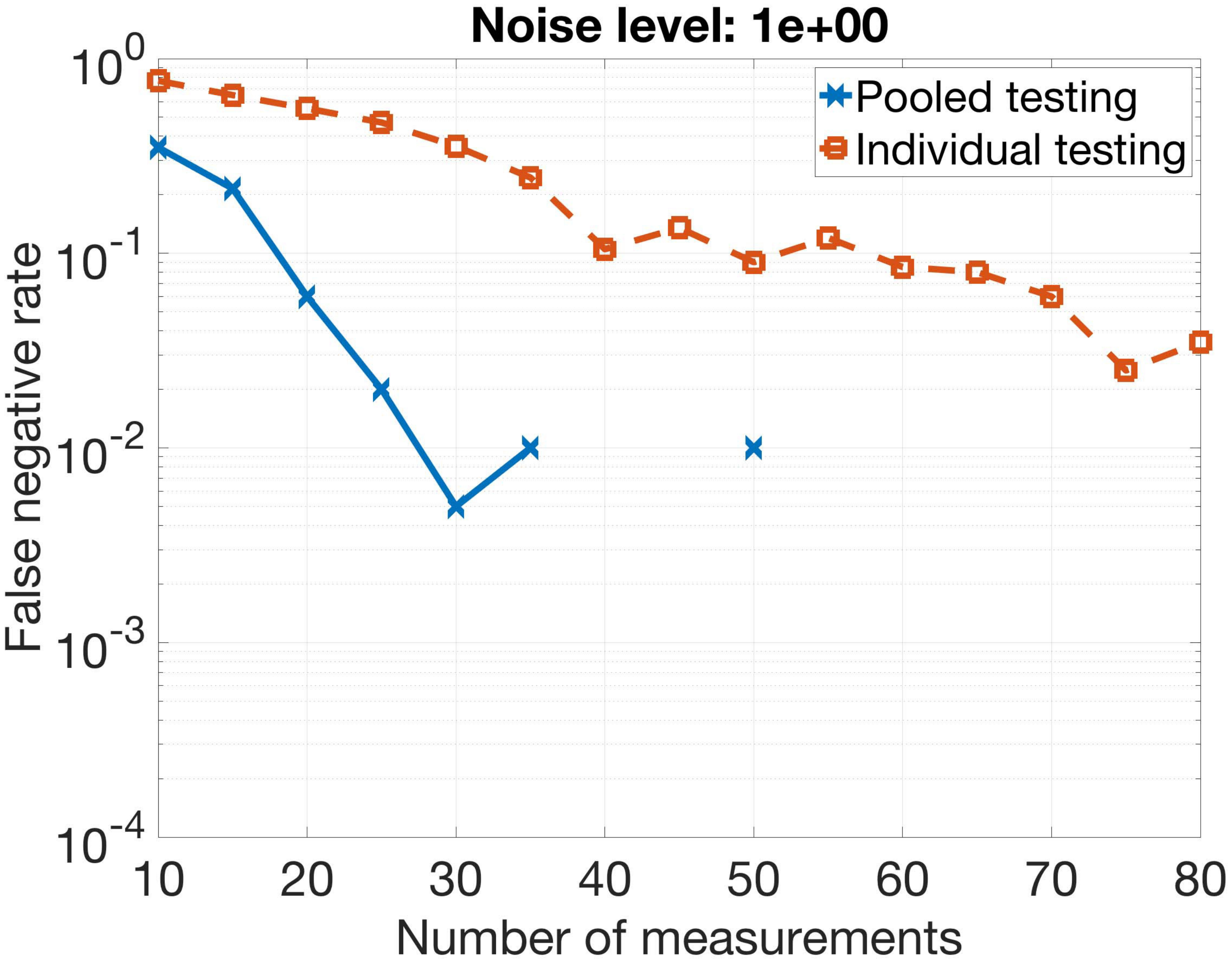}} \\
   \subfloat[FPR ($\Pb_{out} = 0.01$)]{\includegraphics[scale=0.14]{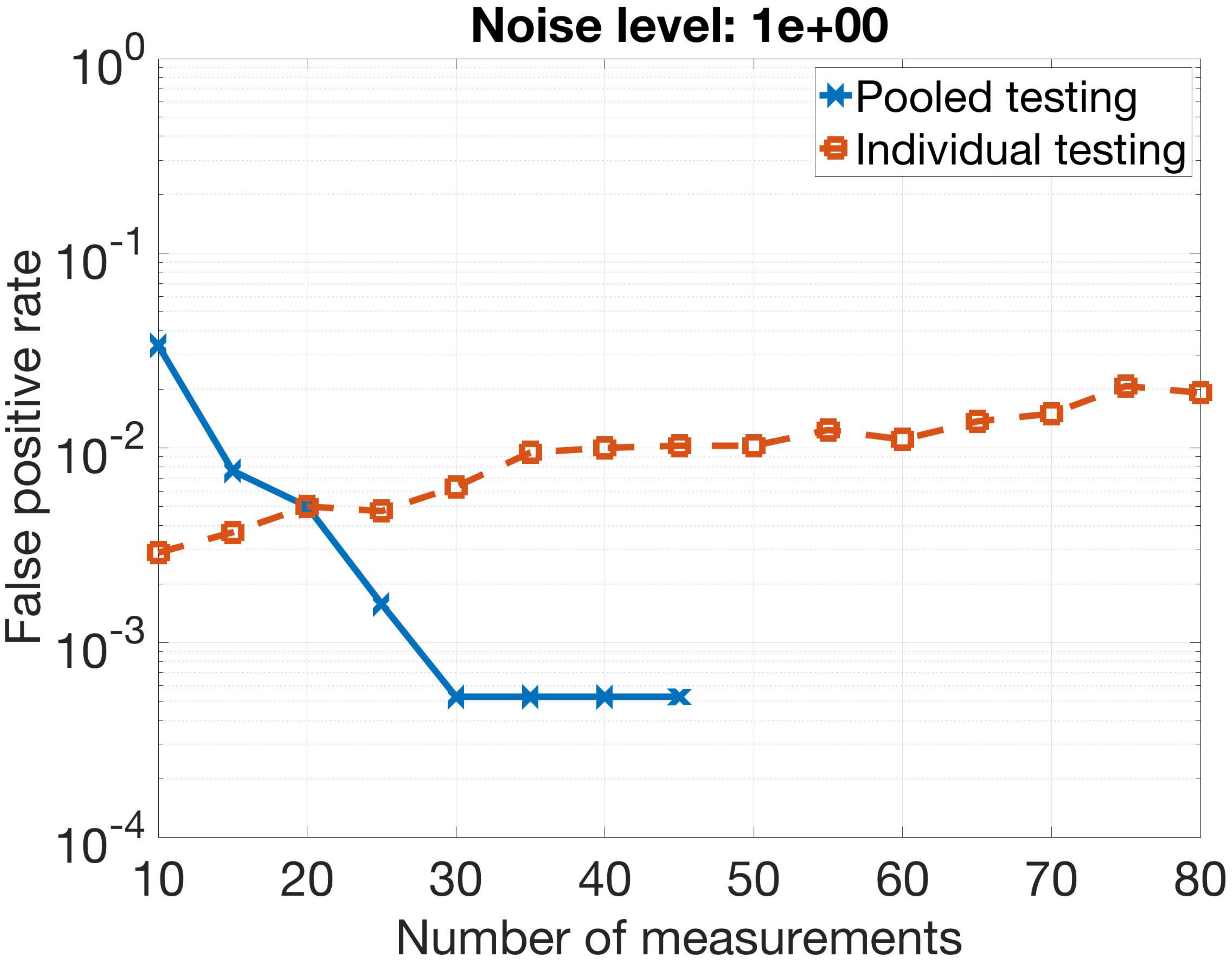}}
   \subfloat[FPR ($\Pb_{out} = 0.05$)]{\includegraphics[scale=0.14]{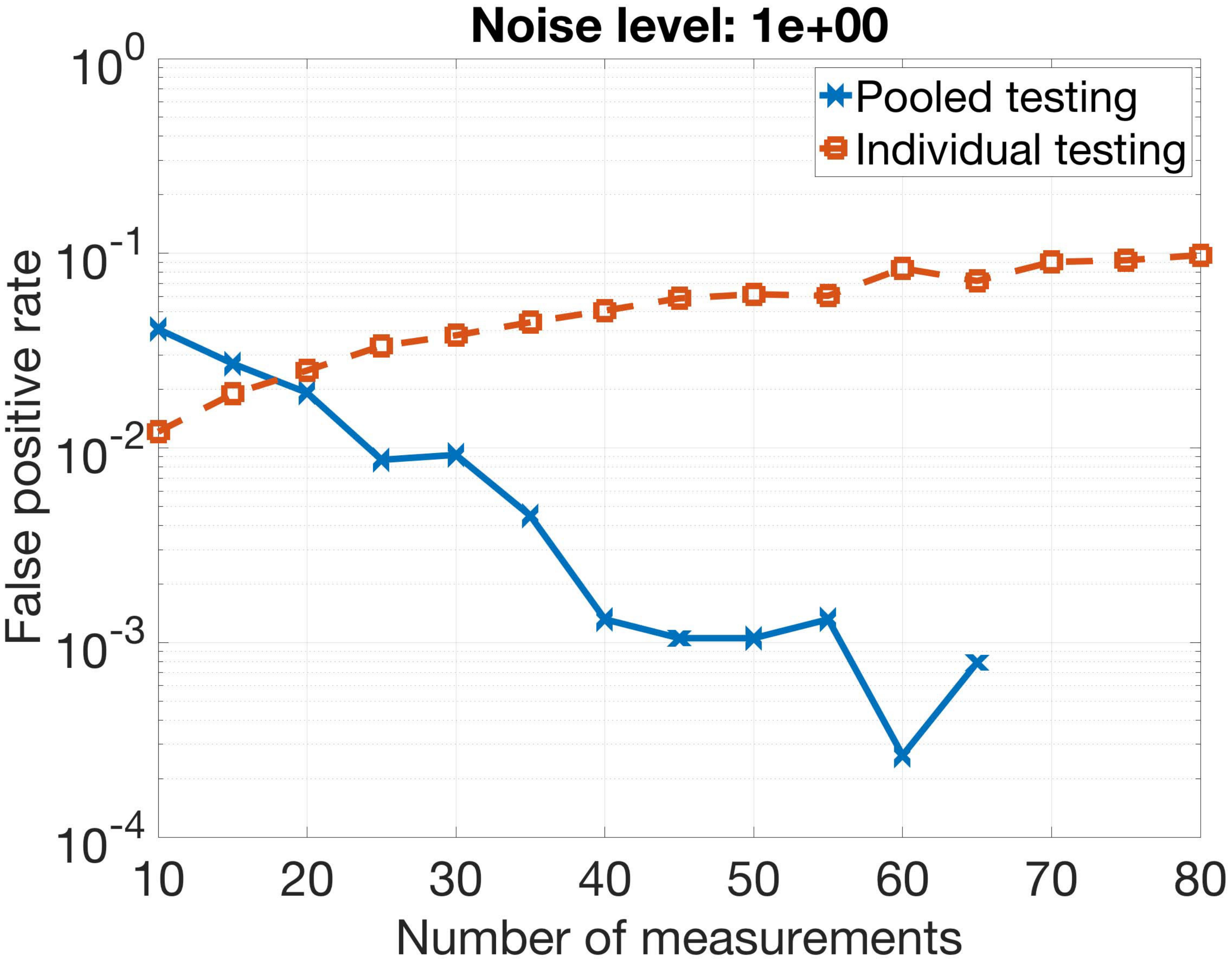}}
   \subfloat[FPR ($\Pb_{out} = 0.15$)]{\includegraphics[scale=0.14]{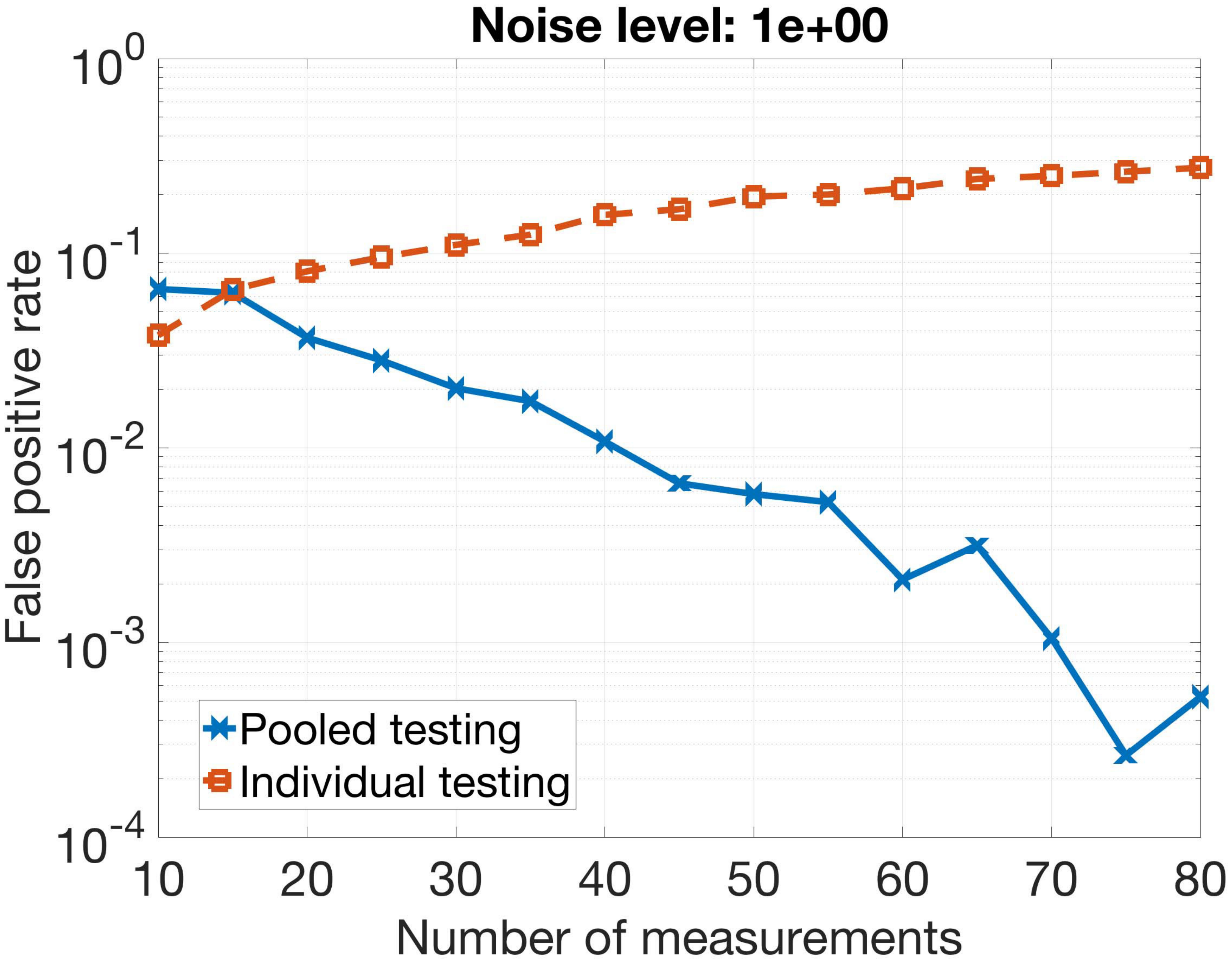}}
    \caption{\small Simulations for different probabilities of outlier errors. False Negative Rate (FNR) and the corresponding False Positive Rate (FPR) with $n=40$, $k=2$, and Gaussian noise level 1e0.}
    \label{fig:FNR_FPR_N40_K2}
\end{figure*}

\begin{figure*}[!htb]
    \centering
   \subfloat[FNR ($\Pb_{out} = 0.01$)]{\includegraphics[scale=0.14]{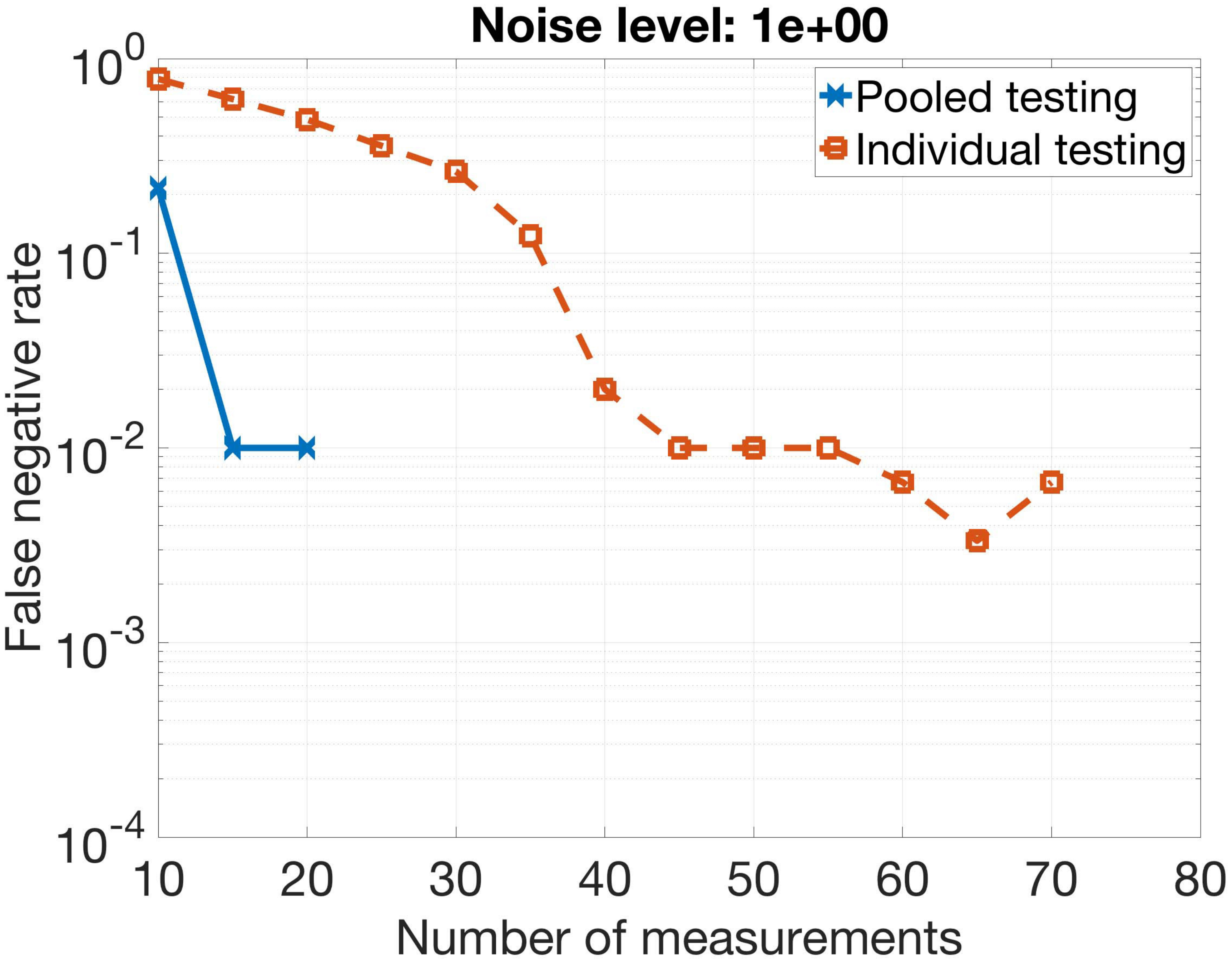}}
   \subfloat[FNR ($\Pb_{out} = 0.05$)]{\includegraphics[scale=0.14]{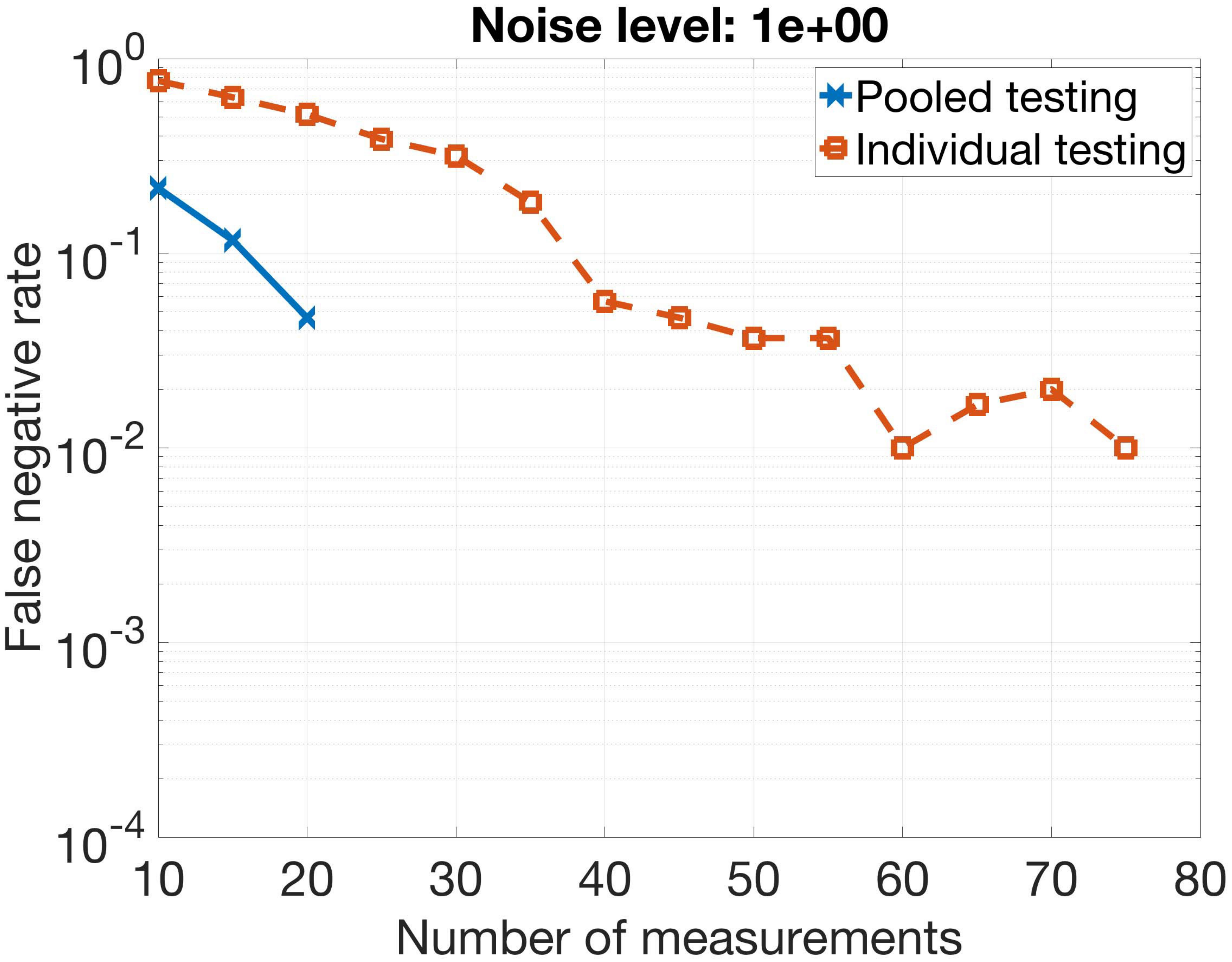}}
   \subfloat[FNR ($\Pb_{out} = 0.15$)]{\includegraphics[scale=0.14]{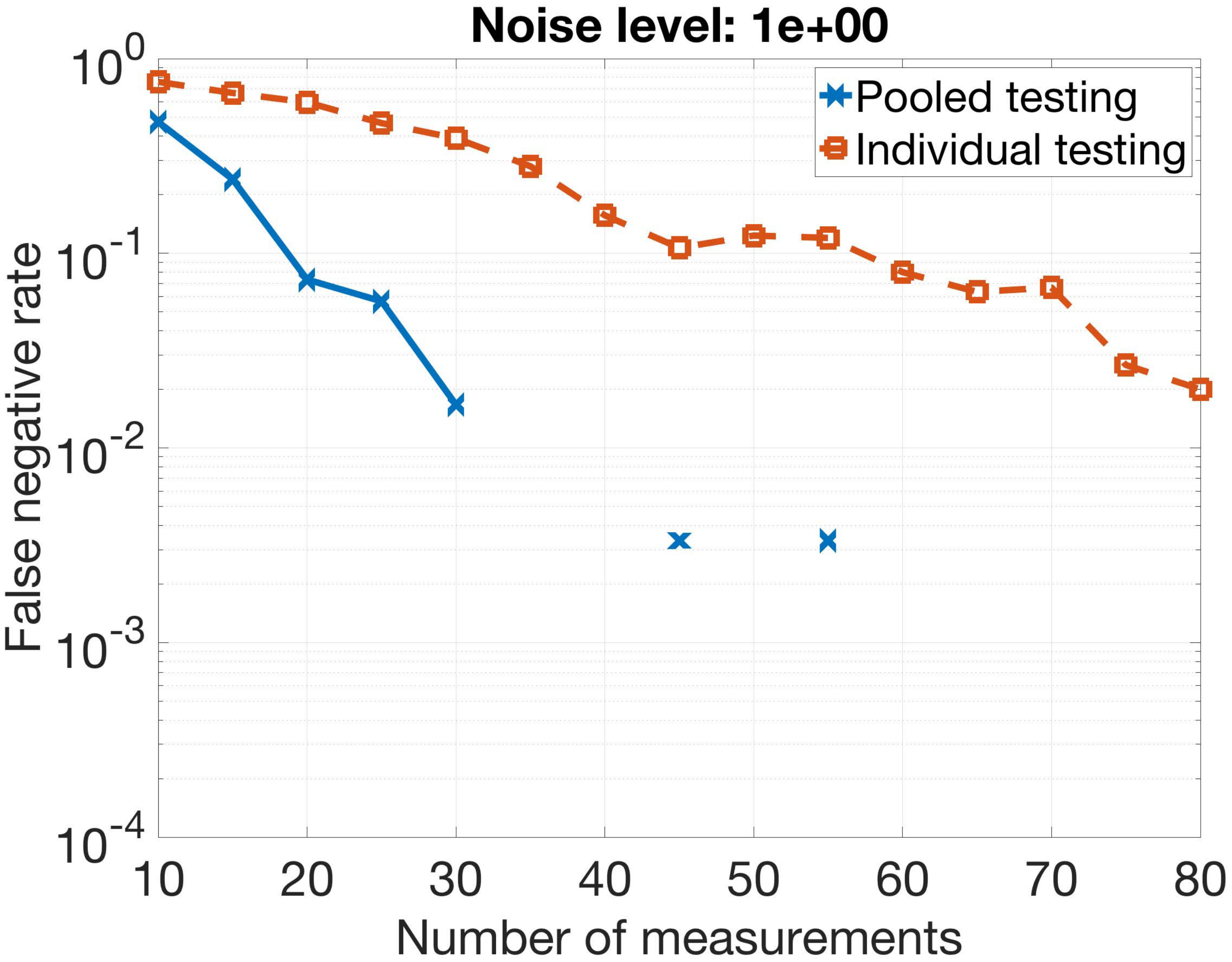}} \\
   \subfloat[FPR ($\Pb_{out} = 0.01$)]{\includegraphics[scale=0.14]{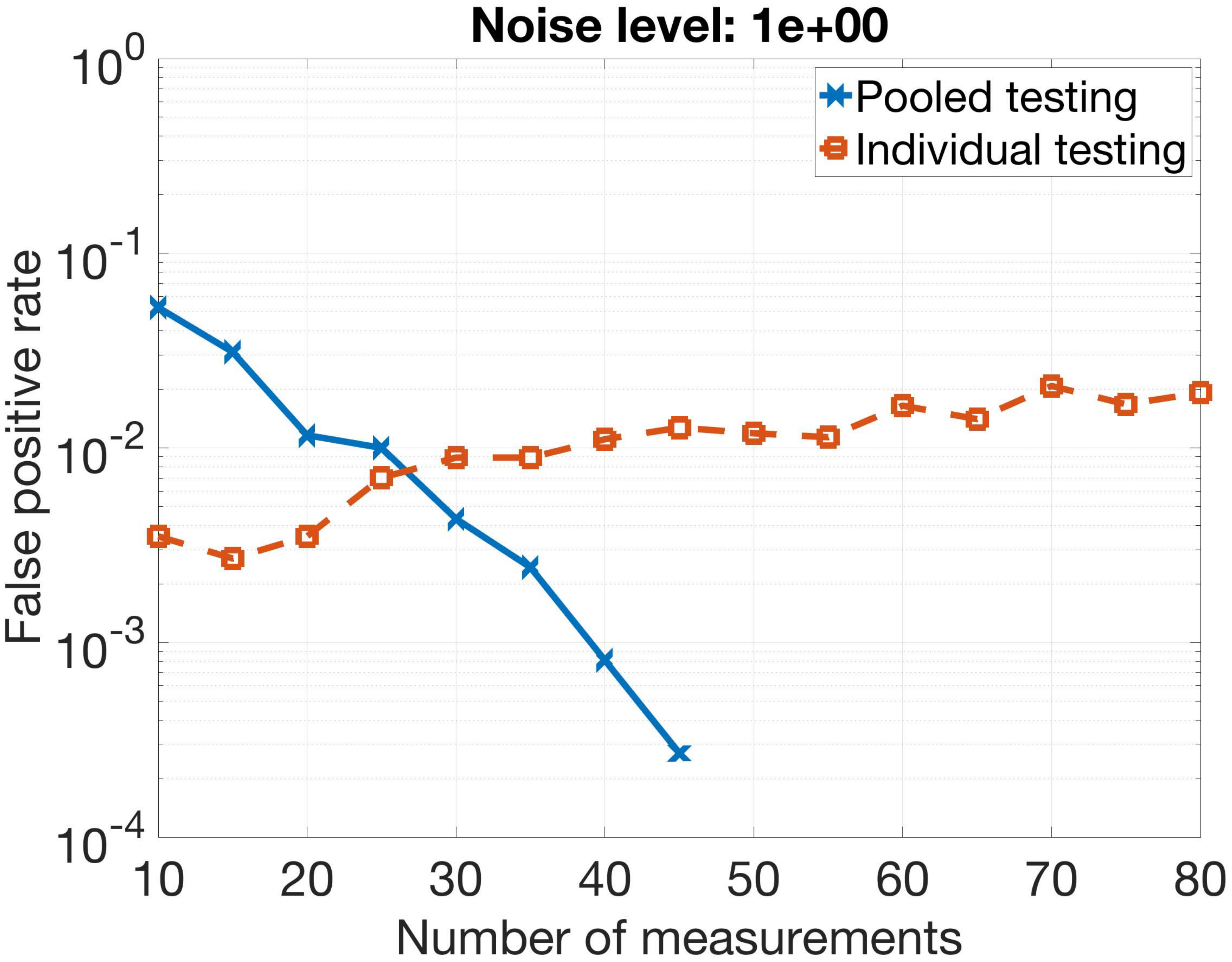}}
   \subfloat[FPR ($\Pb_{out} = 0.05$)]{\includegraphics[scale=0.14]{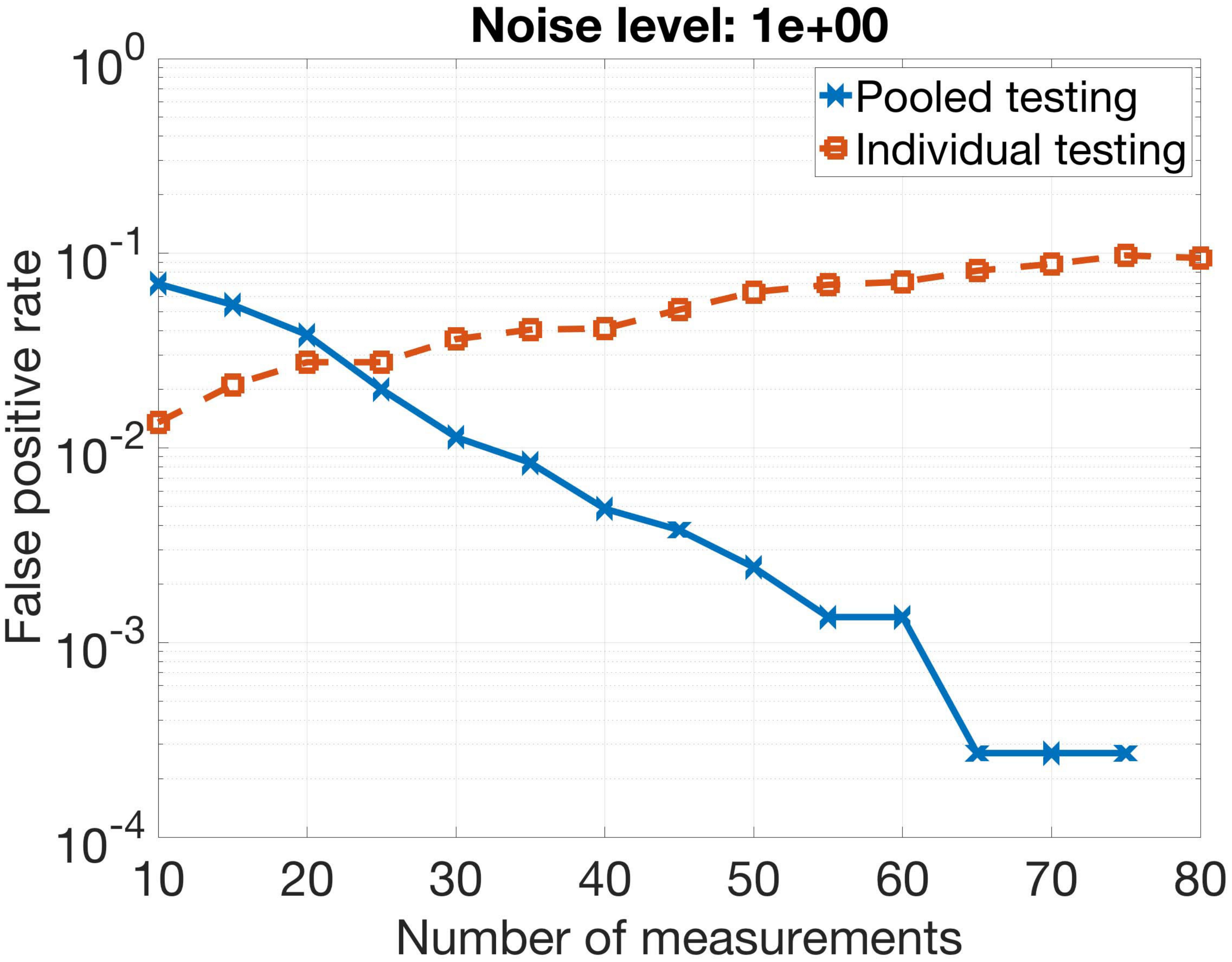}}
   \subfloat[FPR ($\Pb_{out} = 0.15$)]{\includegraphics[scale=0.14]{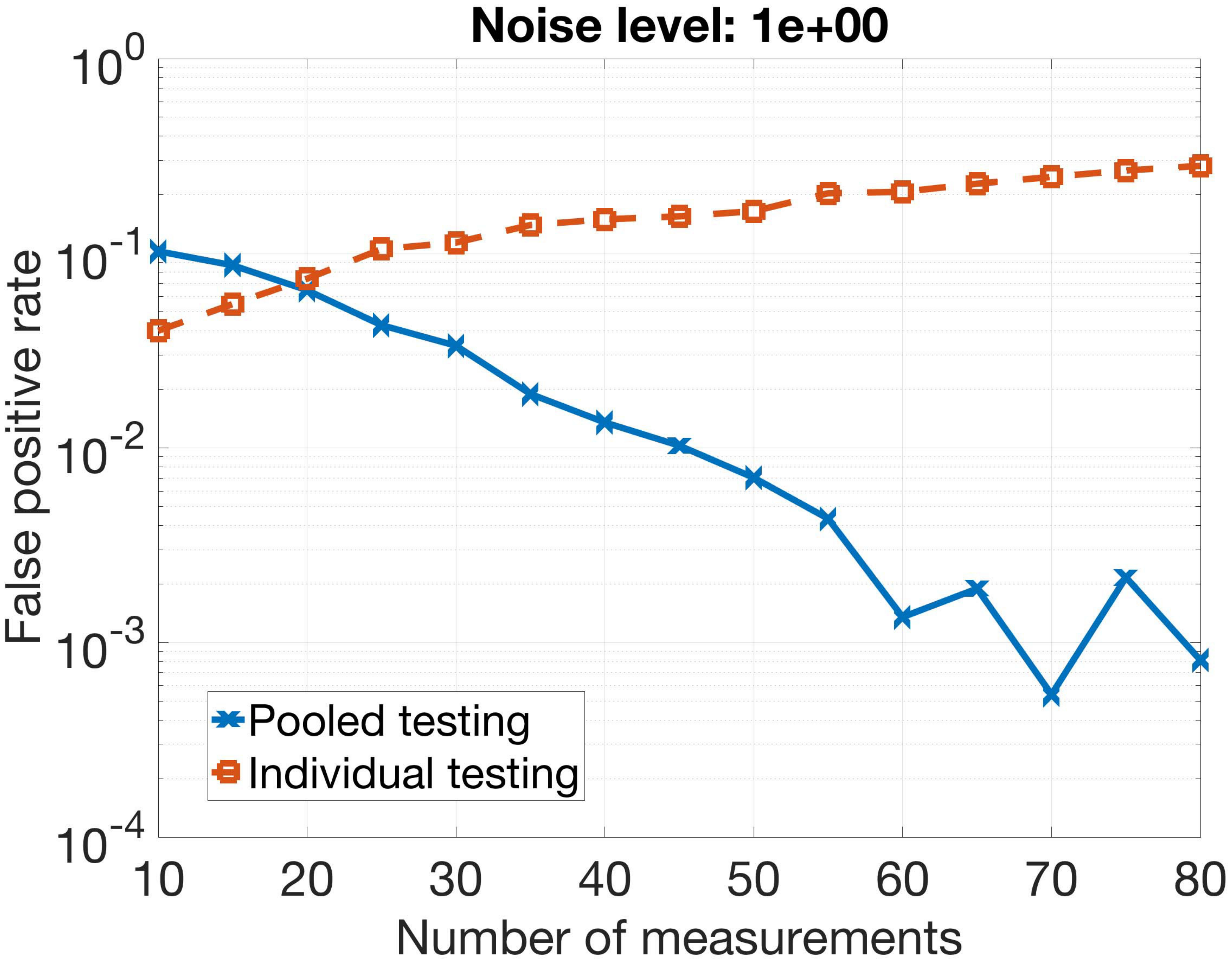}}
    \caption{\small Simulations for different probabilities of outlier errors. False Negative Rate (FNR) and False Positive Rate (FPR) with $n=40$, $k=3$, and Gaussian noise level 1e0.}
    \label{fig:FNR_FPR_N40_K3}
\end{figure*}

\begin{figure*}[!htb]
    \centering
   \subfloat[FNR ($\Pb_{out} = 0.01$)]{\includegraphics[scale=0.14]{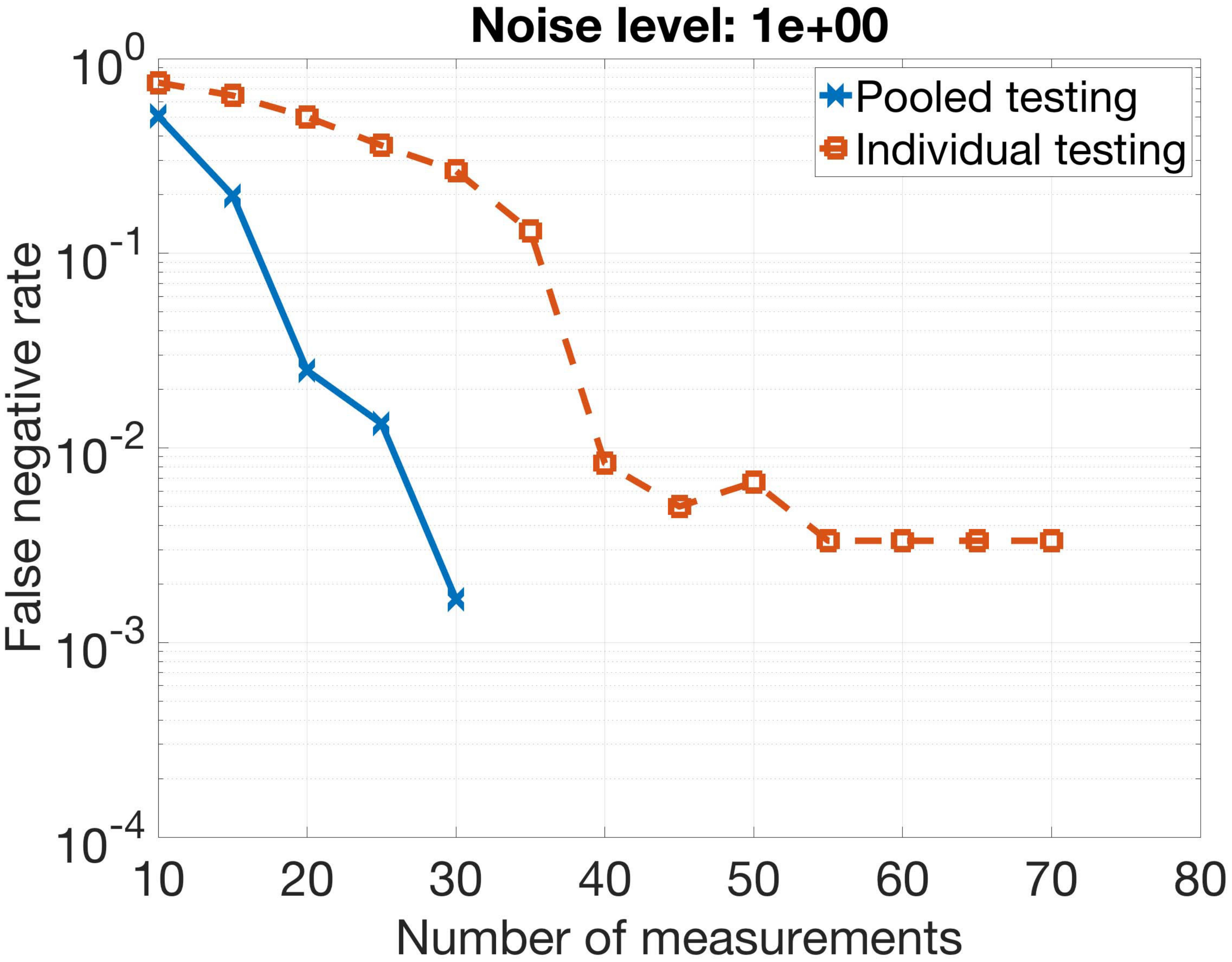}}
   \subfloat[FNR ($\Pb_{out} = 0.05$)]{\includegraphics[scale=0.14]{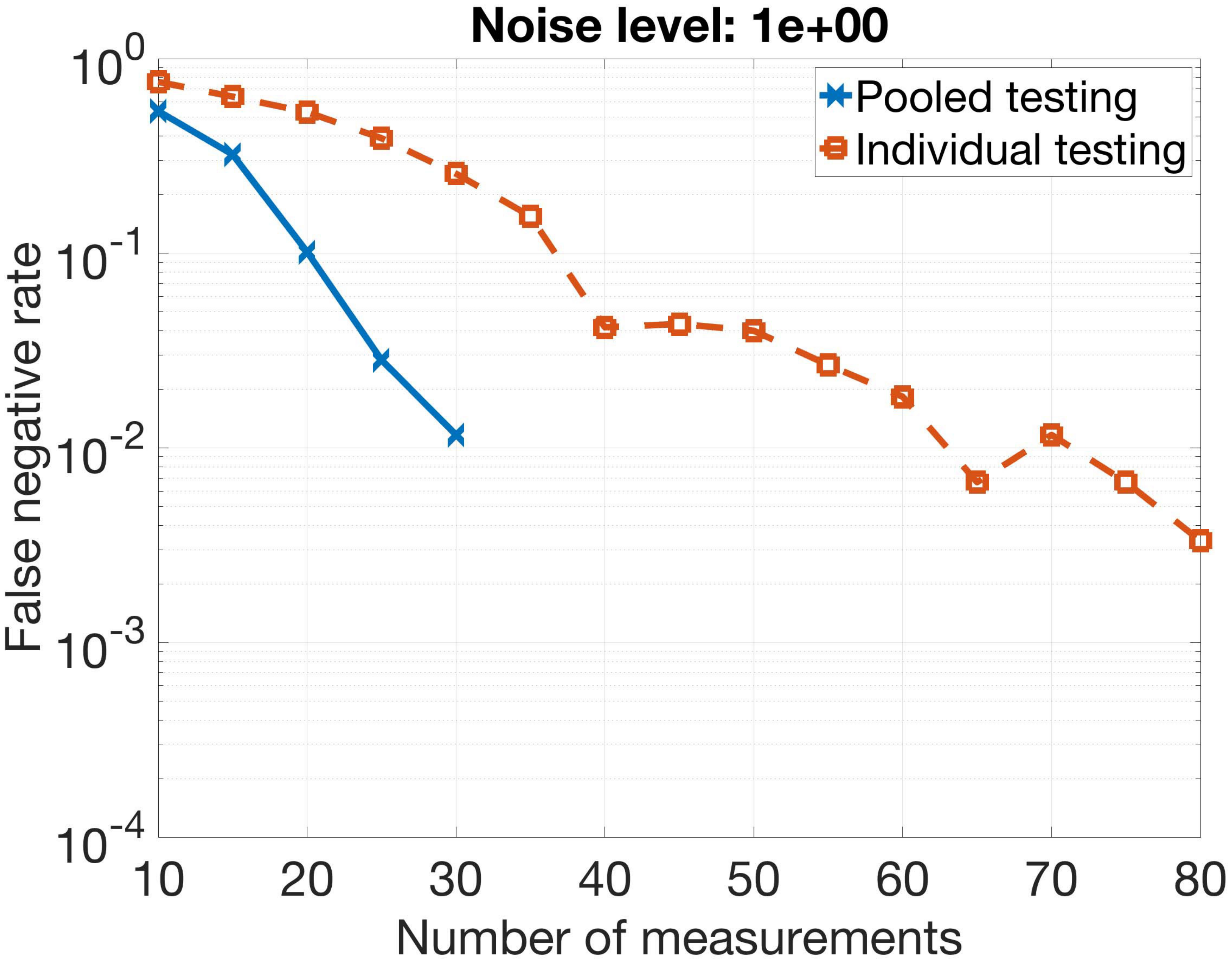}}
   \subfloat[FNR ($\Pb_{out} = 0.15$)]{\includegraphics[scale=0.14]{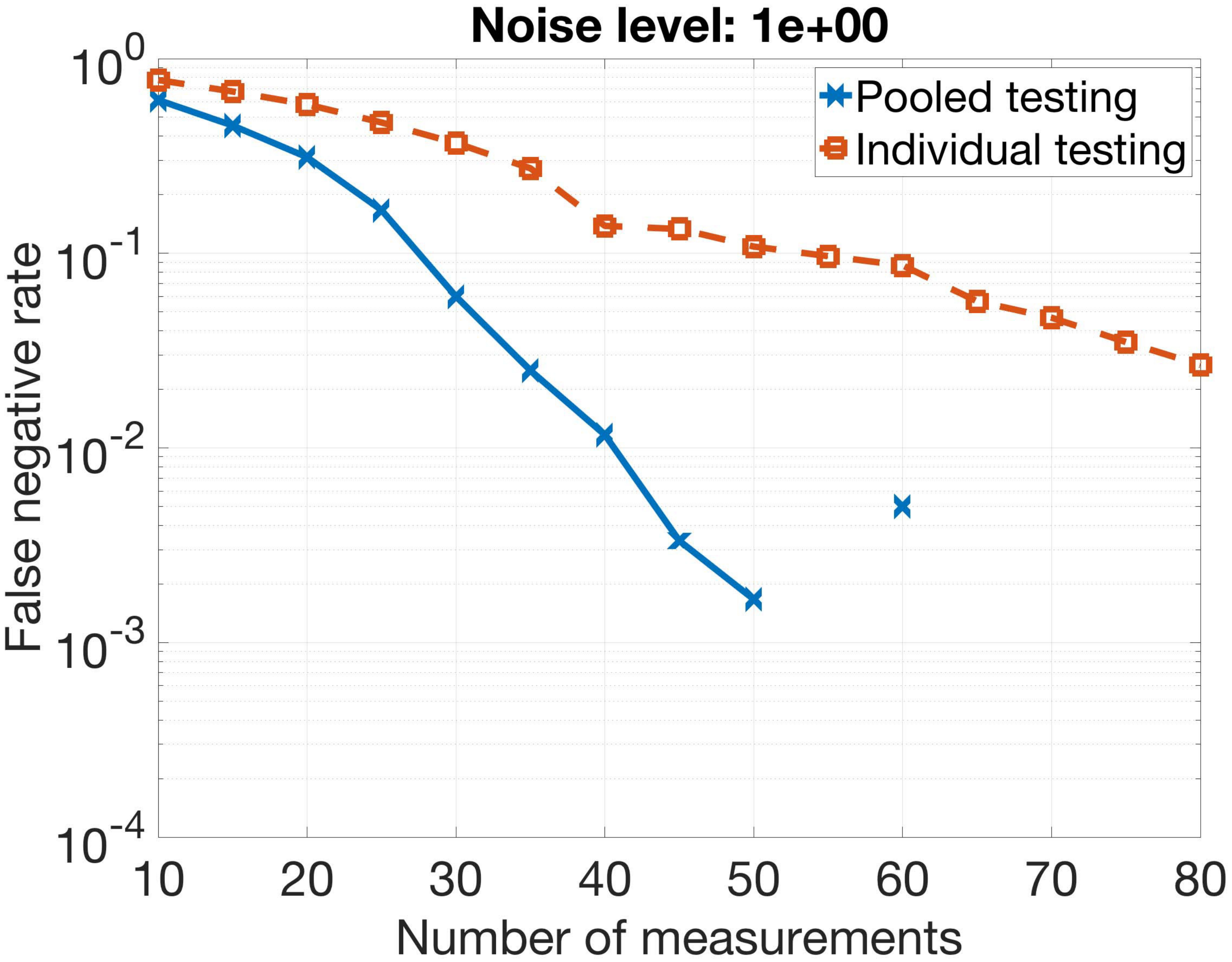}} \\
   \subfloat[FPR ($\Pb_{out} = 0.01$)]{\includegraphics[scale=0.14]{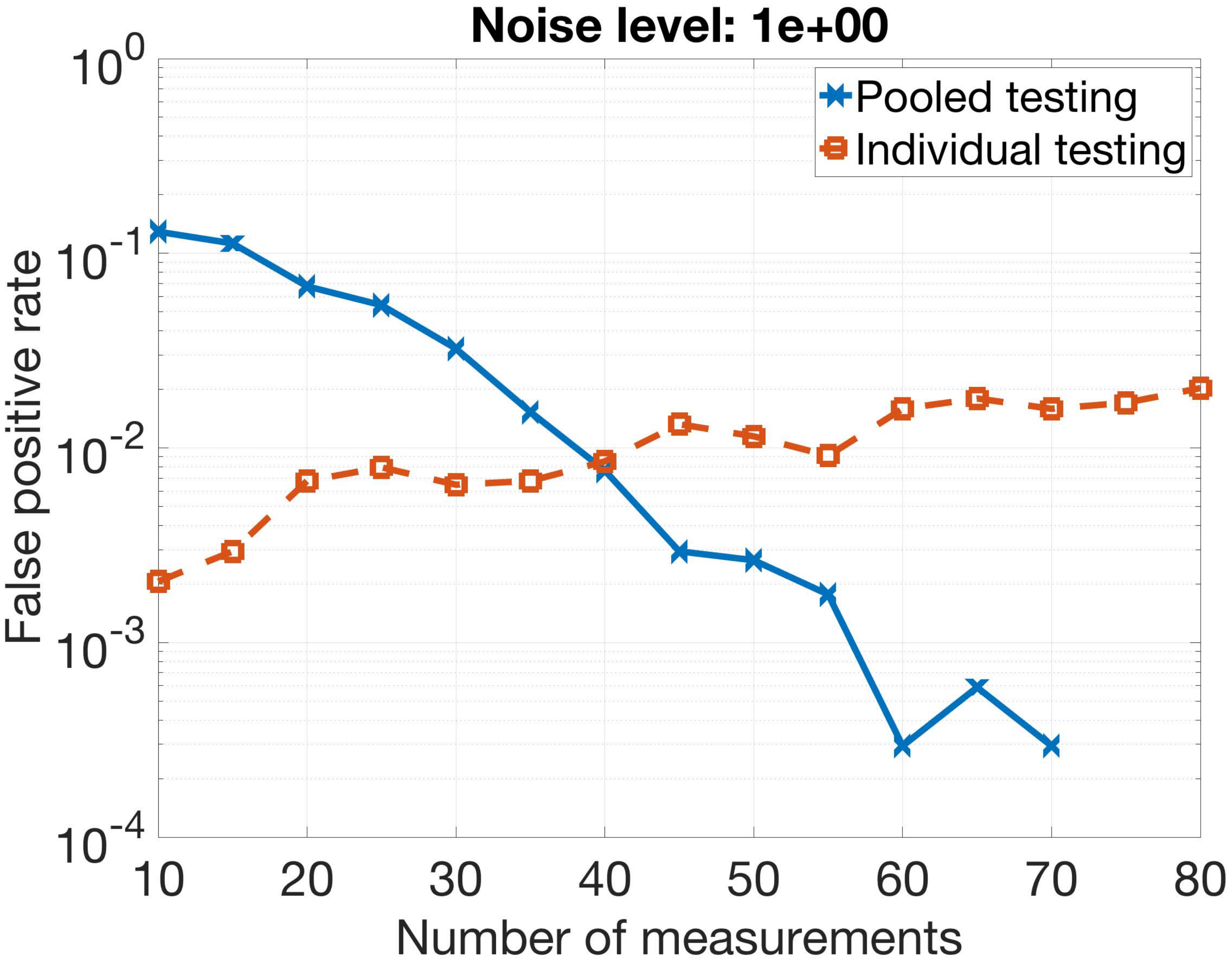}}
   \subfloat[FPR ($\Pb_{out} = 0.05$)]{\includegraphics[scale=0.14]{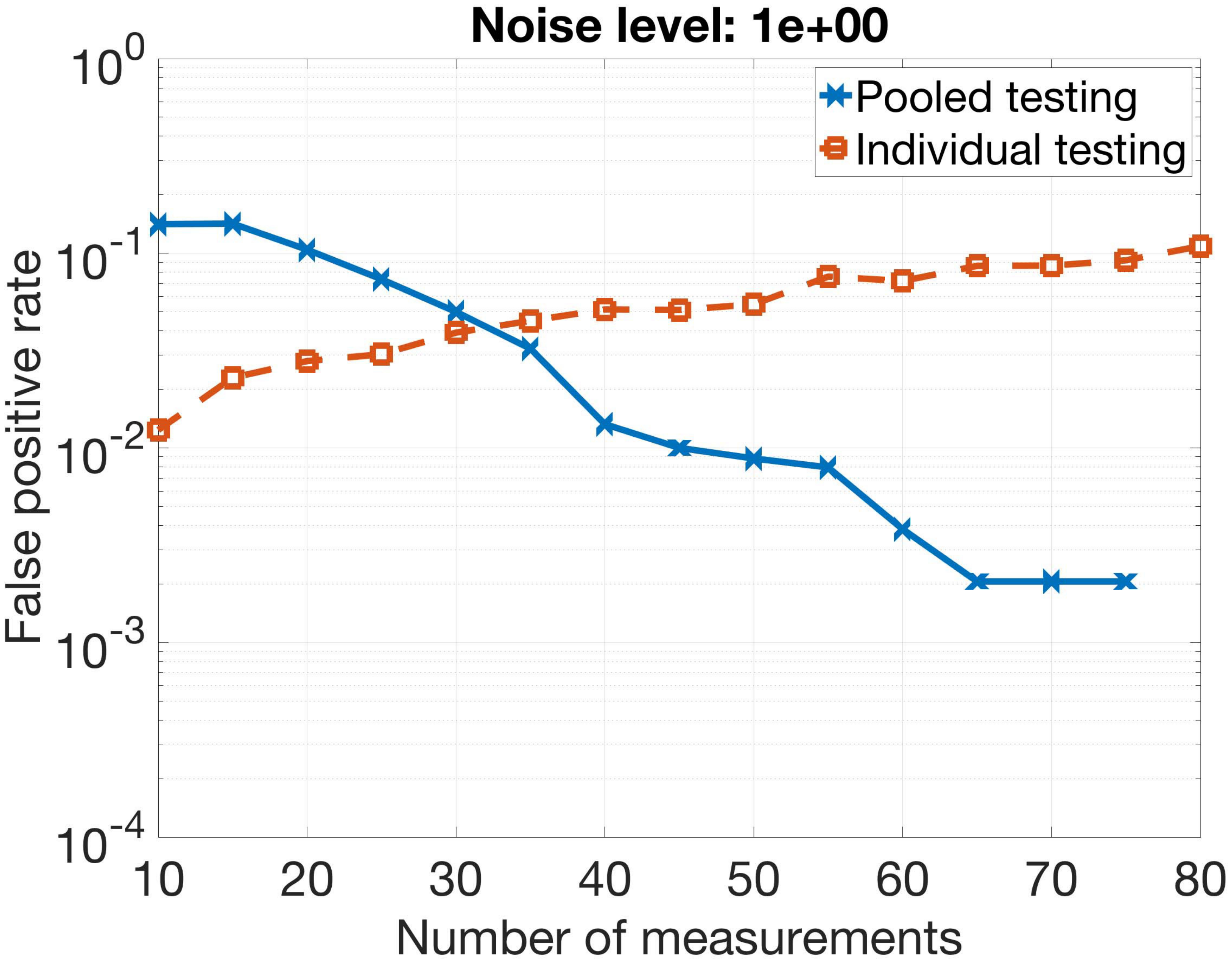}}
   \subfloat[FPR ($\Pb_{out} = 0.15$)]{\includegraphics[scale=0.14]{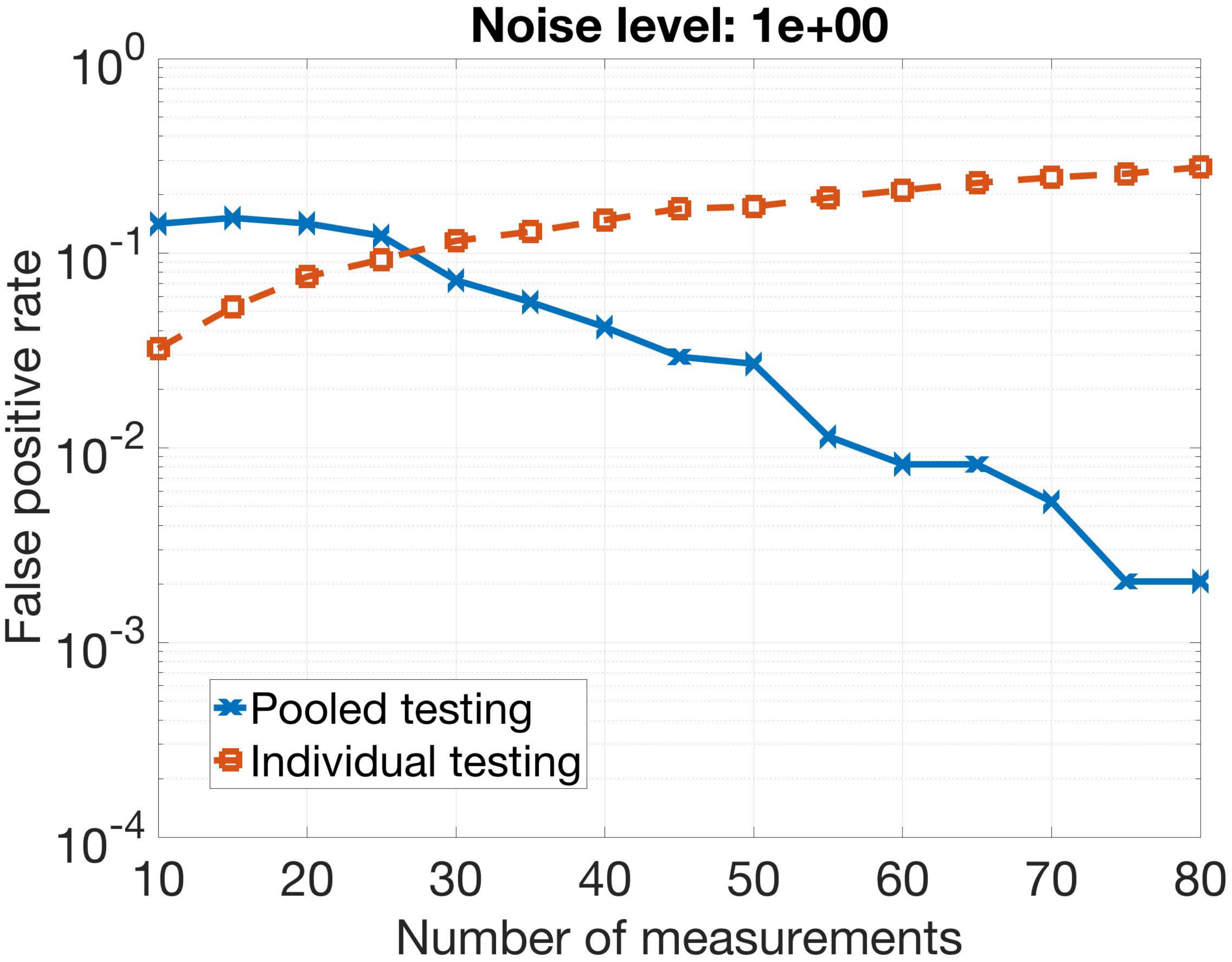}}
    \caption{\small Simulations for different probabilities of outlier errors. False Negative Rate (FNR) and the corresponding False Positive Rate (FPR) with $n=40$, $k=6$, and Gaussian noise level 1e0.}
    \label{fig:FNR_FPR_N40_K6}
\end{figure*}

Furthermore, we demonstrate the outperformance of the pooled testing in the COVID-19 testing against the individual testing with more test subjects. Figures from \ref{fig:FNR_FPR_N40_K1} to \ref{fig:FNR_FPR_N40_K6} show the comparison in both FNR and FPR as the number of measurements increases, between the pooling testing and the individual testing,  for $n=40$. In Figures from \ref{fig:FNR_FPR_N40_K1} to \ref{fig:FNR_FPR_N40_K6}, (a), (b), and (c) show the FNR of the pooled testing and the individual testing with different probabilities of outlier error from $1\%$ to $15\%$ and different sparsity level from $k=1$ to $k=6$. Correspondingly, in Figures from \ref{fig:FNR_FPR_N40_K1} to \ref{fig:FNR_FPR_N40_K6}, (d), (e), and (f) indicate the FPR of the both testing. Through the simulation results shown in Figures from \ref{fig:FNR_FPR_N40_K1} to \ref{fig:FNR_FPR_N40_K6}, with even larger $n$, it is shown that the pooled testing can identify people having COVID-19 virus more accurately than the individual testing with small number of measurements. Therefore, the pooled testing can simultaneously have higher throughput and higher accuracy than the individual testing.  


\subsection{Different noise levels}
\begin{figure*}[!htb]
    \centering
   \subfloat[FNR (Noise level: 5e-1)]{\includegraphics[scale=0.14]{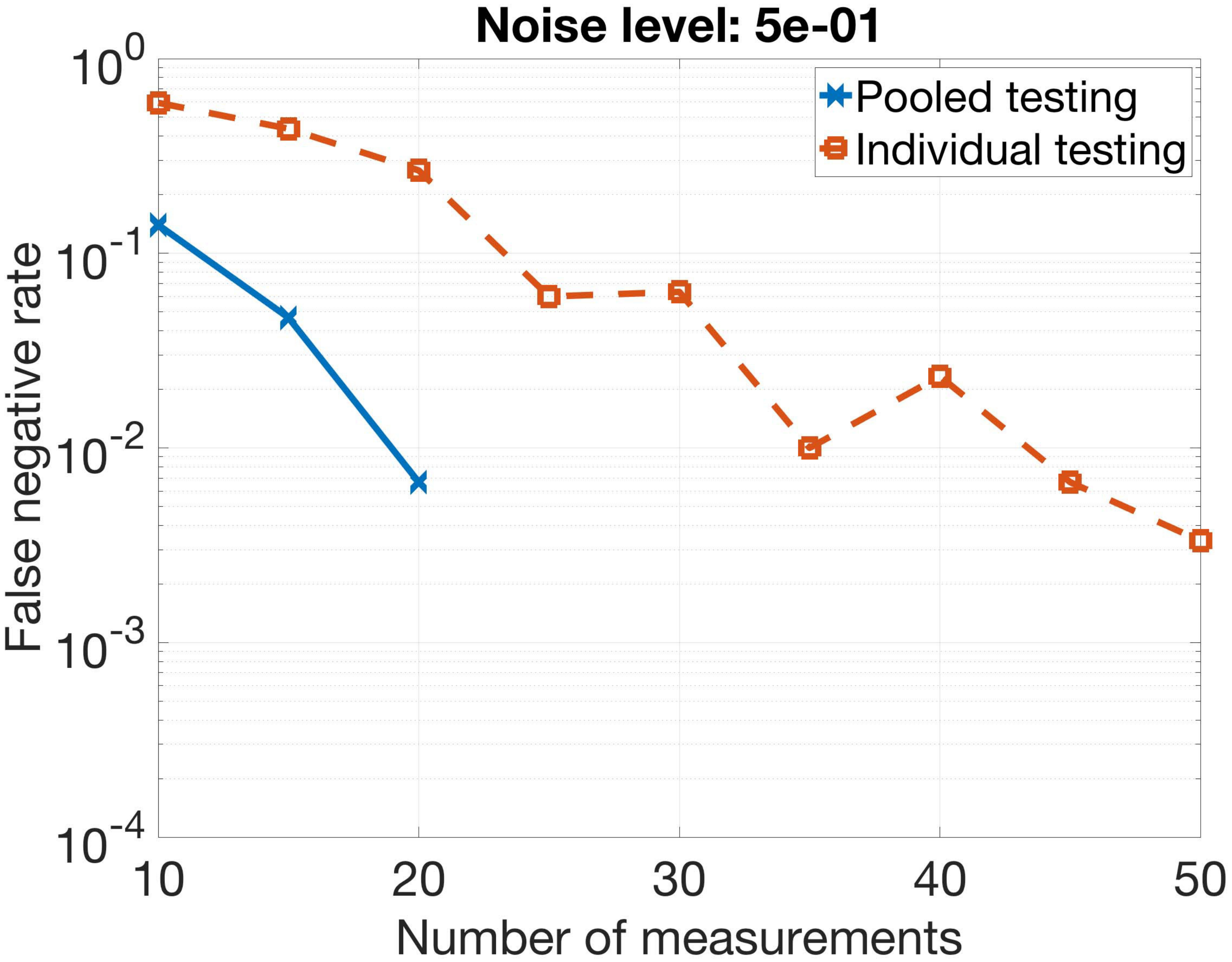}}
   \subfloat[FNR (Noise level: 1e0)]{\includegraphics[scale=0.14]{img/FNR_N25_K3_Bernoulli_Optimization_GaussianTrunc_Pout5e-02_Noise1e00-eps-converted-to}}
   \subfloat[FNR (Noise level: 2e0)]{\includegraphics[scale=0.14]{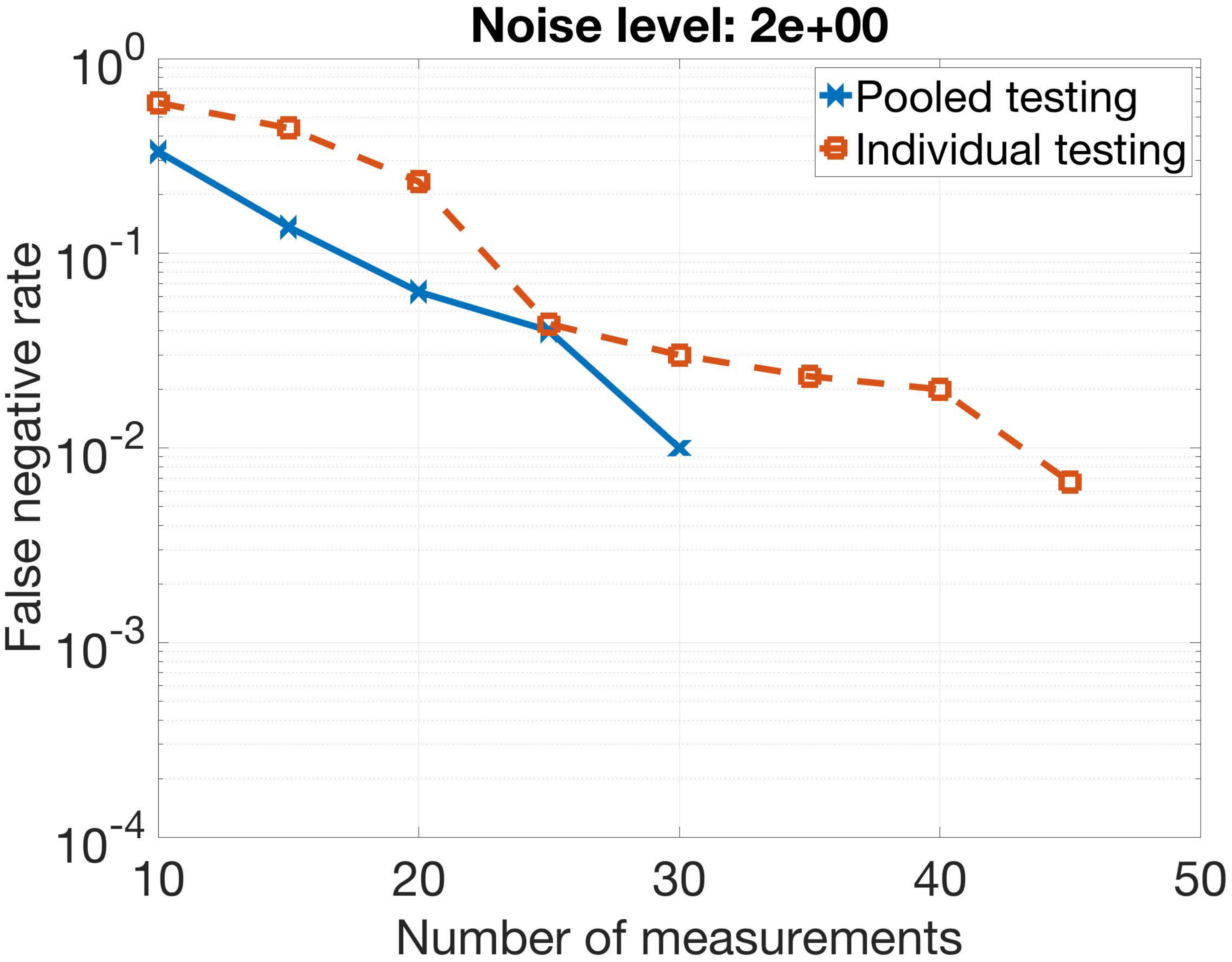}} \\
   \subfloat[FPR (Noise level: 5e-1)]{\includegraphics[scale=0.14]{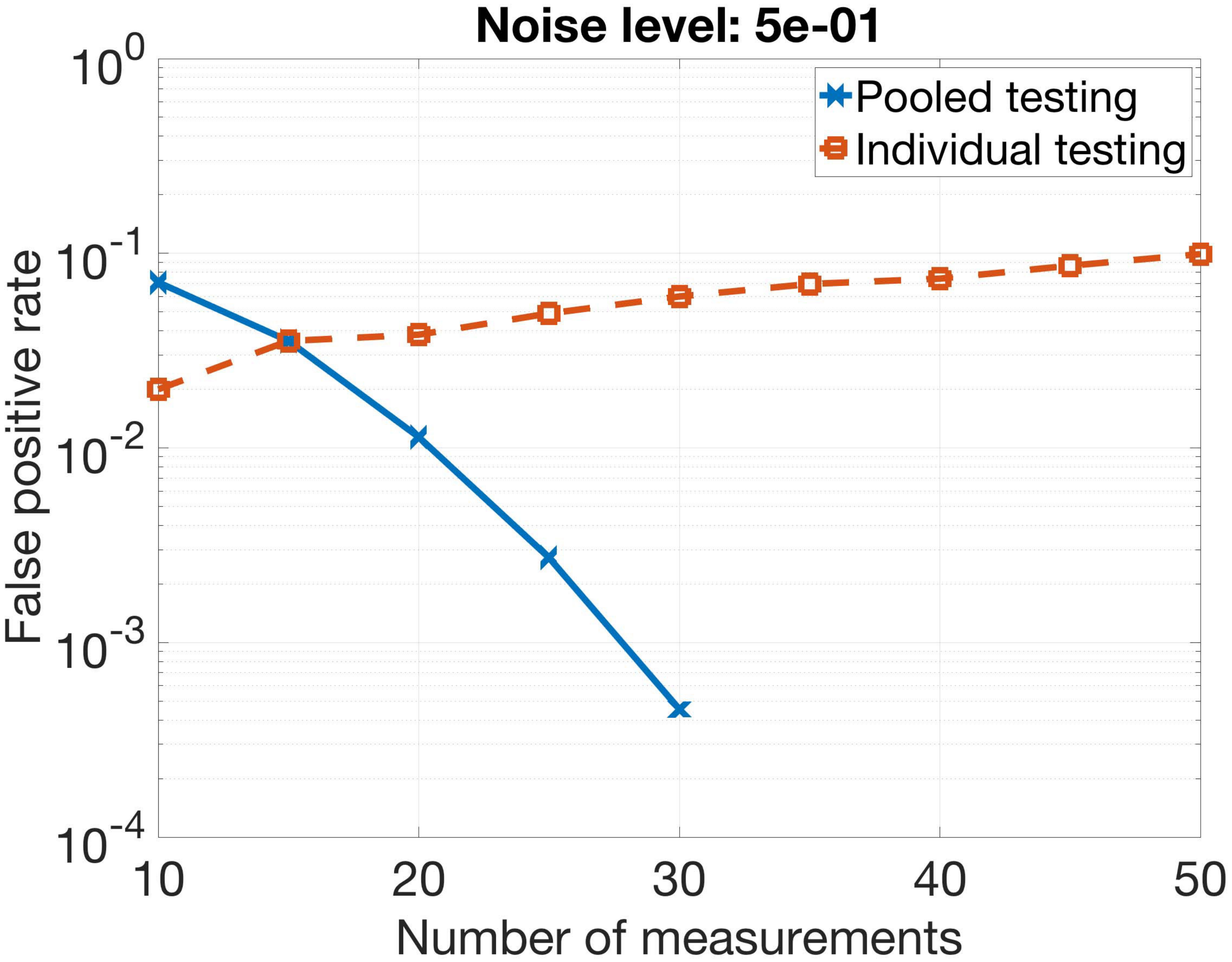}}
   \subfloat[FPR (Noise level: 1e0)]{\includegraphics[scale=0.14]{img/FPR_N25_K3_Bernoulli_Optimization_GaussianTrunc_Pout5e-02_Noise1e00-eps-converted-to}}
   \subfloat[FPR (Noise level: 2e0)]{\includegraphics[scale=0.14]{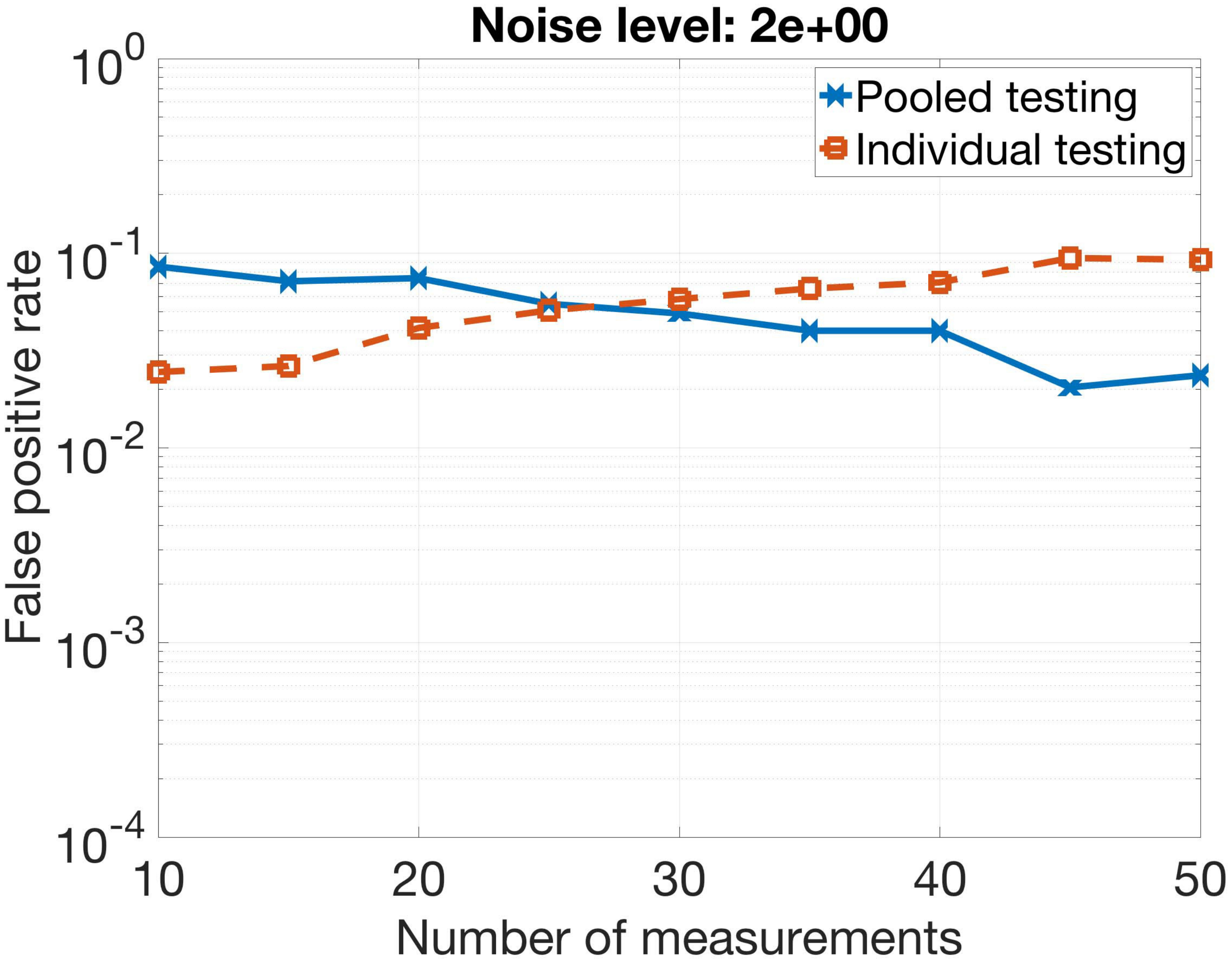}}
    \caption{\small Simulations for different noise levels. False Negative Rate (FNR) and the corresponding False Positive Rate (FPR) with $n=25$, $k=3$, $\Pb_{out}=0.05$, and noise level varied from 5e-1 to 2e0.}
    \label{fig:FNR_FPR_N25_K3_Pout5-02}
\end{figure*}

\begin{figure*}[!htb]
    \centering
   \subfloat[FNR (Noise level: 5e-1)]{\includegraphics[scale=0.14]{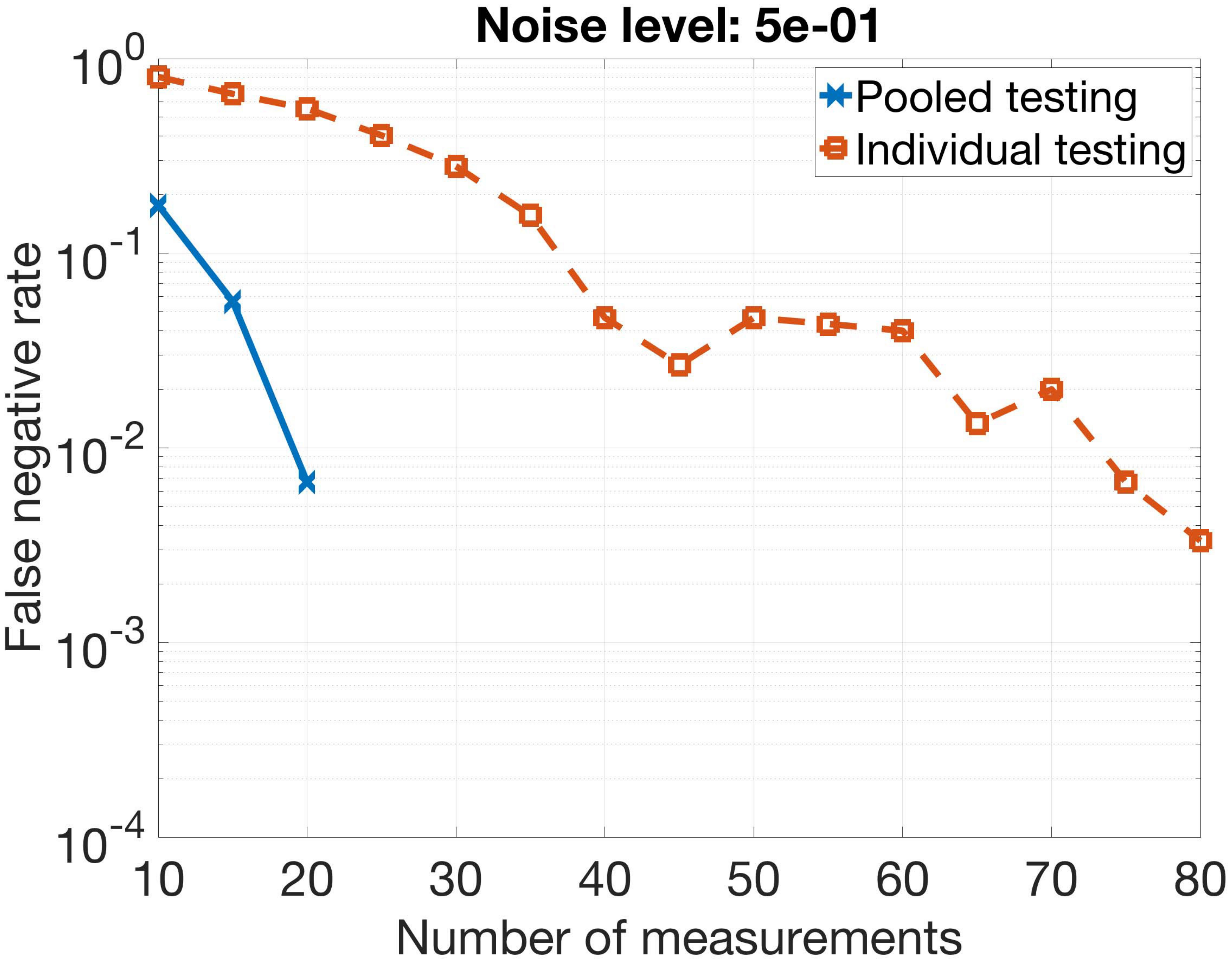}}
   \subfloat[FNR (Noise level: 1e0)]{\includegraphics[scale=0.14]{img/FNR_N40_K3_Bernoulli_Optimization_GaussianTrunc_Pout5e-02_Noise1e00-eps-converted-to}}
   \subfloat[FNR (Noise level: 2e0)]{\includegraphics[scale=0.14]{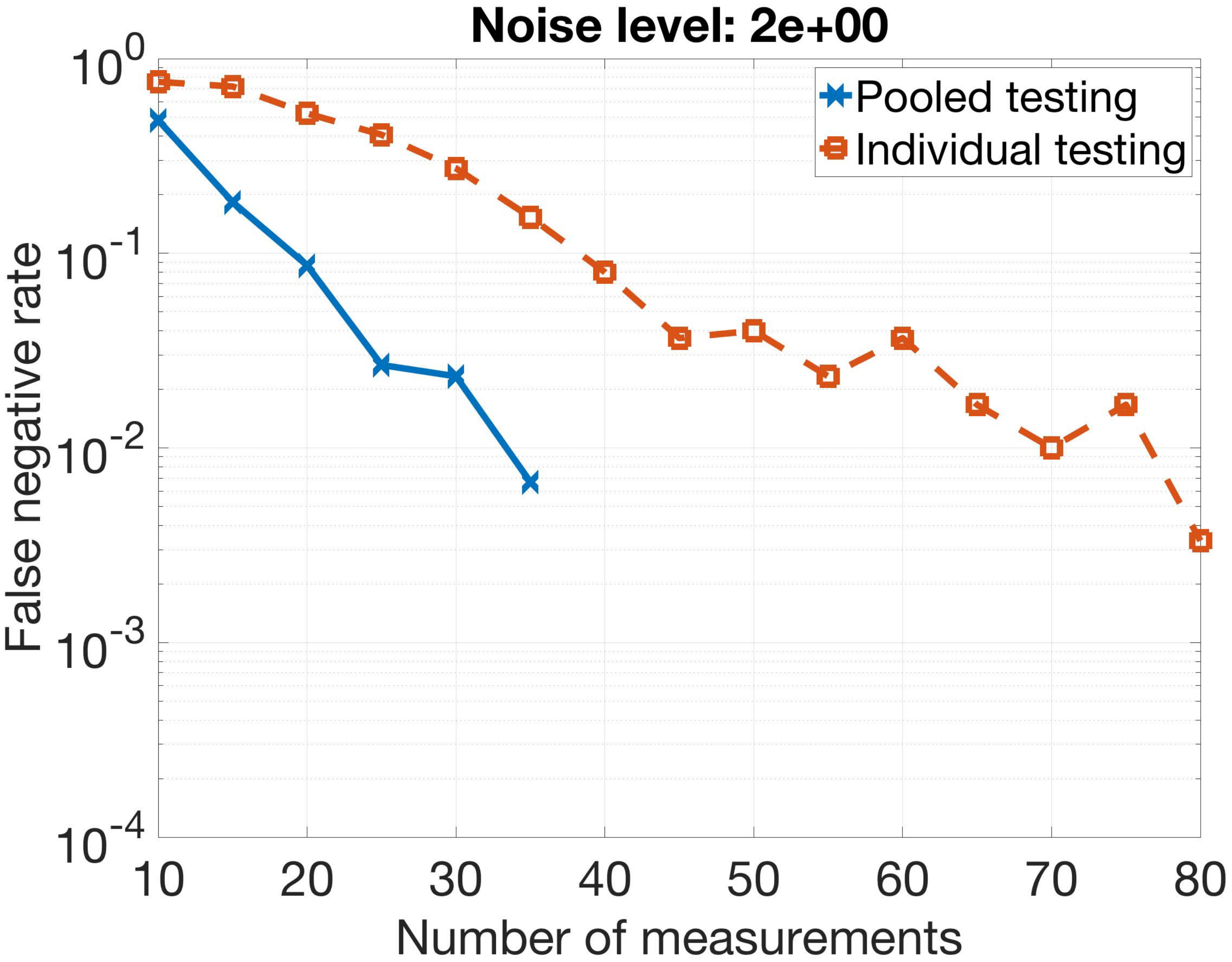}} \\
   \subfloat[FPR (Noise level: 5e-1)]{\includegraphics[scale=0.14]{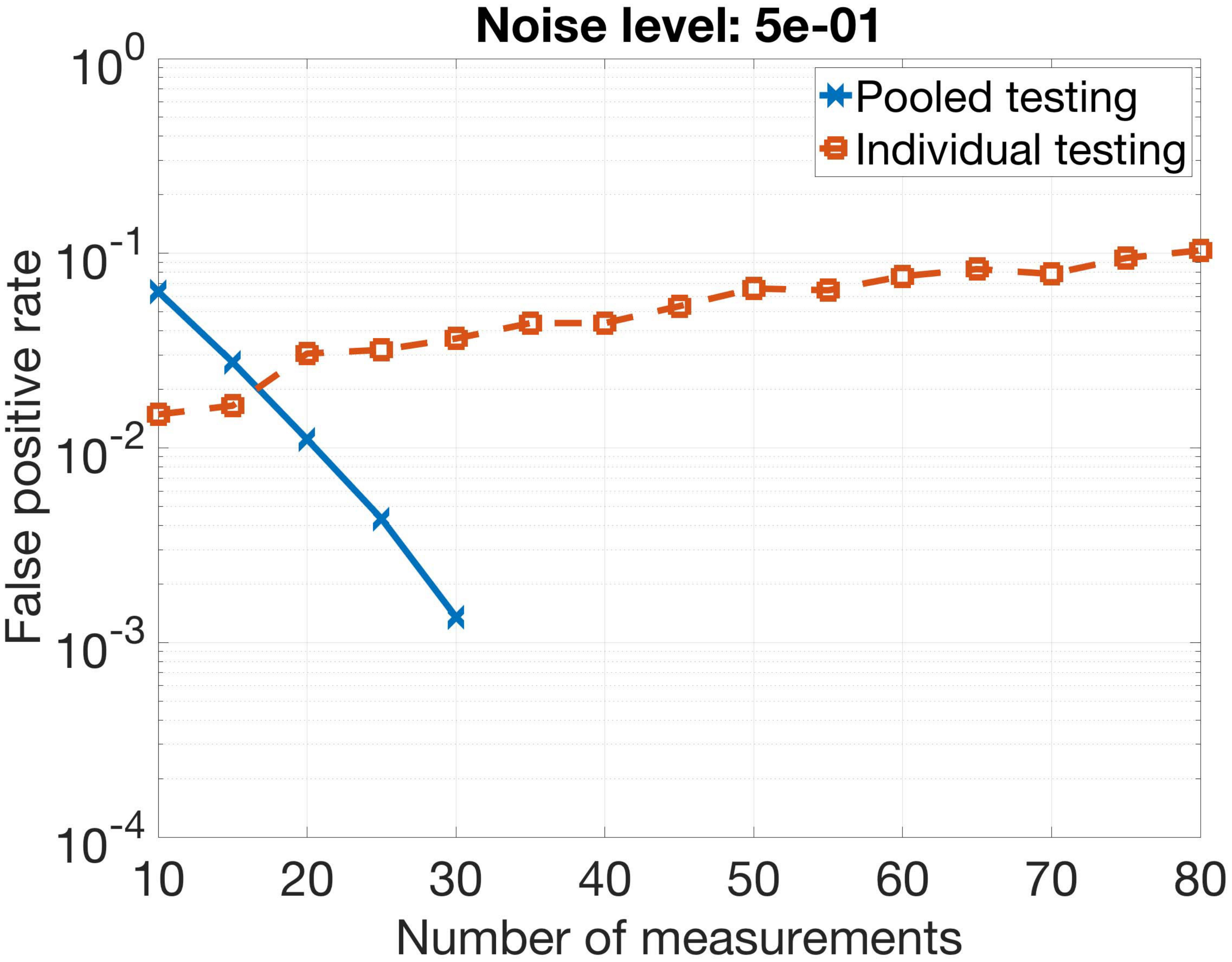}}
   \subfloat[FPR (Noise level: 1e0)]{\includegraphics[scale=0.14]{img/FPR_N40_K3_Bernoulli_Optimization_GaussianTrunc_Pout5e-02_Noise1e00-eps-converted-to}}
   \subfloat[FPR (Noise level: 2e0)]{\includegraphics[scale=0.14]{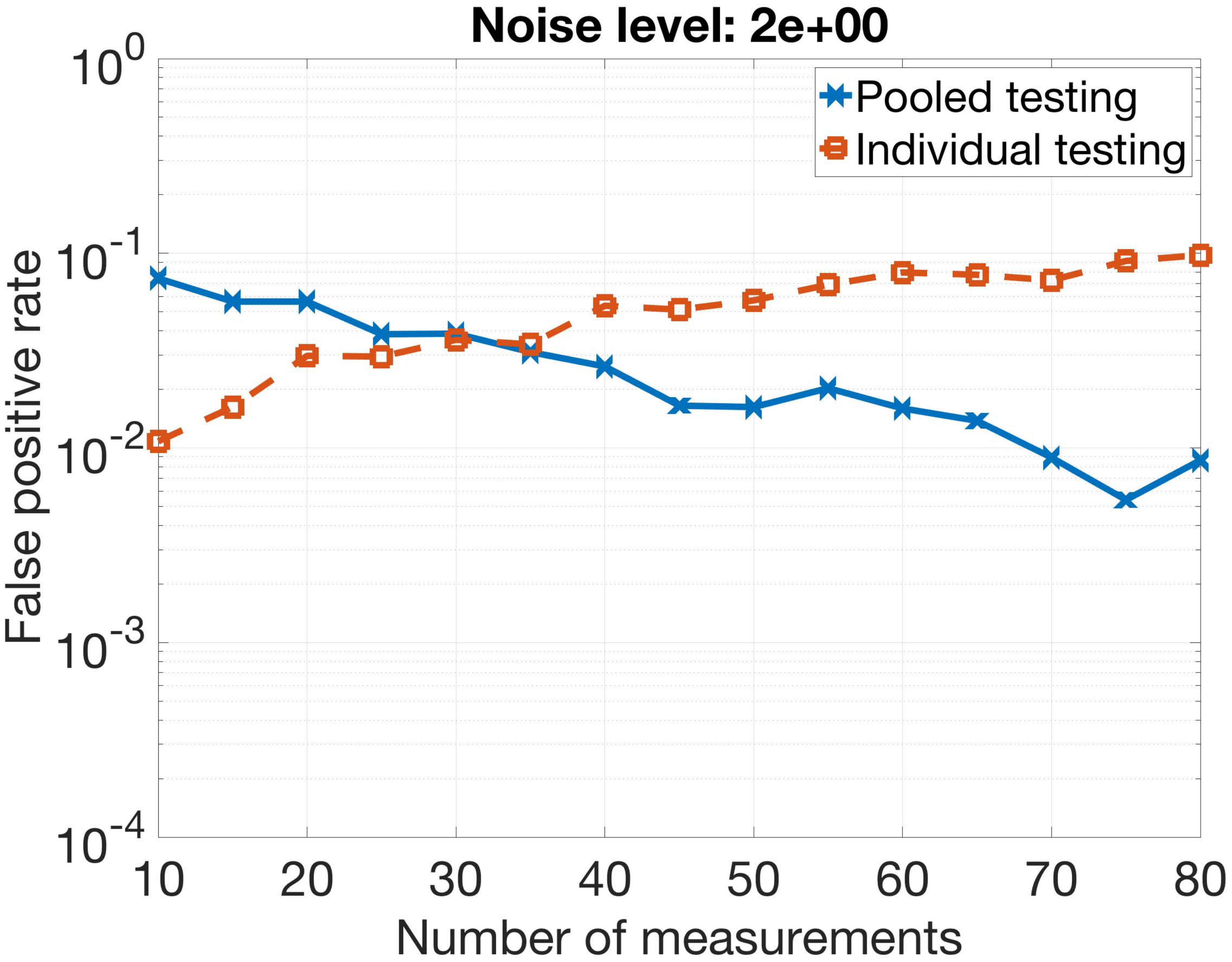}}
    \caption{\small Simulations for different noise levels. False Negative Rate (FNR) and the corresponding False Positive Rate (FPR) with $n=40$, $k=3$, $\Pb_{out}=0.05$, and noise level varied from 5e-1 to 2e0.}
    \label{fig:FNR_FPR_N40_K3_Pout5-02}
\end{figure*}

\begin{figure*}[!htb]
    \centering
   \subfloat[FNR (Noise level: 5e-1)]{\includegraphics[scale=0.14]{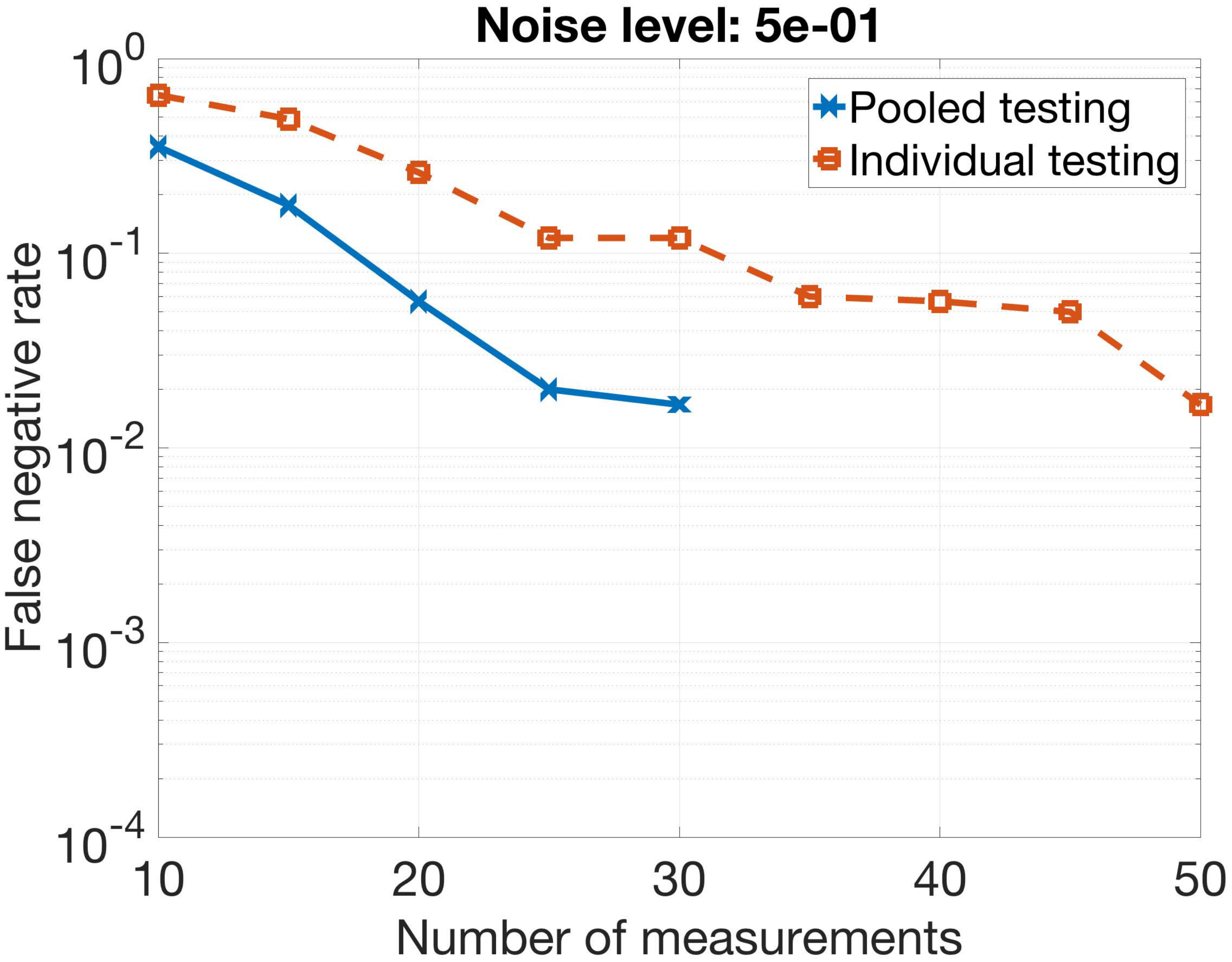}}
   \subfloat[FNR (Noise level: 1e0)]{\includegraphics[scale=0.14]{img/FNR_N25_K3_Bernoulli_Optimization_GaussianTrunc_Pout15e-02_Noise1e00-eps-converted-to}}
   \subfloat[FNR (Noise level: 2e0)]{\includegraphics[scale=0.14]{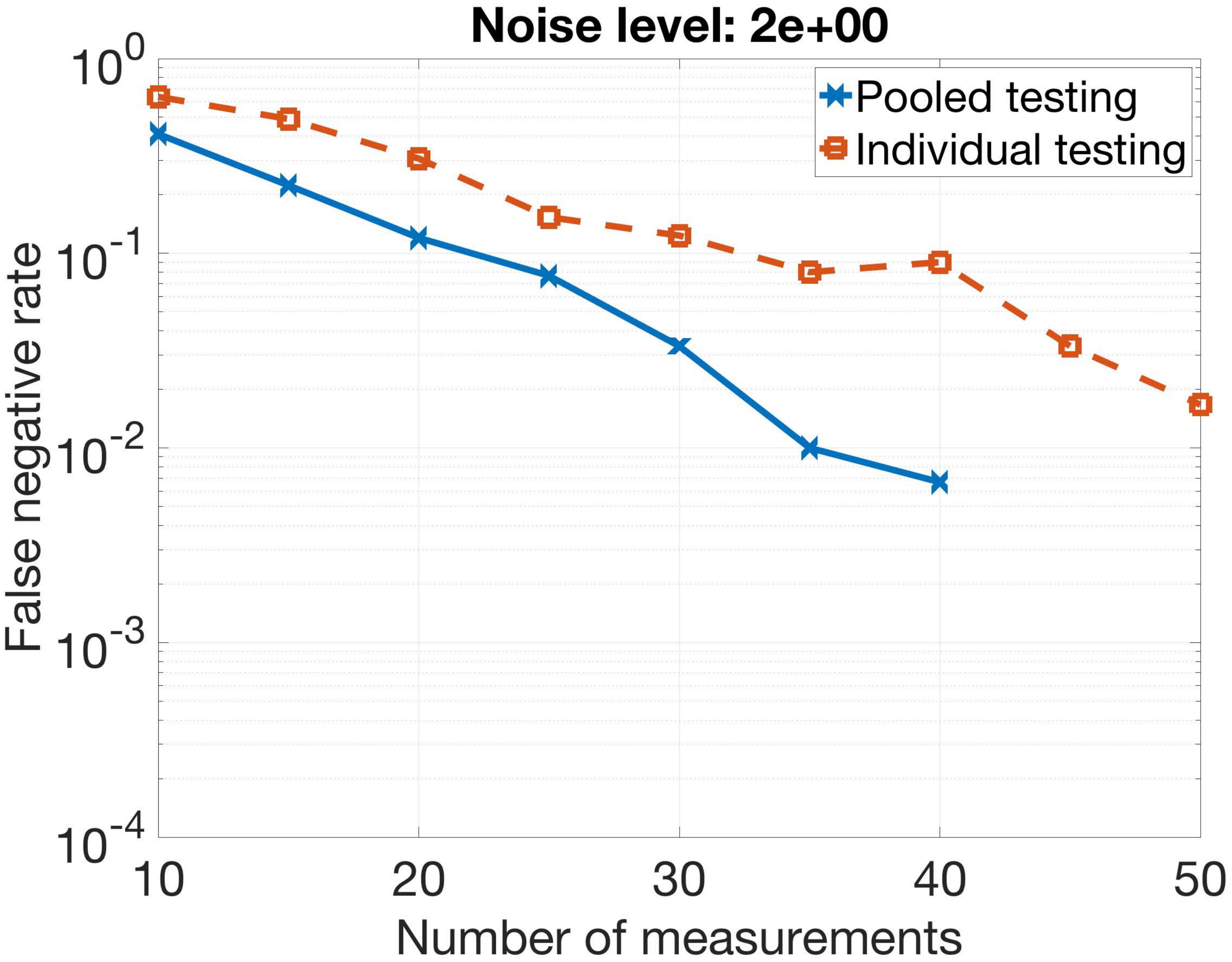}} \\
   \subfloat[FPR (Noise level: 5e-1)]{\includegraphics[scale=0.14]{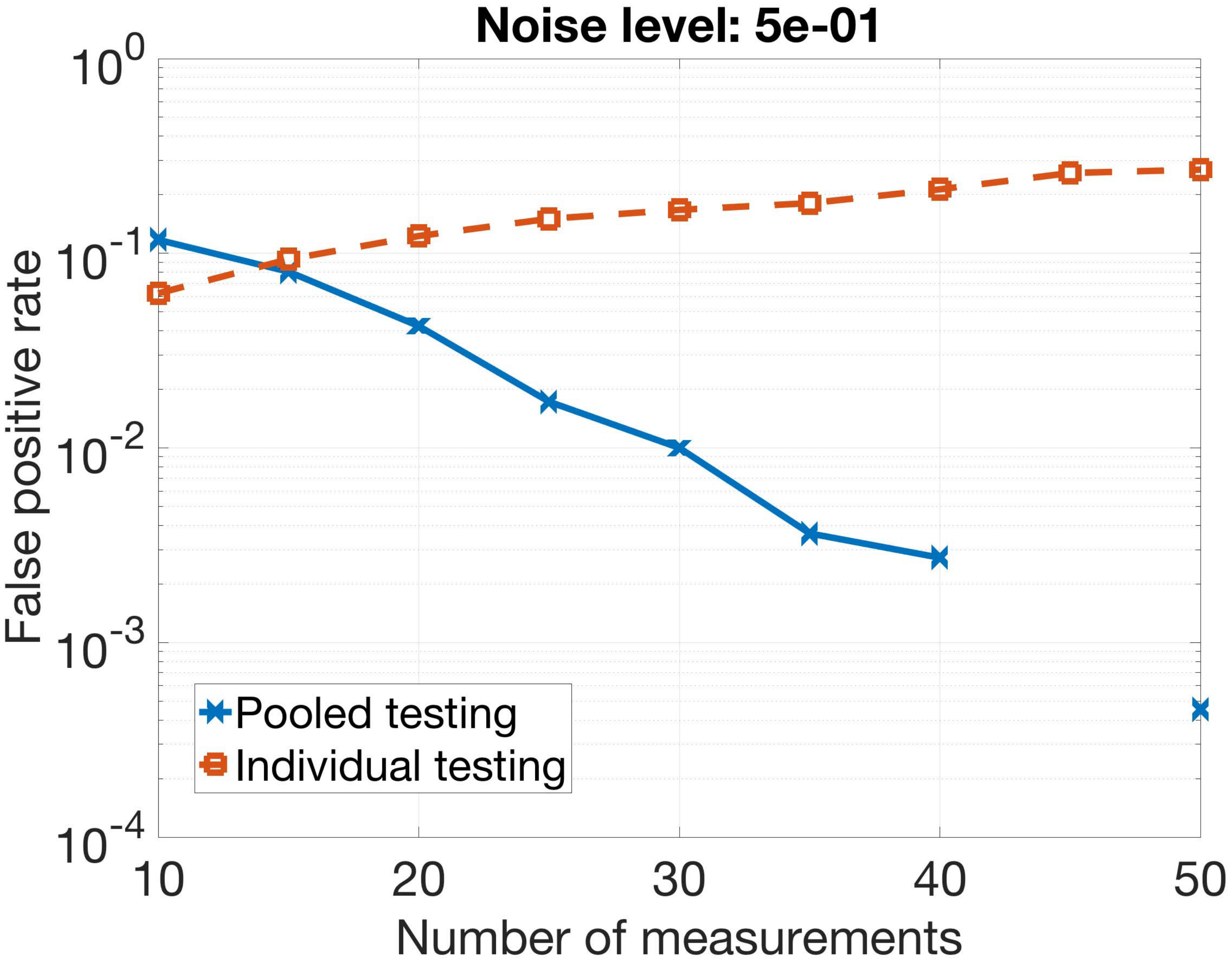}}
   \subfloat[FPR (Noise level: 1e0)]{\includegraphics[scale=0.14]{img/FPR_N25_K3_Bernoulli_Optimization_GaussianTrunc_Pout15e-02_Noise1e00-eps-converted-to}}
   \subfloat[FPR (Noise level: 2e0)]{\includegraphics[scale=0.14]{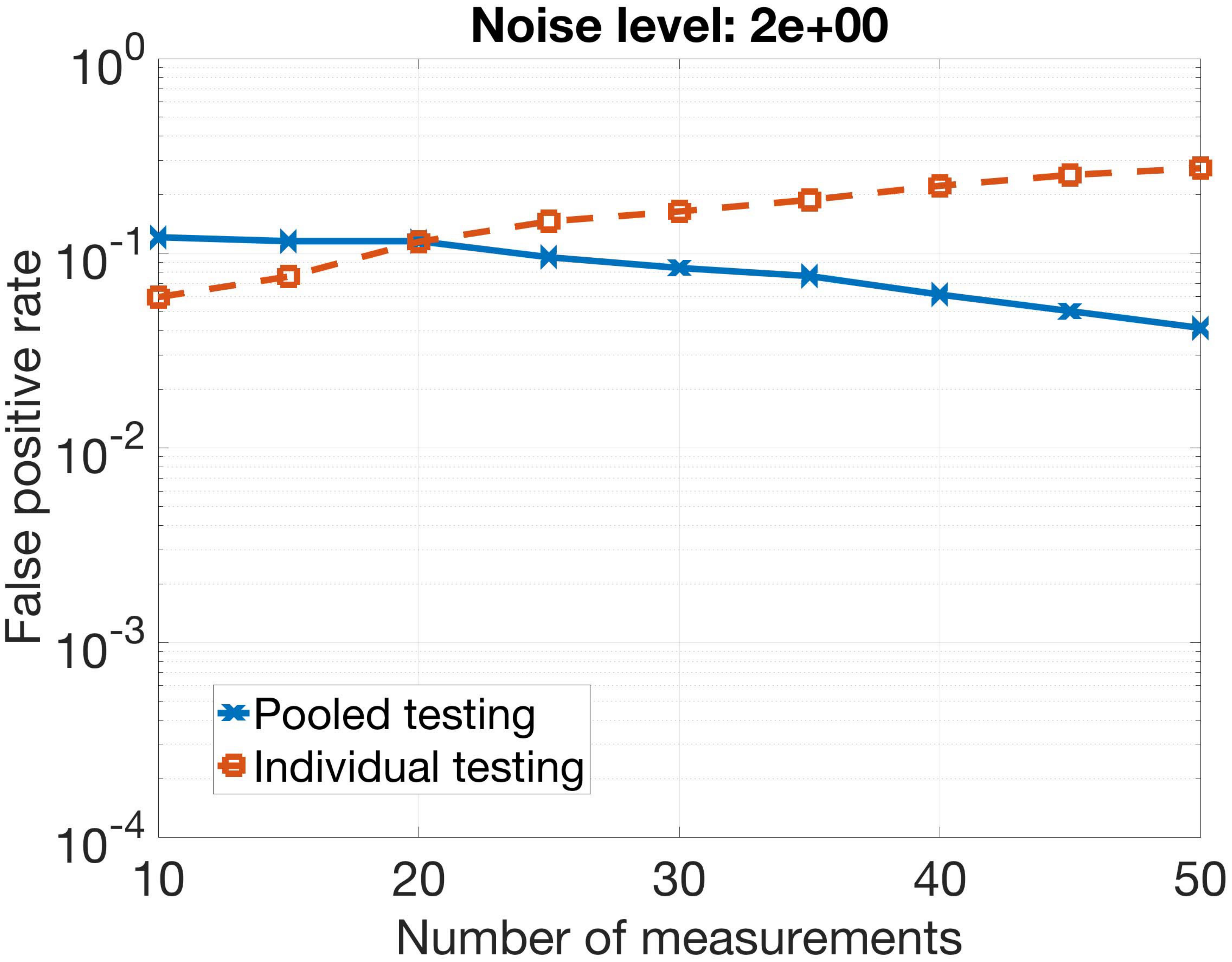}}
    \caption{\small Simulations for different noise levels. False Negative Rate (FNR) and the corresponding False Positive Rate (FPR) with $n=25$, $k=3$, $\Pb_{out}=0.15$, and noise level varied from 5e-1 to 2e0.}
    \label{fig:FNR_FPR_N25_K3_Pout15-02}
\end{figure*}

\begin{figure*}[!htb]
    \centering
   \subfloat[FNR (Noise level: 5e-1)]{\includegraphics[scale=0.14]{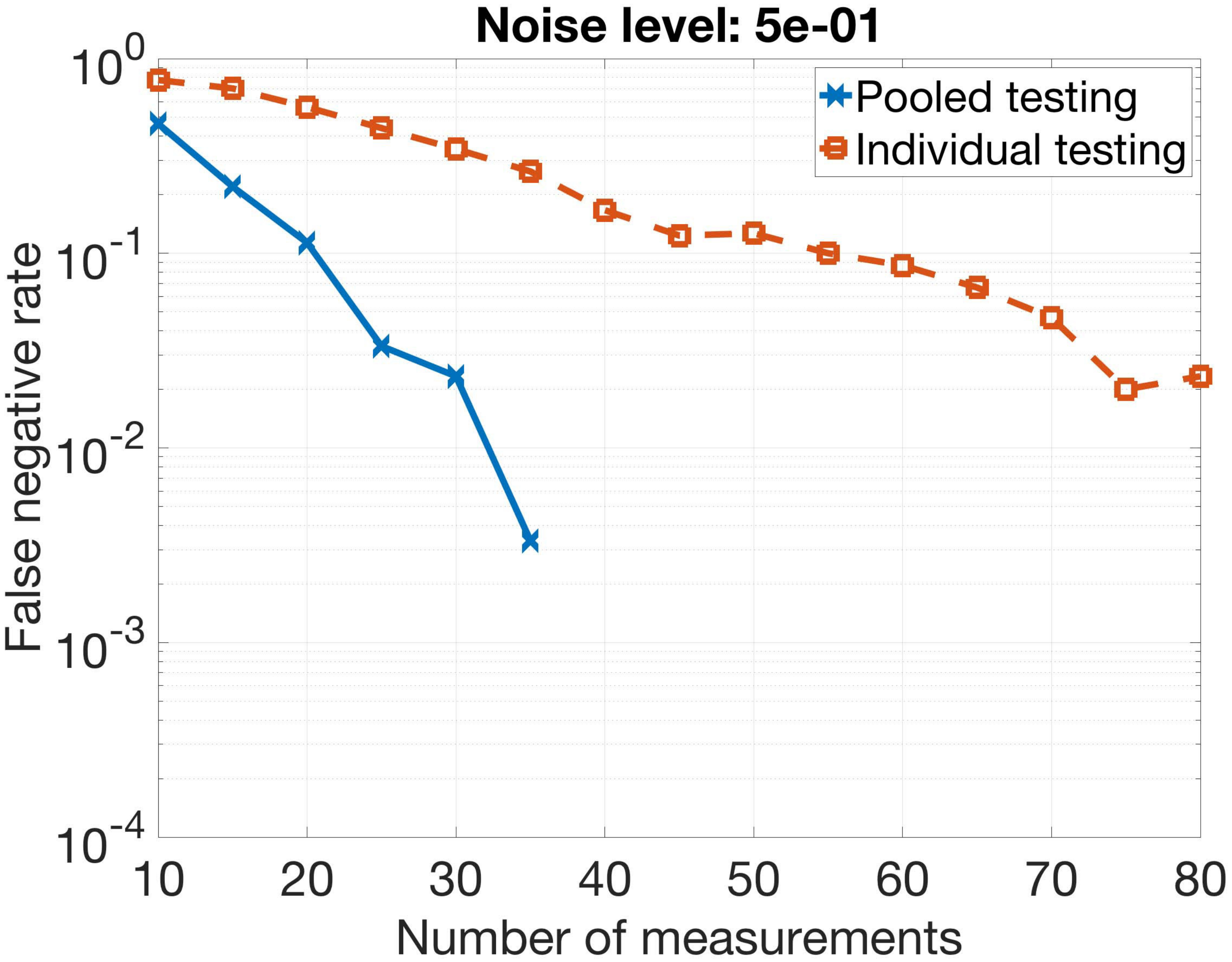}}
   \subfloat[FNR (Noise level: 1e0)]{\includegraphics[scale=0.14]{img/FNR_N40_K3_Bernoulli_Optimization_GaussianTrunc_Pout15e-02_Noise1e00-eps-converted-to}}
   \subfloat[FNR (Noise level: 2e0)]{\includegraphics[scale=0.14]{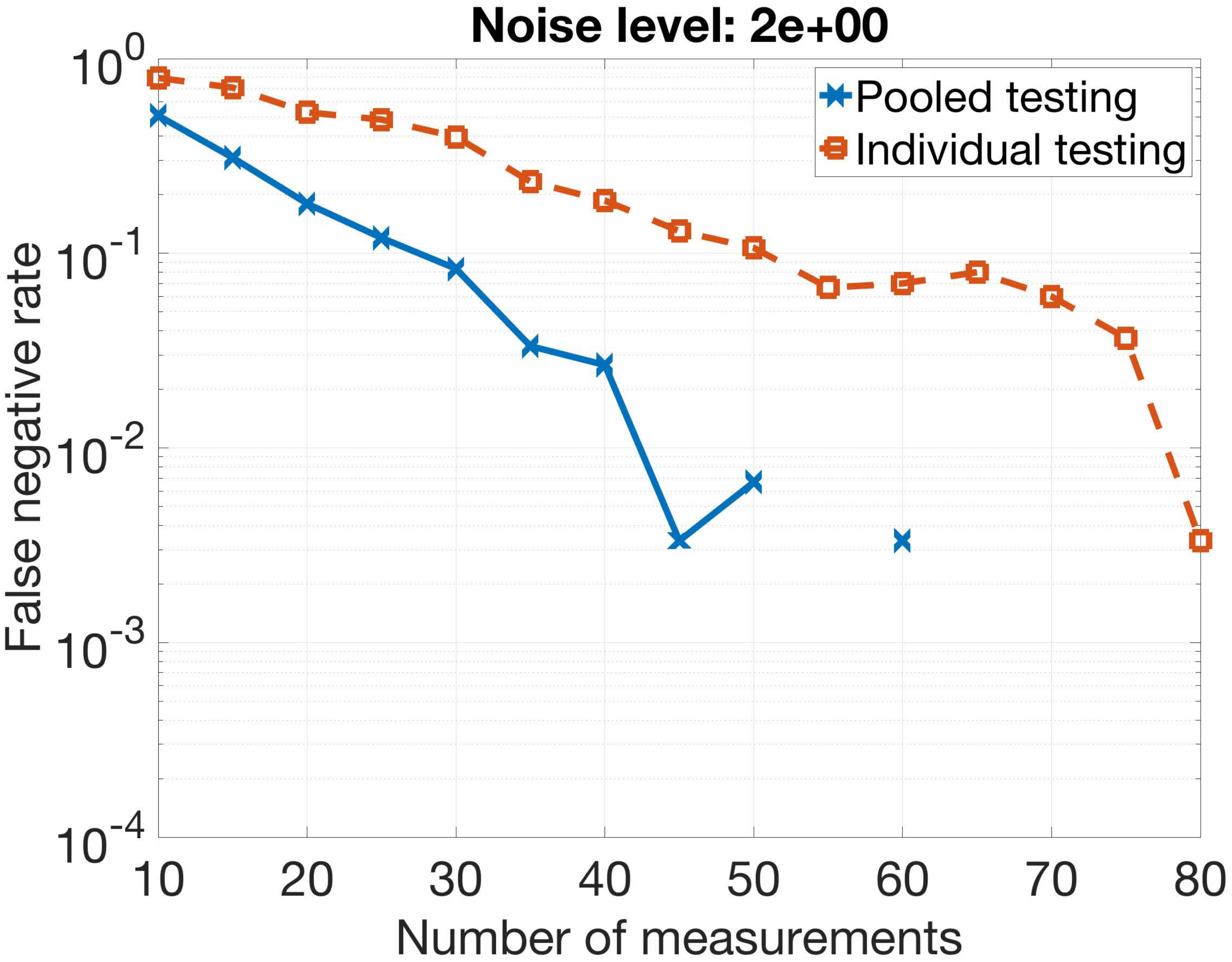}} \\
   \subfloat[FPR (Noise level: 5e-1)]{\includegraphics[scale=0.14]{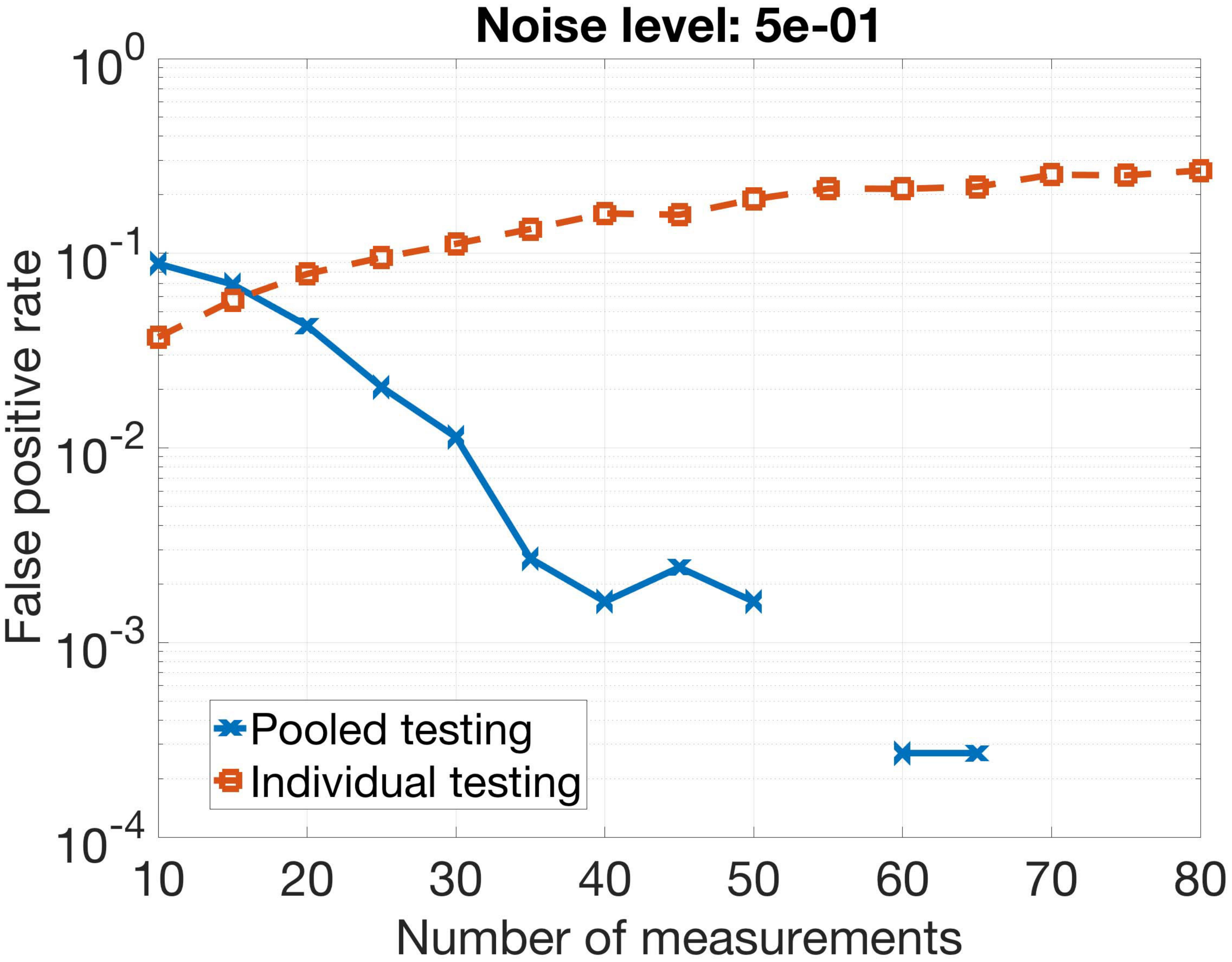}}
   \subfloat[FPR (Noise level: 1e0)]{\includegraphics[scale=0.14]{img/FPR_N40_K3_Bernoulli_Optimization_GaussianTrunc_Pout15e-02_Noise1e00-eps-converted-to}}
   \subfloat[FPR (Noise level: 2e0)]{\includegraphics[scale=0.14]{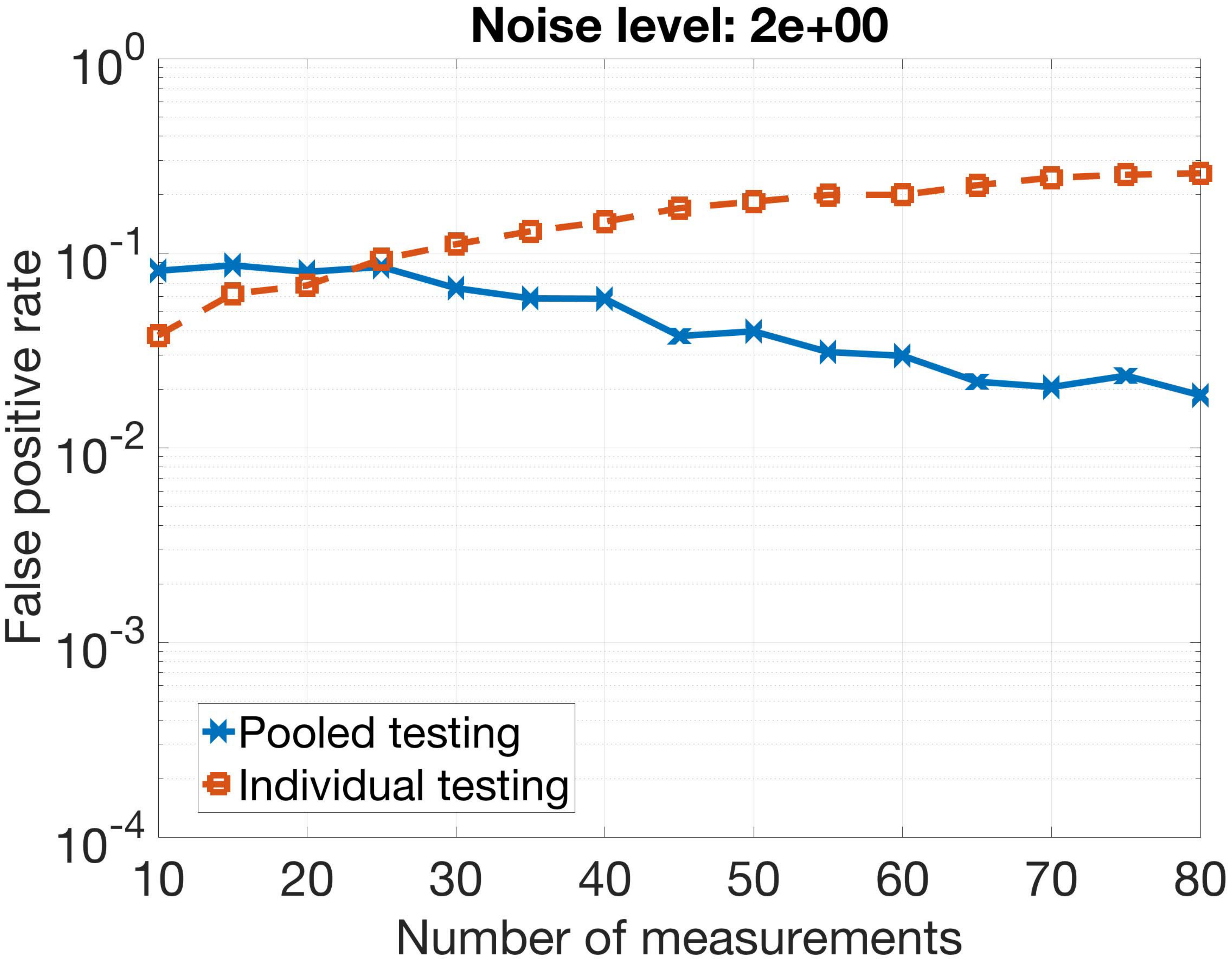}}
    \caption{\small Simulations for different noise levels. False Negative Rate (FNR) and the corresponding False Positive Rate (FPR) with $n=40$, $k=3$, $\Pb_{out}=0.15$, and noise level varied from 5e-1 to 2e0.}
    \label{fig:FNR_FPR_N40_K3_Pout15-02}
\end{figure*}

In order to check the impact of Gaussian noises on the detection performance, we further run simulations by varying noise levels. We vary the Gaussian noise level from 5e-1 to 2e0. We randomly choose 100 trials and record the FNR and the FPR of the pooled testing and the individual testing. Here in the simulations, we set the sparsity level to $3$, i.e., $k=3$, and consider the two probability of outlier error $5\%$ and $15\%$. Figures \ref{fig:FNR_FPR_N25_K3_Pout5-02} and \ref{fig:FNR_FPR_N40_K3_Pout5-02} illustrate the simulation results in log-scale with $\Pb_{out} = 0.05$ when $n=25$ and $n=40$ respectively. In addition, Figures \ref{fig:FNR_FPR_N25_K3_Pout15-02} and \ref{fig:FNR_FPR_N40_K3_Pout15-02}  show the simulation results in log-scale with $\Pb_{out}=0.15$ when $n=25$ and $n=40$ respectively. In those figures, (a) and (d) are for noise level $5e-1$, and (b) and (e) are for noise level $1e0$, and (c) and (f) are for noise level $2e0$. As shown in the figures, as the noise level increases, both FNR and FPR of the pooled testing and the individual testing become worse. However, the pooled testing still outperforms the individual testing with various noise levels in term of the FNR for every examined value of $m$, and in term of the FPR for $m \geq n$.

\subsection{Different sparsity levels}
\begin{figure*}[!htb]
    \centering
   \subfloat[FNR ($k=1$)]{\includegraphics[scale=0.14]{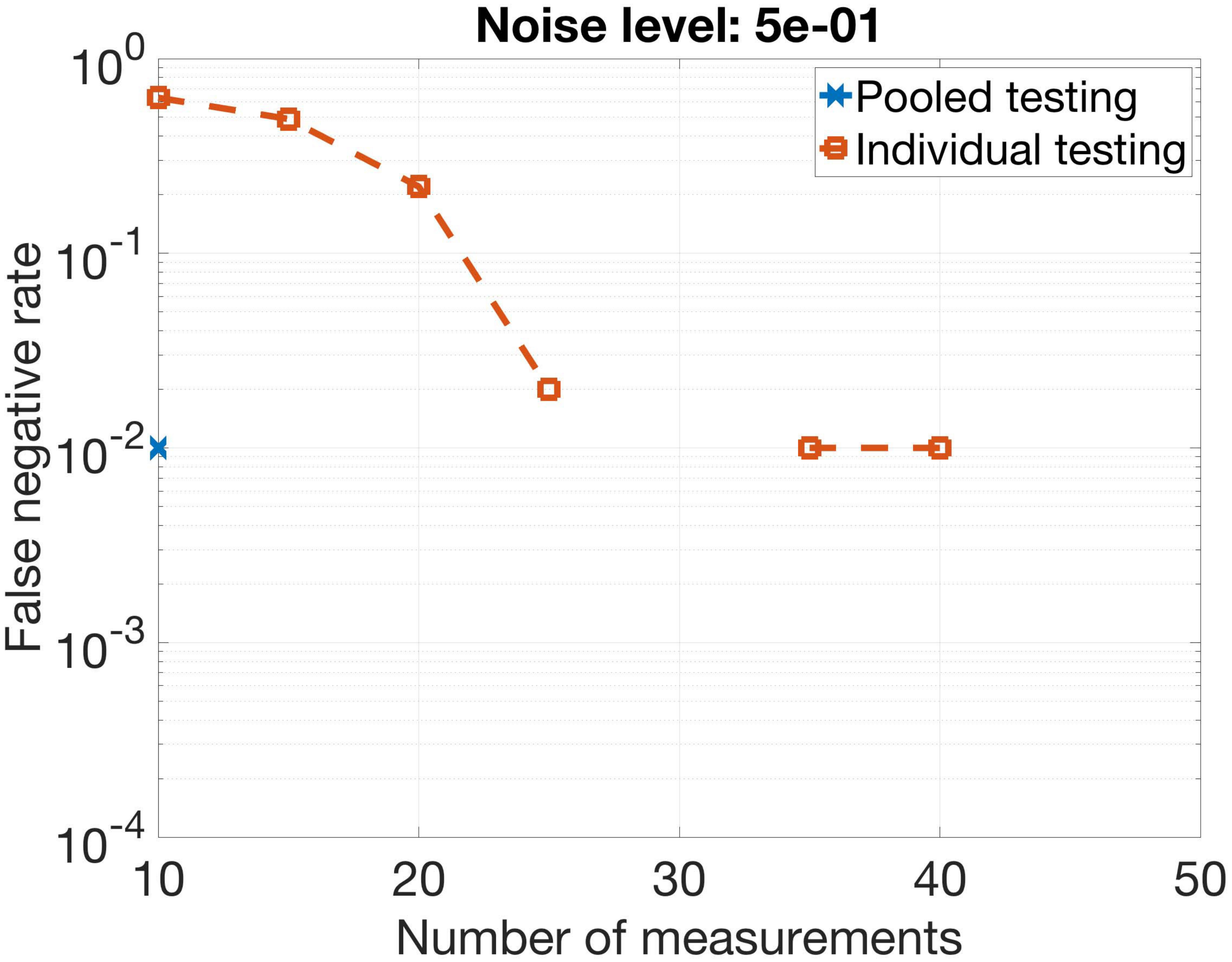}}
   \subfloat[FNR ($k=3$)]{\includegraphics[scale=0.14]{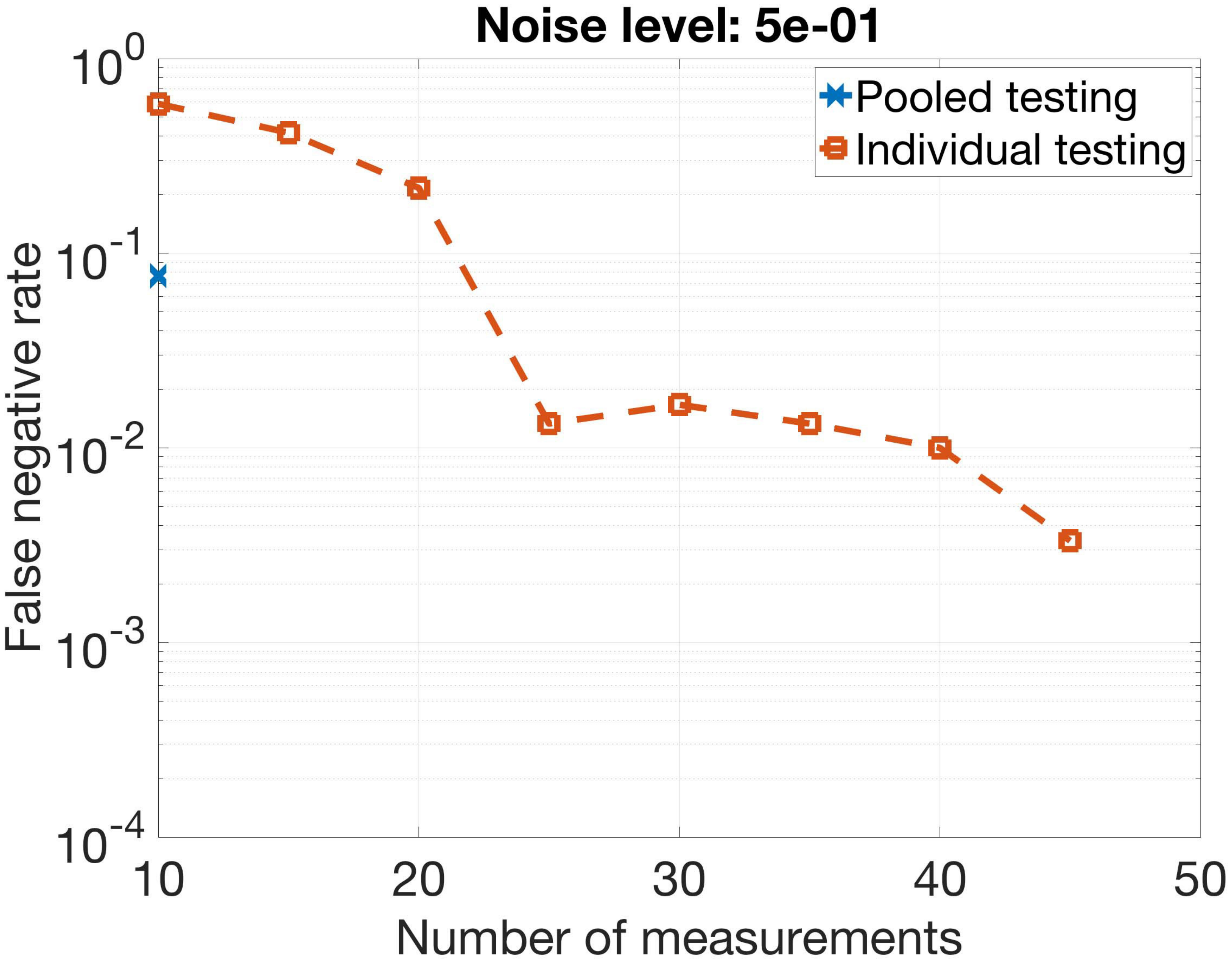}}
   \subfloat[FNR ($k=6$)]{\includegraphics[scale=0.14]{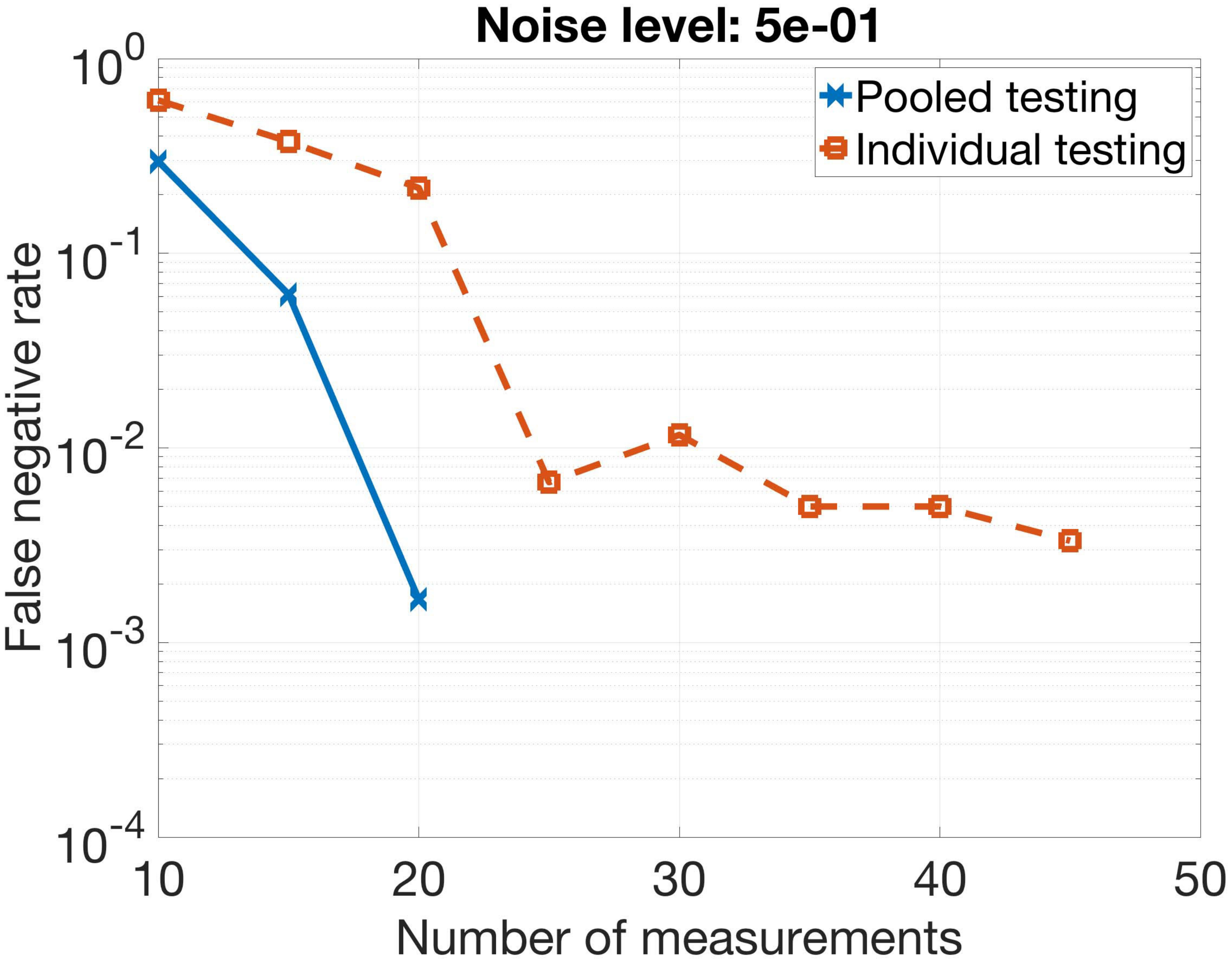}} \\
   \subfloat[FPR ($k=1$)]{\includegraphics[scale=0.14]{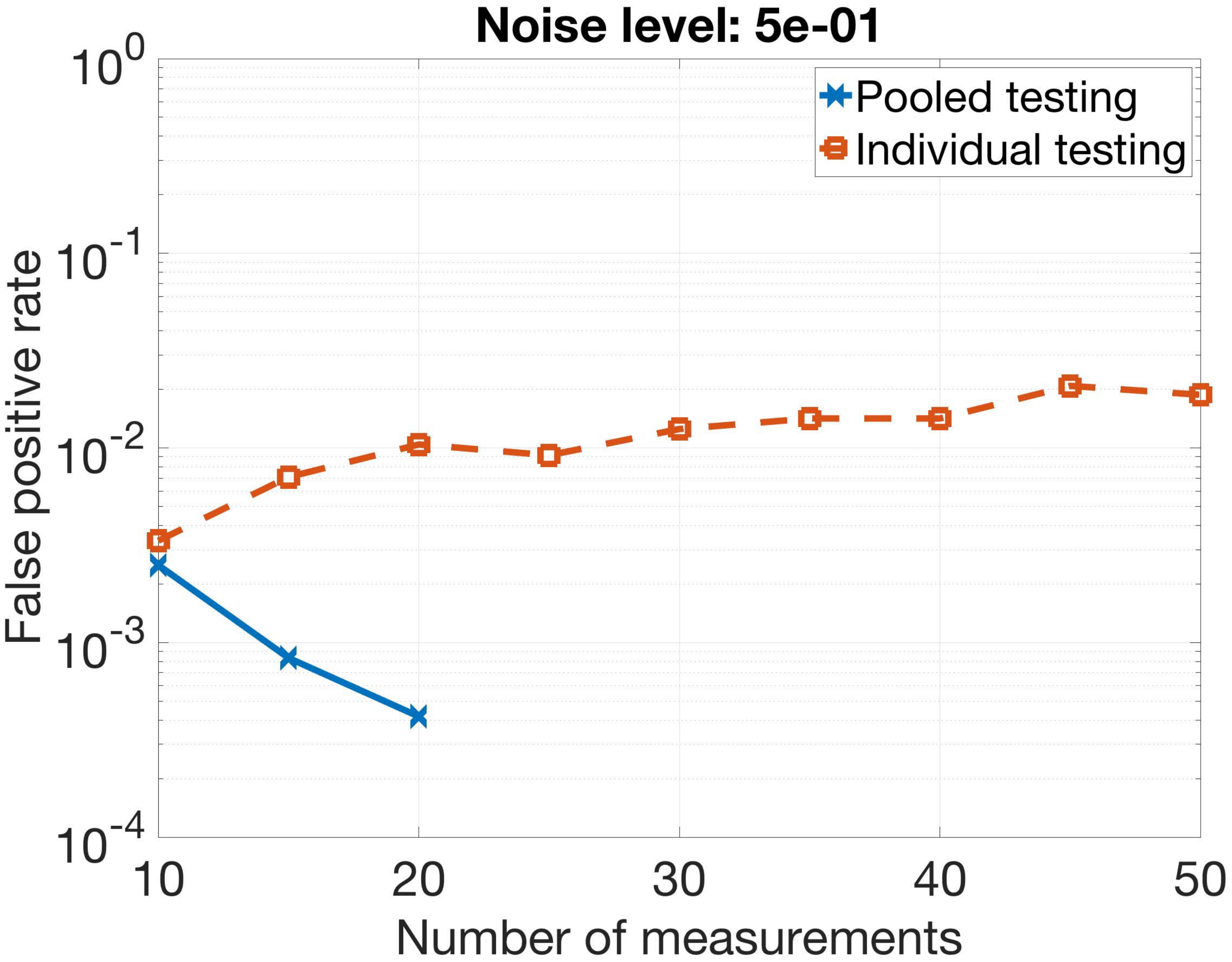}}
   \subfloat[FPR ($k=3$)]{\includegraphics[scale=0.14]{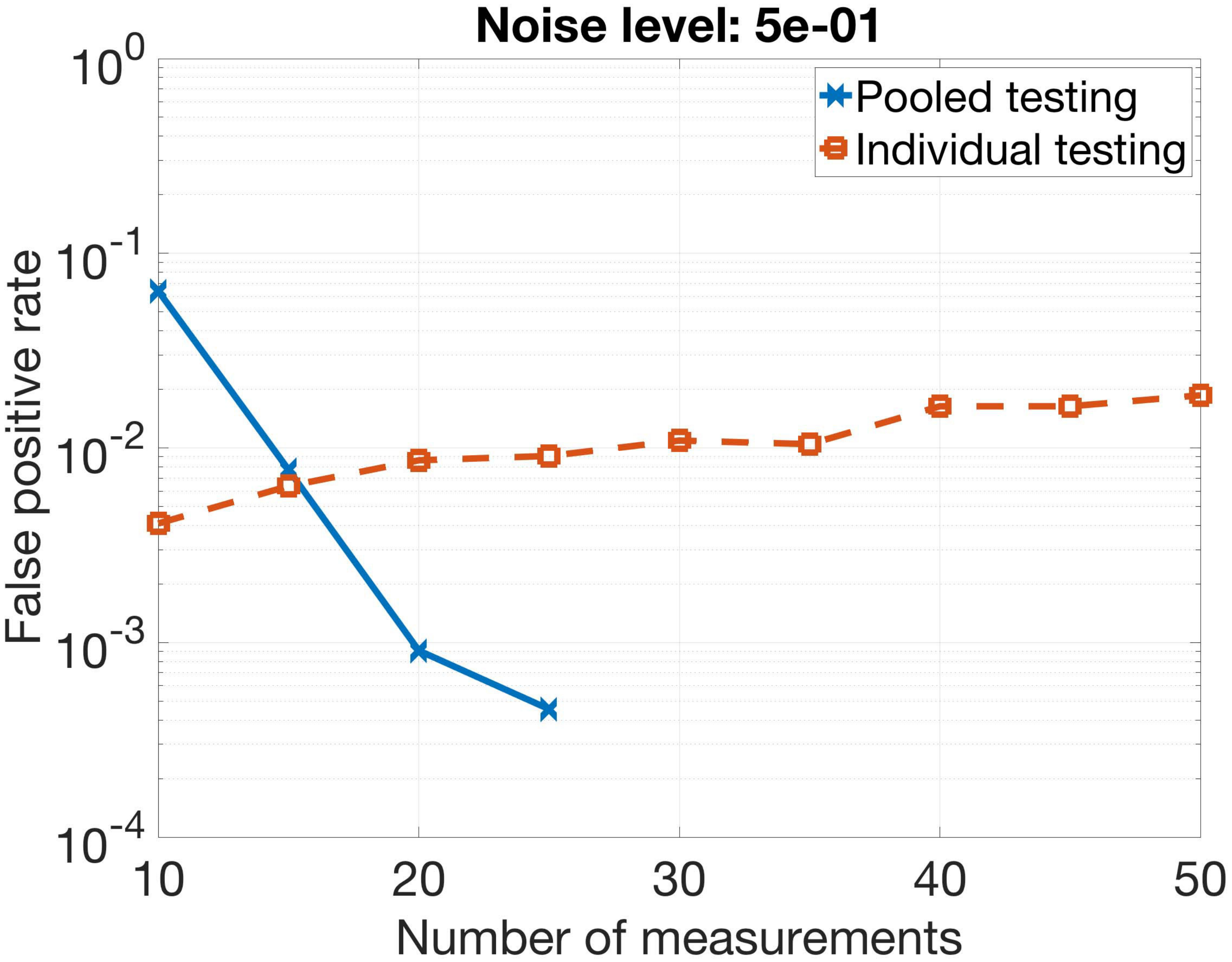}}
   \subfloat[FPR ($k=6$)]{\includegraphics[scale=0.14]{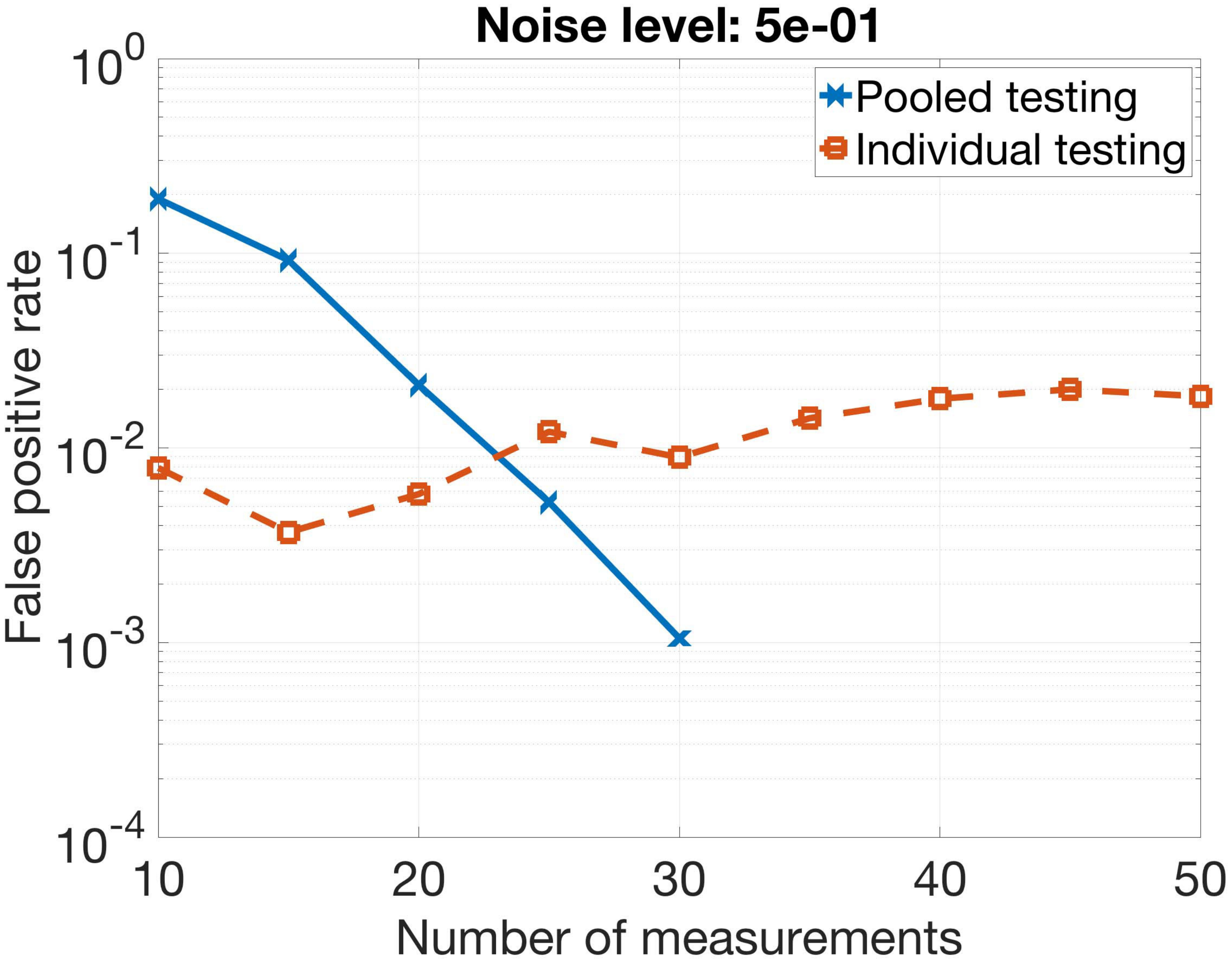}}
    \caption{\small Simulations for different sparsity levels. False Negative Rate (FNR) and the corresponding False Positive Rate (FPR) with $n=25$, $\Pb_{out}=0.01$, and noise level $5e-1$, and $k$ varied from $1$ to $6$.}
    \label{fig:FNR_FPR_N25_Pout1-02_Noise5-01}
\end{figure*}

\begin{figure*}[!htb]
    \centering
   \subfloat[FNR ($k=1$)]{\includegraphics[scale=0.14]{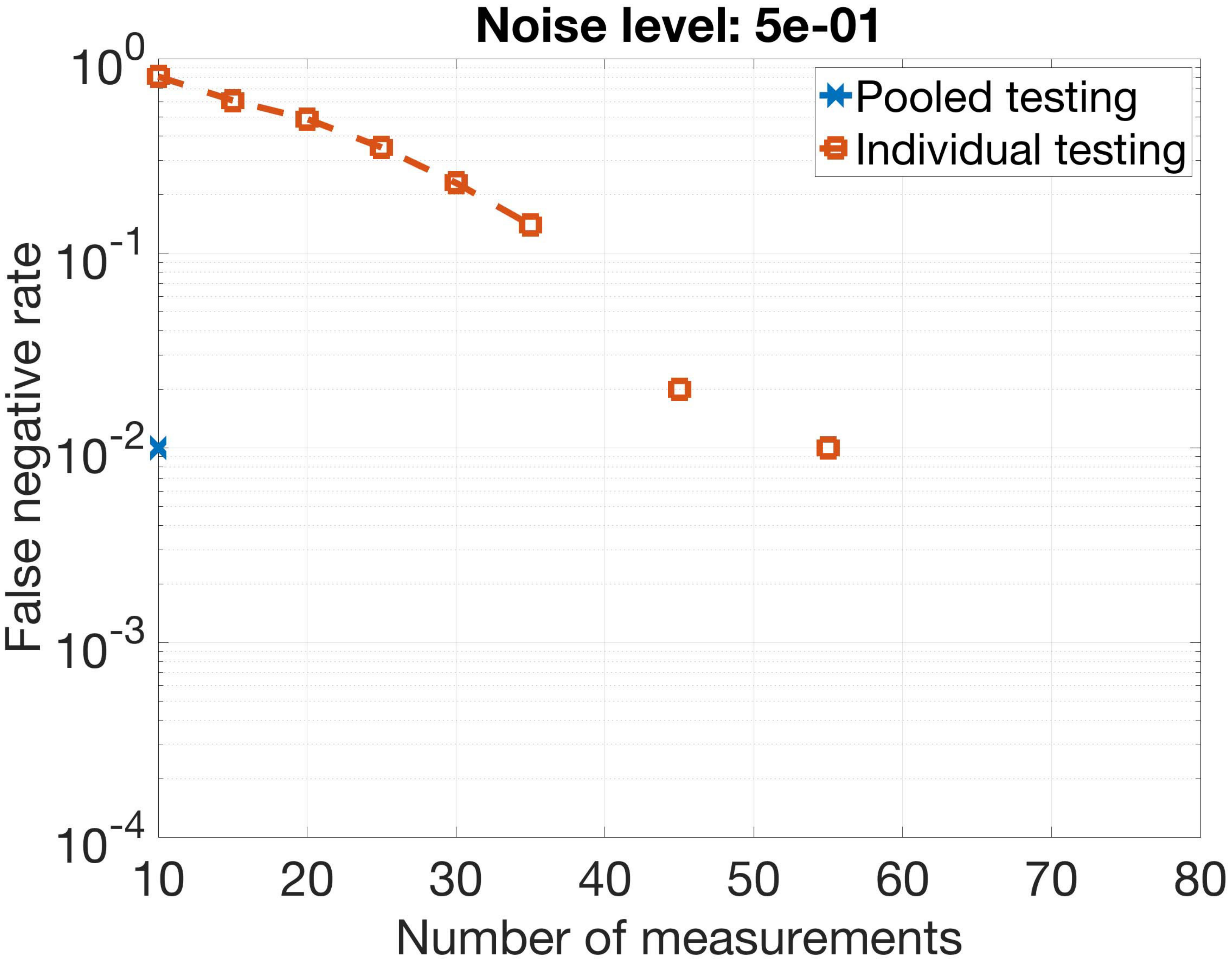}}
   \subfloat[FNR ($k=3$)]{\includegraphics[scale=0.14]{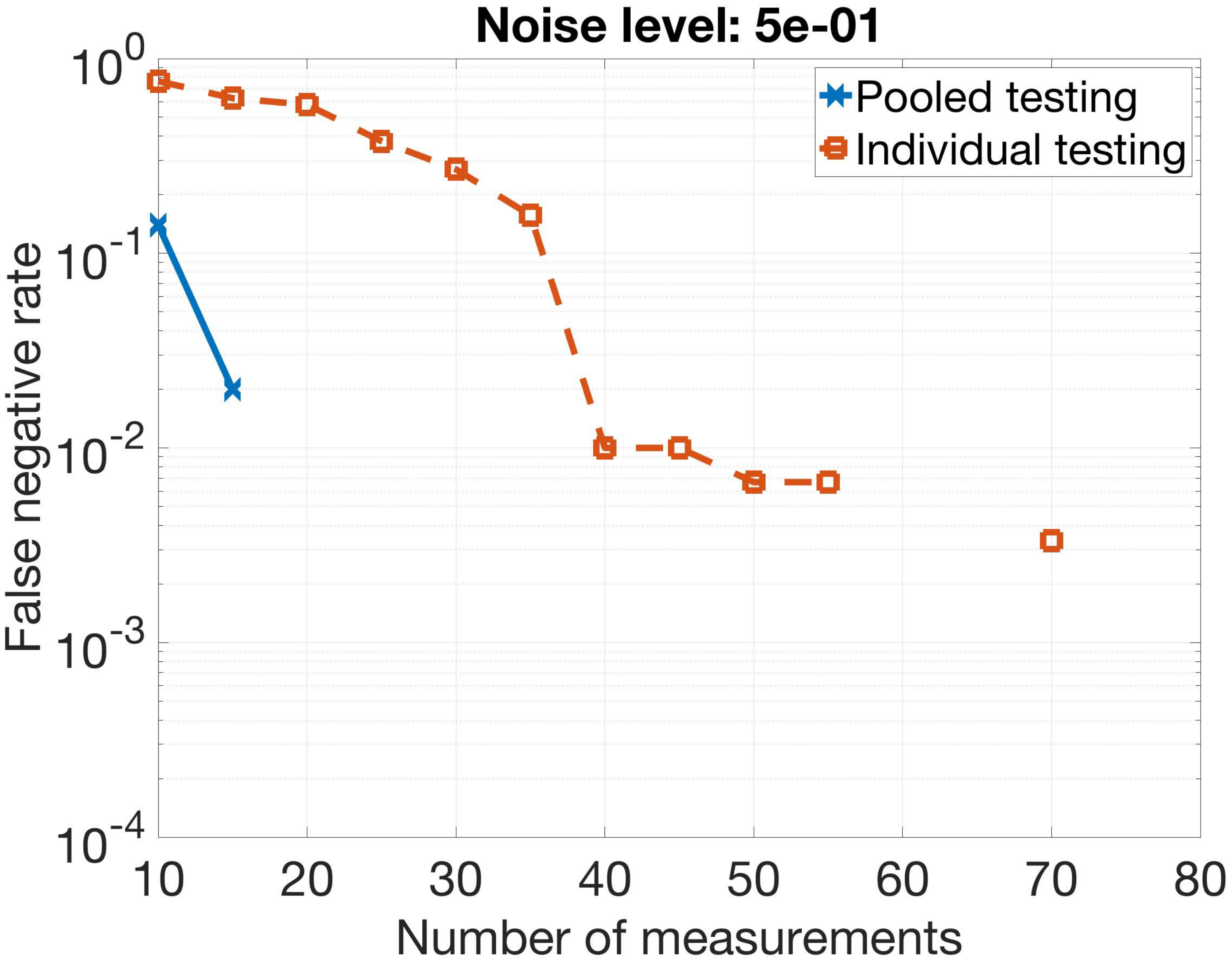}}
   \subfloat[FNR ($k=6$)]{\includegraphics[scale=0.14]{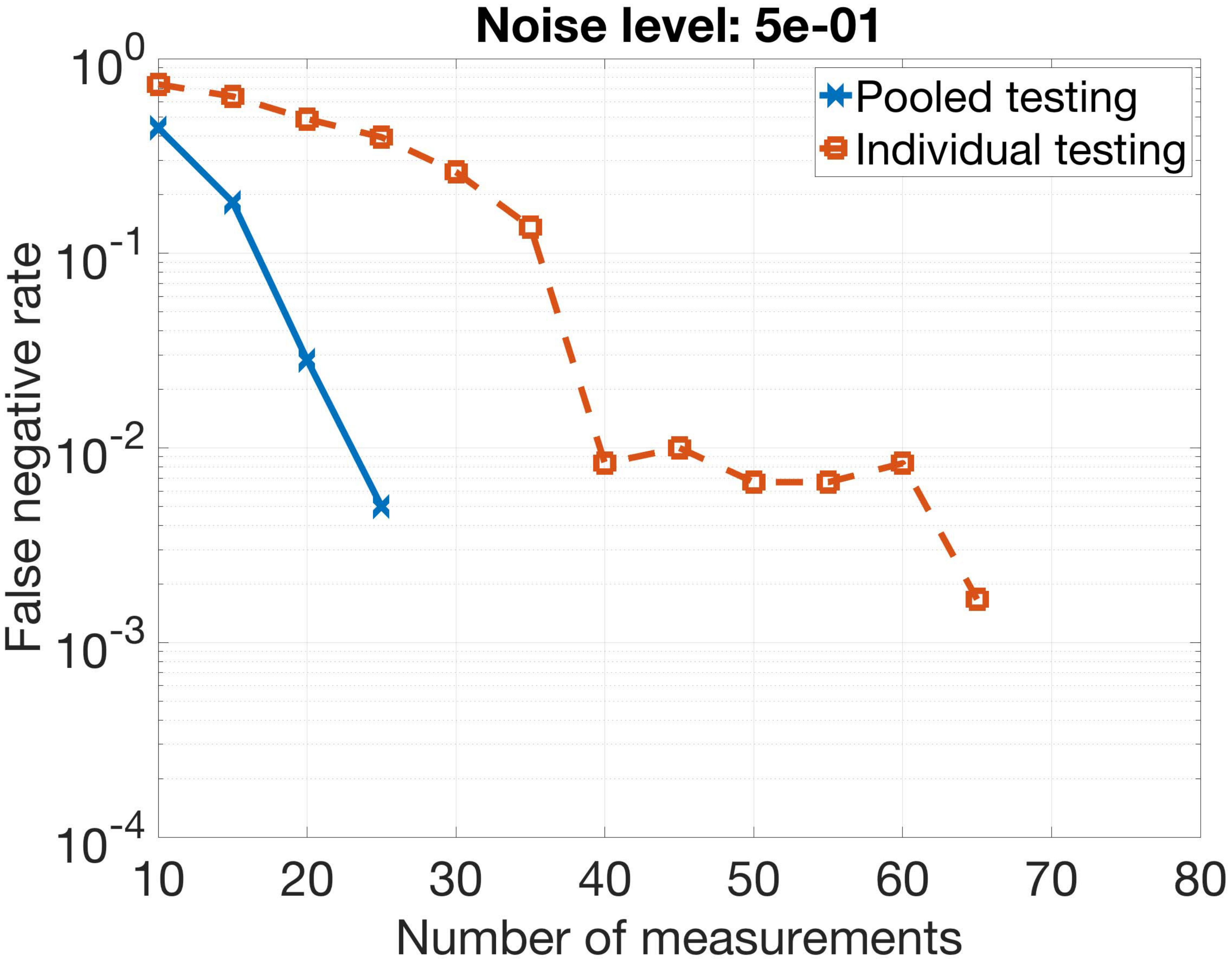}} \\
   \subfloat[FPR ($k=1$)]{\includegraphics[scale=0.14]{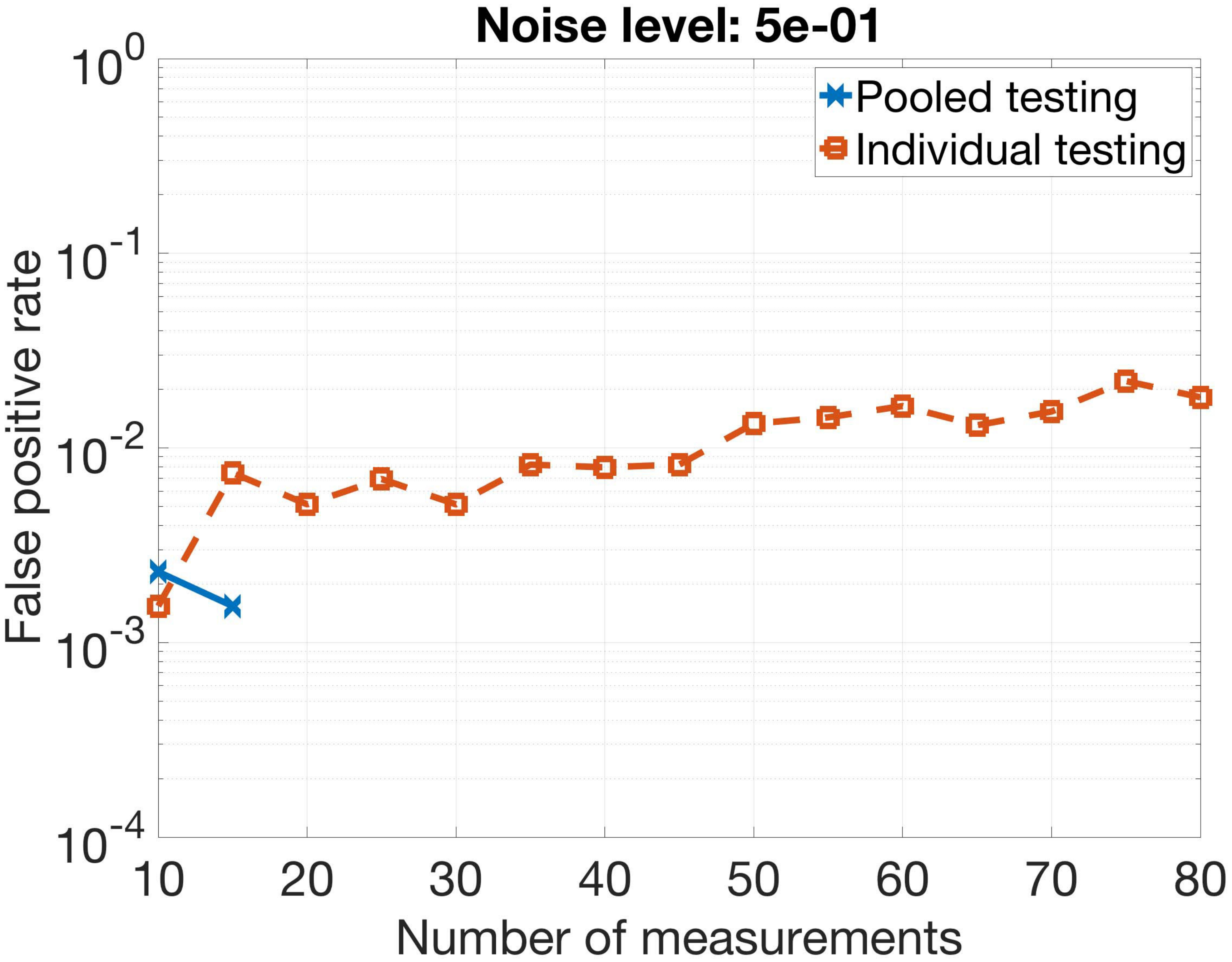}}
   \subfloat[FPR ($k=3$)]{\includegraphics[scale=0.14]{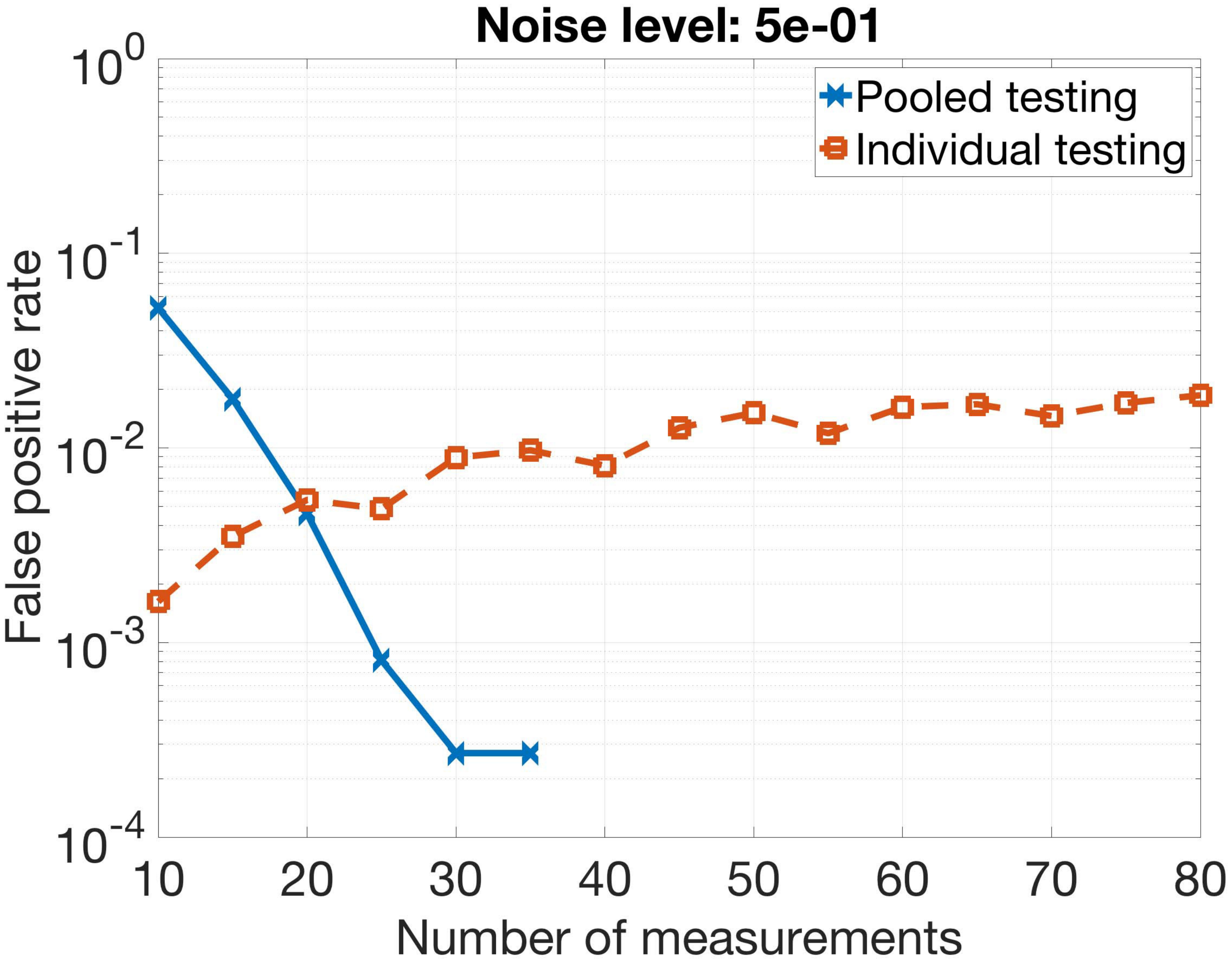}}
   \subfloat[FPR ($k=6$)]{\includegraphics[scale=0.14]{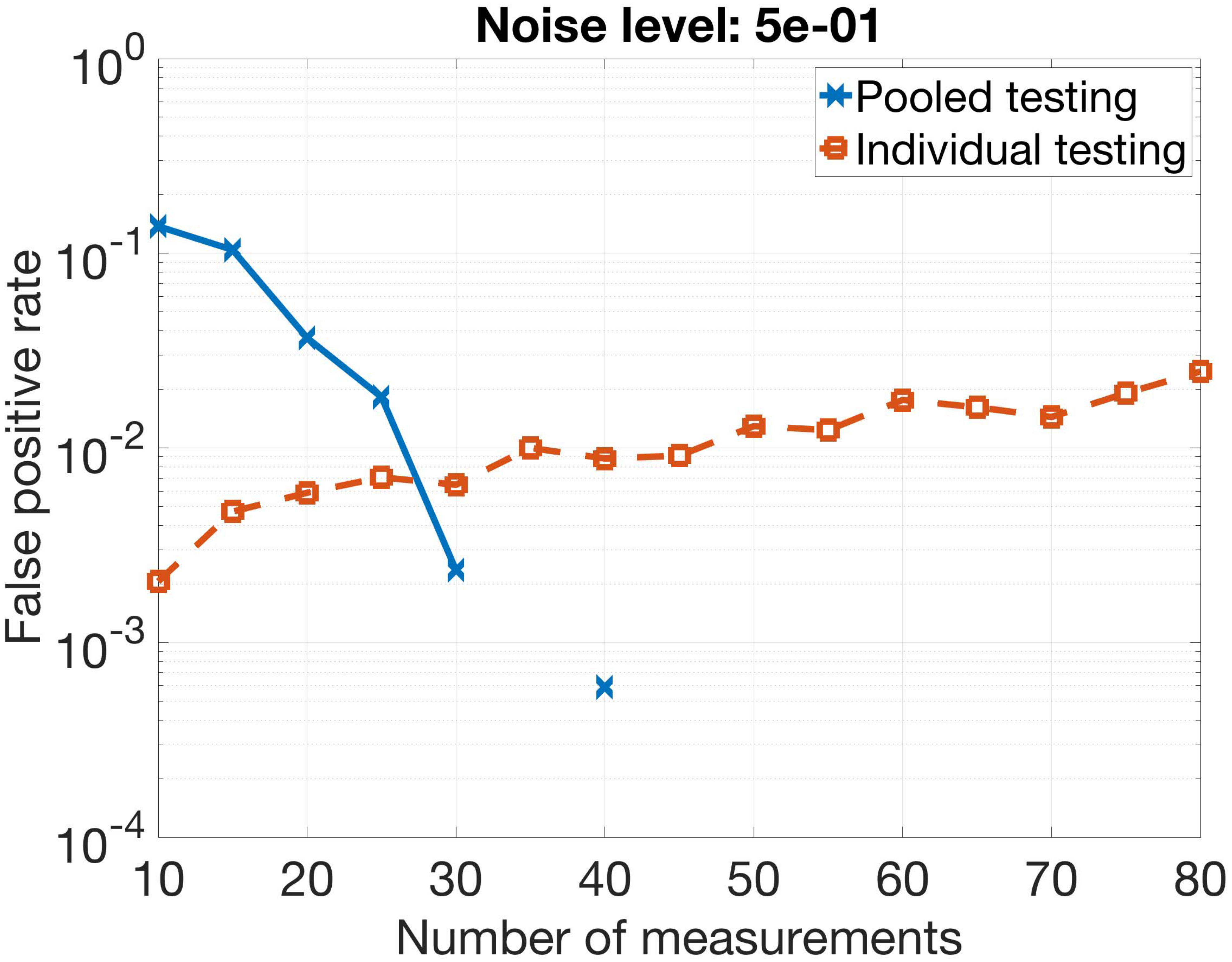}}
    \caption{\small Simulations for different sparsity levels. False Negative Rate (FNR) and the corresponding False Positive Rate (FPR) with $n=40$, $\Pb_{out}=0.01$, and noise level $5e-1$, and $k$ varied from $1$ to $6$.}
    \label{fig:FNR_FPR_N40_Pout1-02_Noise5-01}
\end{figure*}

In this subsection, we further run simulations by varying the sparsity level, i.e., the number of people having COVID-19 viruses. For these simulations, we set the noise level to $5e-1$ and the probability of outlier error $\Pb_{out}$ to $0.01$. We vary the sparsity level $k$ from 1 to 6. Figures \ref{fig:FNR_FPR_N25_Pout1-02_Noise5-01} and \ref{fig:FNR_FPR_N40_Pout1-02_Noise5-01} show the FNR and FPR of both the pooled testing and individual testing with different sparsity level when $n=25$ and $n=40$ respectively.

\subsection{Discussion}

The overall takeaway from the Figures \ref{fig:FNR_FPR_N40_K1} to \ref{fig:FNR_FPR_N40_K6} is that  the pooled sampling method achieves significantly higher accuracy compared to individual testing. Also in absolute terms, the pooled sampling method is able to provide accurate diagnostic results even when individual test results are highly noisy.  Some specific observations from the simulations are as follows.
\begin{itemize}

\item In most of the simulations, the pooled sampling method simultaneously achieves lower FPR and FNR than individual sampling. {\it We did not observe even a single instance when the opposite was true i.e. where individual testing outperformed the pooled sampling method in both FPR and FNR.}

\item The FPR for the individual sampling method actually gets worse with increased number of measurements. This is simply an artifact of the individual testing method in conservative strategy in order to prevent miss in COVID-19 positive case. The overall accuracy of the individual testing method does always improve with increased number of measurements when FNR is taken into account along with FPR.

\item For the pooled sampling method, both FPR and FNR always monotonically decrease with increased number of measurements. (The apparent non-monotonicity in e.g. Fig. \ref{fig:FNR_FPR_N25_Pout1-02_Noise5-01}(f) is simply an artifact of the randomness in the simulations.)

\end{itemize}

\clearpage
\bibliography{01Ref_CS_Virus_Testing}
\bibliographystyle{IEEEtran}




\end{document}